\documentclass[twocolumn]{aastex63}

\newcommand{\twco}{$^{12}$CO\ } 
\newcommand{\thco}{$^{13}$CO\ } 
\newcommand{\ceighto}{C$^{18}$O\ } 
\newcommand{\msun}{M$_\odot$}
\newcommand{\msunsp}{M$_\odot$\ }

\usepackage{amsmath}

\received{}
\revised{}
\accepted{}
\submitjournal{ApJ}

\shorttitle{$^{13}$CO(2-1) in Andromeda's GMCs}
\shortauthors{Viaene et al.}
\graphicspath{{./}{figures/}}

\begin{document}

\title{Simultaneous Deep Measurements of CO isotopologues and Dust Emission in Giant Molecular Clouds in the Andromeda Galaxy}

\correspondingauthor{Jan Forbrich}
\email{j.forbrich@herts.ac.uk}

\author{Sébastien Viaene}
\affiliation{Sterrenkundig Observatorium, Universiteit Gent, Krijgslaan 281, 9000, Gent, Belgium}

\author{Jan Forbrich}
\affiliation{Centre for Astrophysics Research, University of Hertfordshire, College Lane, Hatfield AL10 9AB, UK}
\affiliation{Center for Astrophysics | Harvard \& Smithsonian, 60 Garden St, MS 72, Cambridge, MA 02138, USA}

\author{Charles J. Lada}
\affiliation{Center for Astrophysics | Harvard \& Smithsonian, 60 Garden St, MS 72, Cambridge, MA 02138, USA}

\author{Glen Petitpas}
\affiliation{Center for Astrophysics | Harvard \& Smithsonian, 60 Garden St, MS 72, Cambridge, MA 02138, USA}

\author{Christopher Faesi}
\affiliation{Department of Astronomy, University of Massachusetts Amherst, 710 North Pleasant Street, Amherst, MA 01003, USA}



\begin{abstract}

We present simultaneous measurements of emission from dust continuum at 230 GHz and the J=2-1 \twco, \thco and \ceighto isotopologues at $\sim$ 15 pc resolution from individual Giant Molecular Clouds (GMCs) in the Andromeda galaxy (M31). These observations were obtained in an ongoing survey of this galaxy being conducted with the Submillimeter Array (SMA). Initial results describing the continuum and \twco emission were published earlier. Here we primarily analyze the observations of  \thco and \ceighto emission and compare them to the measurements of dust continuum and \twco emission. We also report additional  dust continuum and CO measurements from newly added GMCs to the M31 sample. We detect spatially resolved \thco emission with high signal-to-noise in 31 objects. We find the extent of the \thco emission to be  nearly comparable to that of \twco, typically covering 75\% of the area of the \twco emission. We derive \thco and \ceighto  abundances of 2.9 $\times 10^{-6}$ and 4.4 $\times 10^{-7}$ relative to H$_2$, respectively, by comparison with hydrogen column densities of the same regions derived from the dust continuum observations assuming a Milky Way gas-to-dust ratio. We find the isotopic abundance ratio [\thco]/[\ceighto] = 6.7$\pm$2.9 to be consistent with the Milky Way value (8.1). Finally, we derive the mass-to-light conversion factors for all three CO species to be $\alpha_{12} = 8.7 \pm 3.9$, $\alpha_{13} = 48.9 \pm 20.4$ and $\alpha_{18} = 345^{+25}_{-31}$ M$_\odot$ (K km s$^{-1}$pc$^2$)$^{-1}$ for the J=2-1 transitions of \twco, \thco and \ceighto, respectively.   

\end{abstract}




\section{Introduction} \label{sec:intro}

The evolution of galaxies is driven by an ongoing conversion of gas into stars. The process of star formation primarily happens in Giant Molecular Clouds (GMCs) located in galactic discs. While the various steps in this process are qualitatively understood, a quantitative theory of star formation remains elusive. A key challenge confronting star formation research is the accurate determination of the physical properties of the GMCs which spawn stars. 

Tracing and measuring the physical properties of star forming molecular gas is not trivial. Molecular hydrogen is by far the most abundant molecule in GMCs, but unfortunately does not generate easily measurable emission at the temperatures and densities prevalent in these GMCs. The next most abundant molecule, carbon monoxide, emits the strongest emission lines in the cold GMCs  and consequently its lowest lying transitions are used as a proxy for H$_2$ to map molecular gas from the most nearby Galactic GMCs  to high redshift galaxies. 
Already the first resolved CO observations of extragalactic GMCs (in M\,31, as it happens) were carried out more than 30 years ago at resolutions of $\sim25-50$~pc \citep{vog87,lad88}, which is a resolution range that has become available for systematic studies (e.g., \citealp{sch13, hir18}). Recent GMC studies with ALMA in nearby galaxies have routinely reached resolutions of 10~pc or better (e.g., \citealp{sch17,fae18}).
However, uncertainties resulting from variations in molecular abundances, opacities and excitation have impeded use of molecular-line observations for making accurate measurements of basic GMC properties such as size, structure and mass using molecular-line observations. 
Observations of dust extinction and emission have been shown to be the gold standard for tracing H$_2$ and measuring many of the physical properties of GMCs in the local Milky Way \citep[e.g.,][]{2001Natur.409..159A,2007prpl.conf....3L,2009ApJ...692...91G,2010A&A...519L...7L}. However, until now, such observations have been mostly limited to the local Milky Way.

In a previous paper we reported the first detection of resolved continuum emission from dust within individual GMCs in M31, the Andromeda galaxy (\citealt{Forbrich2020}, hereafter Paper I). These novel results were obtained as part of an ongoing large program using deep interferometric observations made with the Submillimeter Array (SMA) utilizing its wide-band continuum capability at 230 GHz. Simultaneous observations of CO(2-1) emission (with identical {\it u,v} coverage) provided the first direct measurements of the CO conversion factor, $\alpha_{CO}$, within individual GMCs across this galaxy. The measured values were found to be similar to those of Milky Way GMCs and did not appear to vary significantly across the disk of M31. 

The depth of these SMA observations was determined by the need to detect dust continuum emission from individual GMCs with good signal-to-noise. The sensitivity achieved in these observations also resulted in the very high signal-to-noise detection of \twco emission  at high velocity resolution. In most GMCs  the dynamic range in the \twco lines exceeded two orders of magnitude. 
Moreover, these observations were deep enough to enable  detection of  the rare \thco and \ceighto  isotopologues as well. 
In this paper, we present these latter observations and report the first resolved, high signal-to-noise, detection of extended  \thco emission from individual GMCs across the Andromeda galaxy. We also describe the first detection of \ceighto emission at $\sim$ 15 pc spatial scales in M31 GMCs. Furthermore we present additional measurements of dust continuum and \twco emission from our continuously growing sample of GMCs in the Andromeda galaxy. This includes both cases where we have since obtained better data when compared to what was available at the time of publication of our previous paper and new targets.

The simultaneous detection of emission from dust and the rarer isotopologues of CO enables us to  directly measure the \thco and \ceighto mass conversion factors in individual GMCs across this galaxy. Comparison of velocity resolved emission spectra from the three CO isotopologues further permits estimates of line opacities and molecular column densities. Combining the derived column densities with the dust measurements allows us to produce determinations of the CO abundances within individual GMCs across M31.  

In Section 2 below we describe the observations, and in Section 3 we present the results of these observations. In Section 4 we report our determinations of the physical properties of the GMCs in M31 including the dust masses and the CO conversion factors for all three main isotopologues, as well as the derived CO abundances for the GMCs in our sample. In Section 5 we discuss some implications of our results and in Section 6 we summarize the primary findings of this paper.

\section{Observations}  \label{sec:data}

This present work is based on the data gathered in the first two seasons (fall 2018 and 2019) of the ongoing SMA large program targeting Andromeda (2017B-S075, PI: J. Forbrich). We provide a brief summary of the observations and data reduction here. For a full description of the observations we refer to Paper I. Briefly, the observations were obtained in the subcompact configuration using all eight antennas. This provides a beam size (FWHM) of 4.5$'' \times $3.8$''$ at 230 GHz, equivalent to a spatial resolution of $\sim15$~pc at the distance of M31. On a few occasions, one or two antennas were missing due to maintenance, degrading the beam size to 8$'' \times $5$''$. The spectrometer was configured to provide 32 GHz of continuous bandwidth between 213.55 and 245.55 GHz with a spectral resolution of 140.0 kHz per channel. Besides the continuum emission from our targeted sources the band also includes emission lines from the three most abundant isotopologues of CO.  We find that these lines can contribute between 10-50\% of the flux in a typical continuum measurement if we integrate the full band. Consequently channels containing the CO lines were removed from the band to obtain an independent CO free measurement of the dust continuum emission as well as separate measurements of the individual CO lines. Continuum contamination was found to be insignificant for the CO emission lines, hence no correction was made to remove continuum from the CO emission cubes.

Targets were selected from the galaxy-wide \textit{Herschel} catalog of Giant Molecular Associations (GMAs) compiled by \citet{Kirk2015}. Source selection was made to ensure ensuring that the sample is characterized by sufficient variation in measured brightness, dust masses, temperatures, emissivity indices $\beta$, and galactocentric radii.  The standard observing strategy amounts to a total of one full  observing  track ($\sim$ 6 hours) allocated per target, typically yielding a relatively uniform sensitivity of $0.25$ mJy or better.

The integration time for each target was set with the goal of detecting and resolving the dust emission at 1.3 mm for individual GMCs in Andromeda. We confirm the earlier finding of Paper I that the required sensitivity to detect these objects is indeed routinely achieved. The observations are sufficiently deep that the simultaneously detected \twco emission traces the full extent of a GMC and consequently we require a continuum detection to overlap with CO emission to be selected as candidate for cloud-related dust emission. 

Out of 25 pointings (15 of which were reported in Paper I), only 5 did not show any significant continuum signal overlapping with CO emission, while 9 show more than one such continuum peak\footnote{The strongest continuum peak in a field will be labeled "A" even if it is the only peak detected in a GMC.}. $^{12}$CO emission was detected in all 25 fields. For the purposes of this paper we will determine individually the properties of each continuum emission peak and refer to these peaks as clouds. Using this terminology a GMC can consist of one or more dust continuum peaks or clouds, since we define a GMC by the full spatial extent of its \twco emission which is always greater than that of the dust. The ten additional SMA observations reported here yield 10 new dust clouds. One cloud from Paper I, K134a, was detected at 2.8$\sigma$ in the present data set and did not meet the significance cut for analysis in this paper. In total, we can therefore amass a sample of 32 (dust) clouds  of which 10 are spatially resolved in continuum emission. These clouds are contained within 20 GMCs.

The raw spectral line data were first binned to a velocity resolution of $1.3$ km s$^{-1}$ and calibrated using the MIR IDL software for the SMA. The calibrated {\it uv}-data were exported to MIRIAD\footnote{\url{https://www.cfa.harvard.edu/sma/miriad}} where additional spectral flagging was performed before inverting (smoothing to 1.50 km s$^{-1}$) and CLEANing.  It was determined early in the project that natural weighting gave the best combination of sensitivity to the dust emission with minimal angular resolution loss. Specifically, the Briggs robust parameter was set to 2.0 for this entire sample of clouds to ensure a uniform data set. Since the continuum and CO data were calibrated and imaged with identical parameters, there should be effectively no calibration differences in the two data sets.

For this paper we make use of the dust continuum, velocity integrated CO emission maps, as well as extracted spectra. The line moment~0 maps were computed in a velocity interval determined by the $^{12}$CO line profiles. For our CO analysis we use integrated information from the moment~0 maps and line profiles spatially integrated from the cubes over a mask derived from the dust continuum map (see Paper I and Sect.~\ref{sec:results}). Appendix A  displays the \twco and \thco spectra integrated over these spatial masks for each of the 32 clouds considered here. The spectra are well resolved in velocity with velocity widths $\sim$ 4-6 km/s typical of individual Milky Way GMCs \citep[e.g.][]{2016ApJ...822...52R}.

\begin{figure*} 
\gridline{\fig{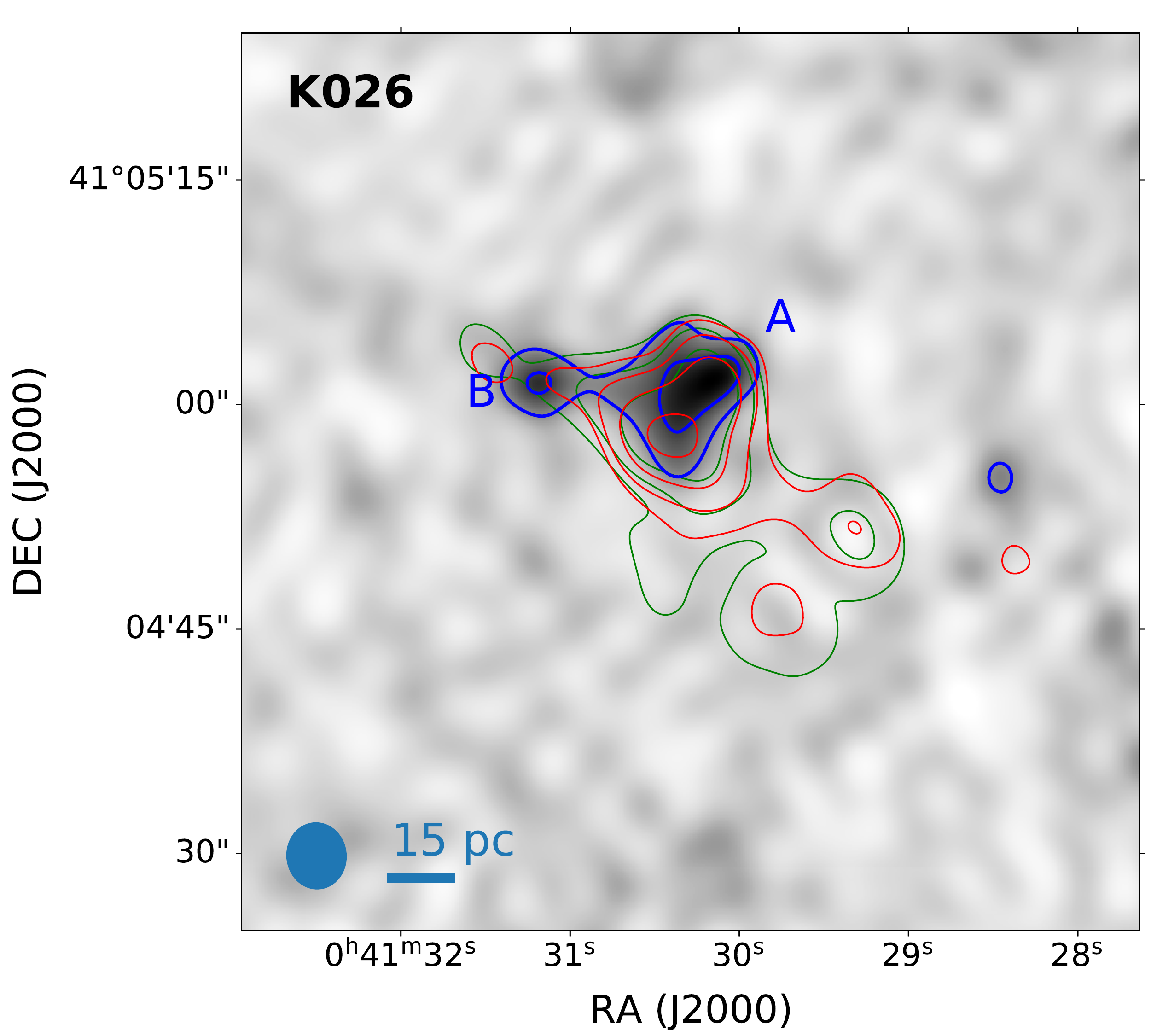}{0.33\textwidth}{}
          \fig{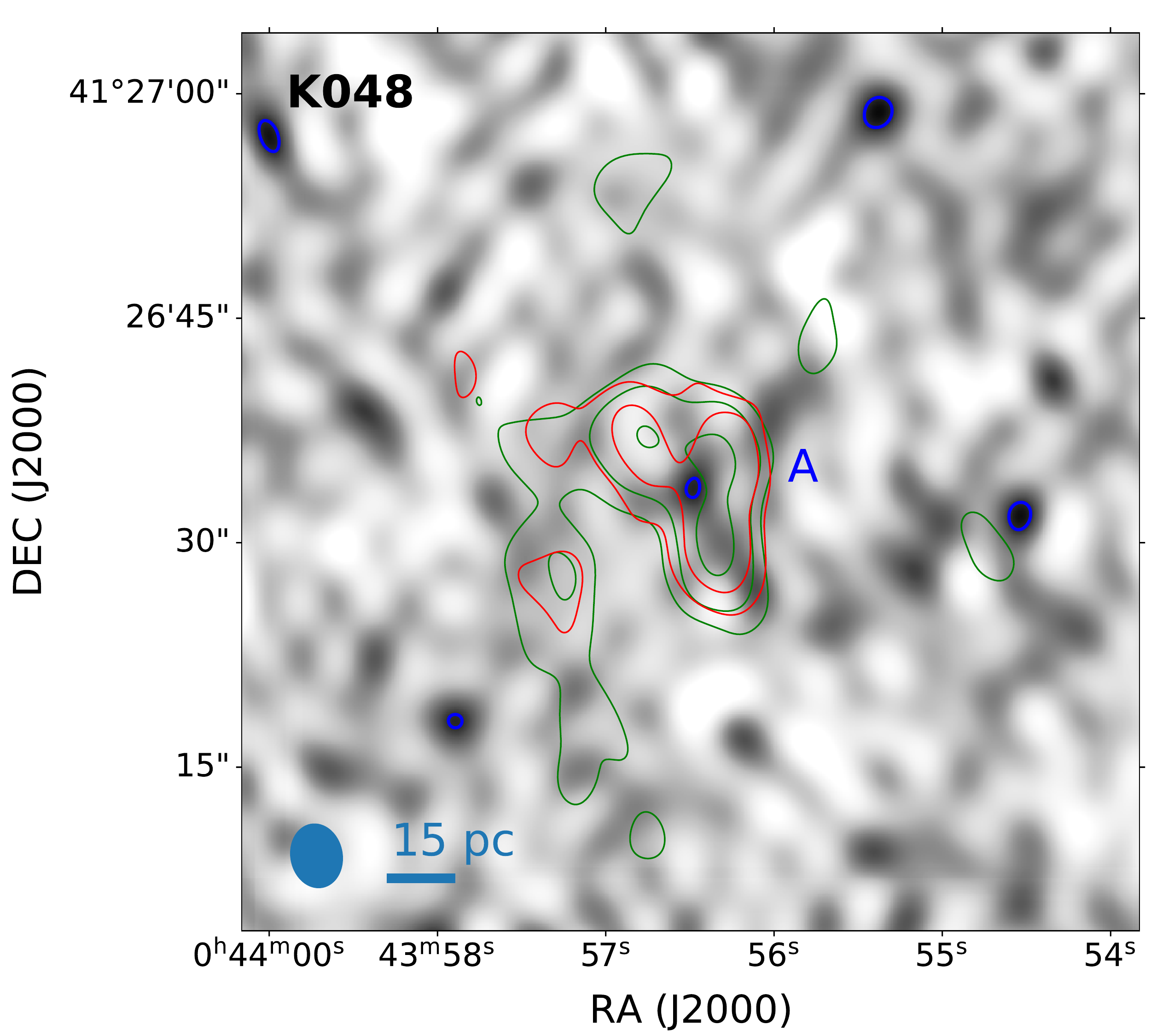}{0.33\textwidth}{}
          \fig{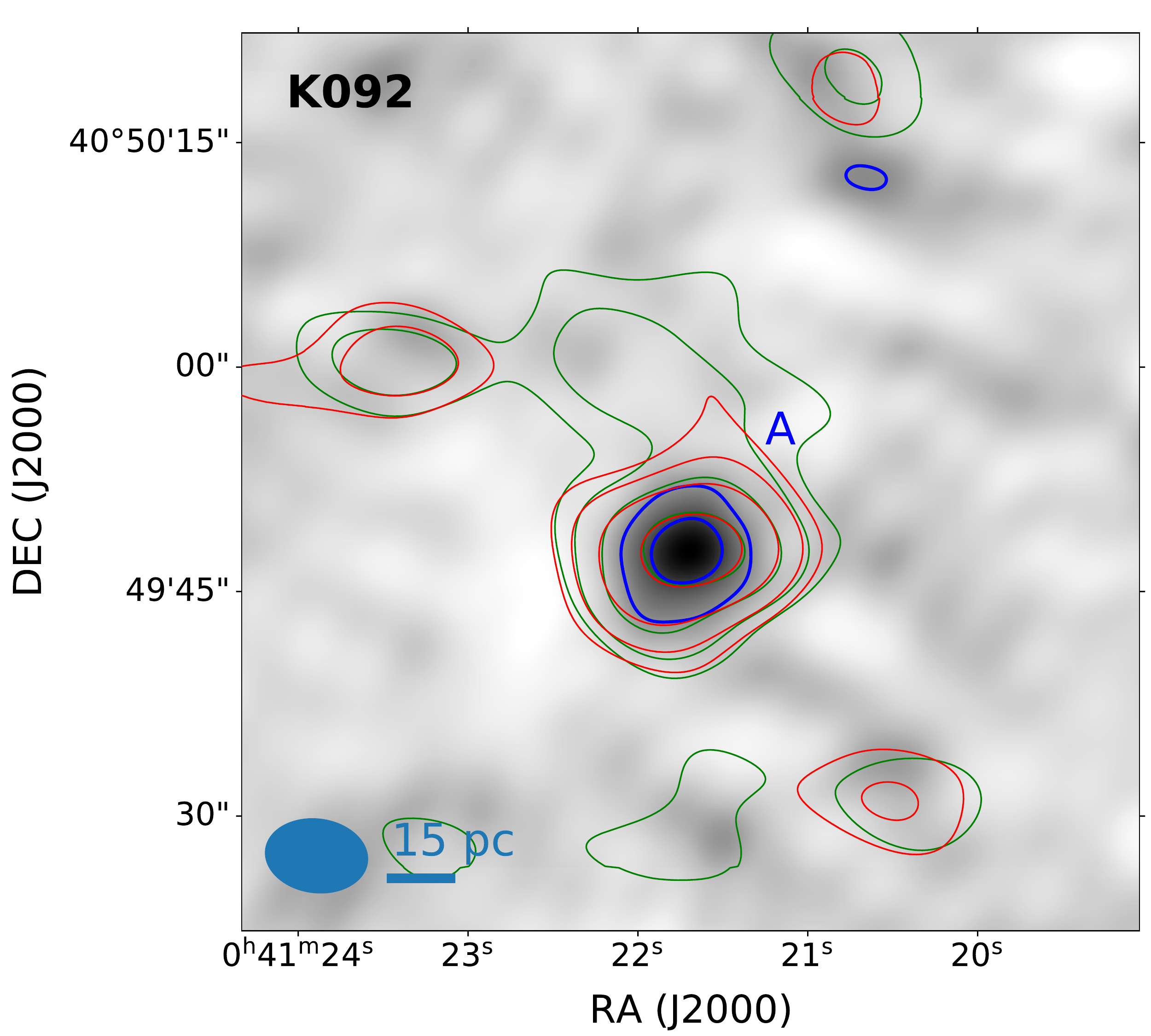}{0.33\textwidth}{}
          }
\gridline{\fig{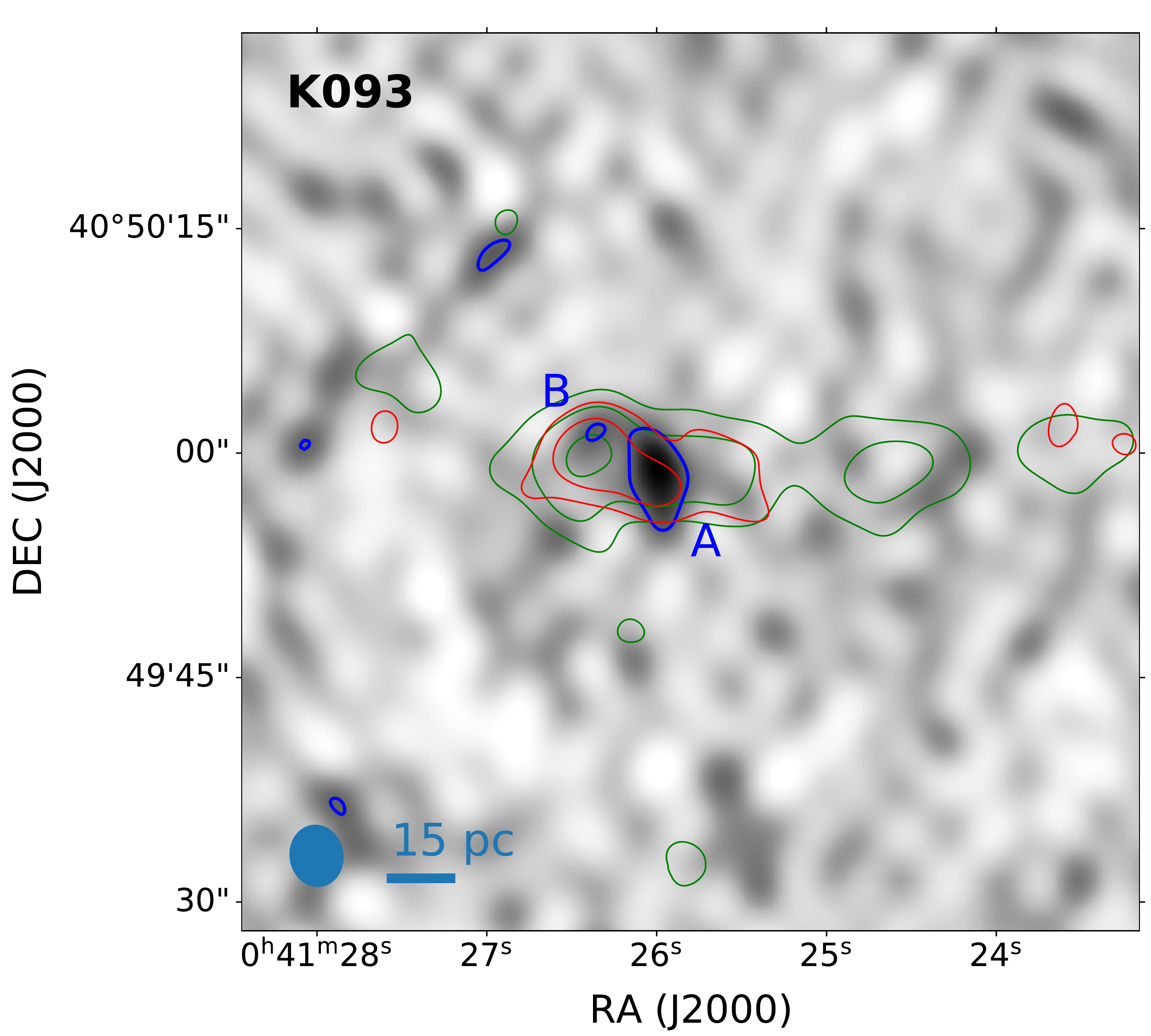}{0.33\textwidth}{}
          \fig{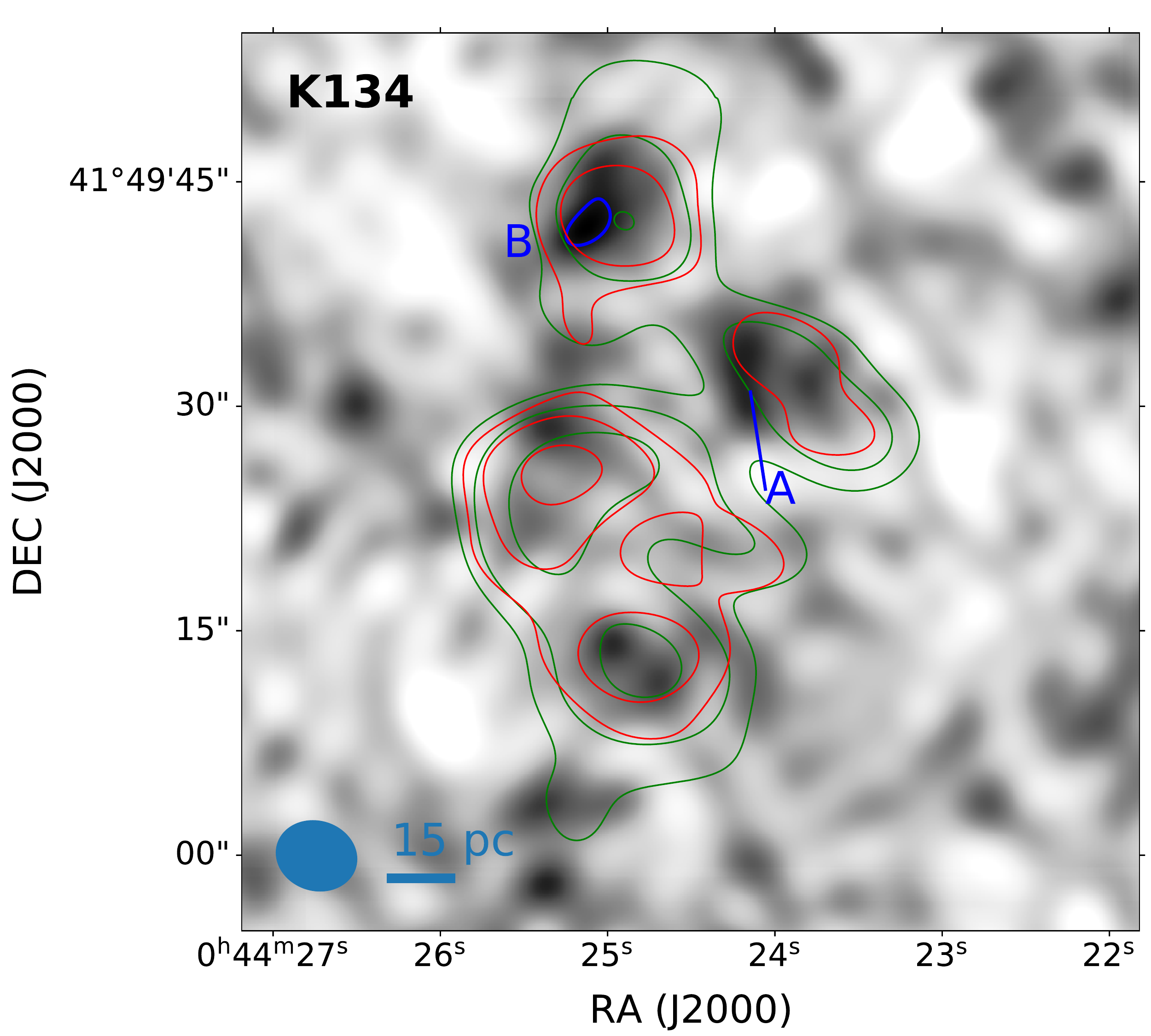}{0.33\textwidth}{}
          \fig{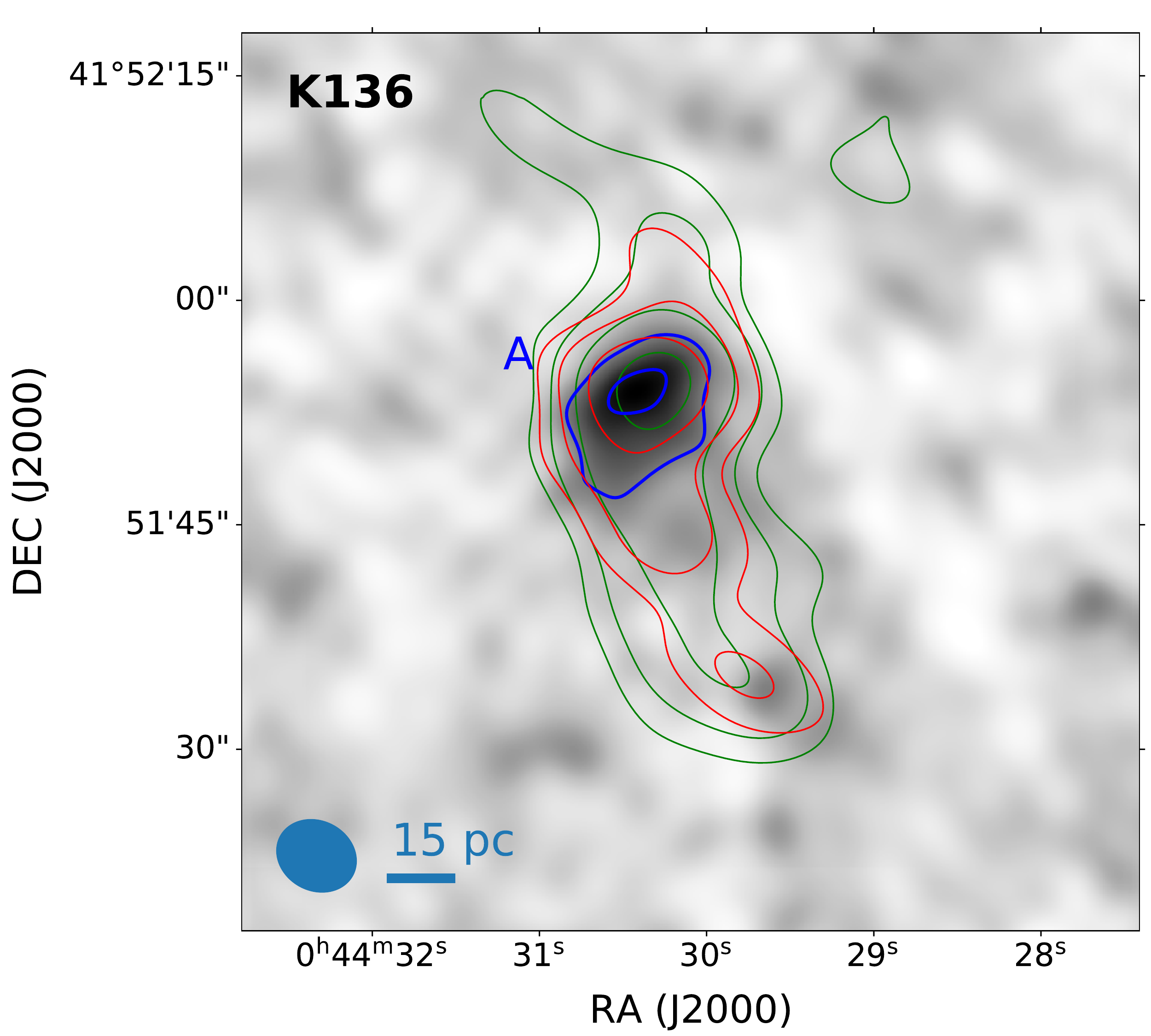}{0.33\textwidth}{}
          }
\gridline{\fig{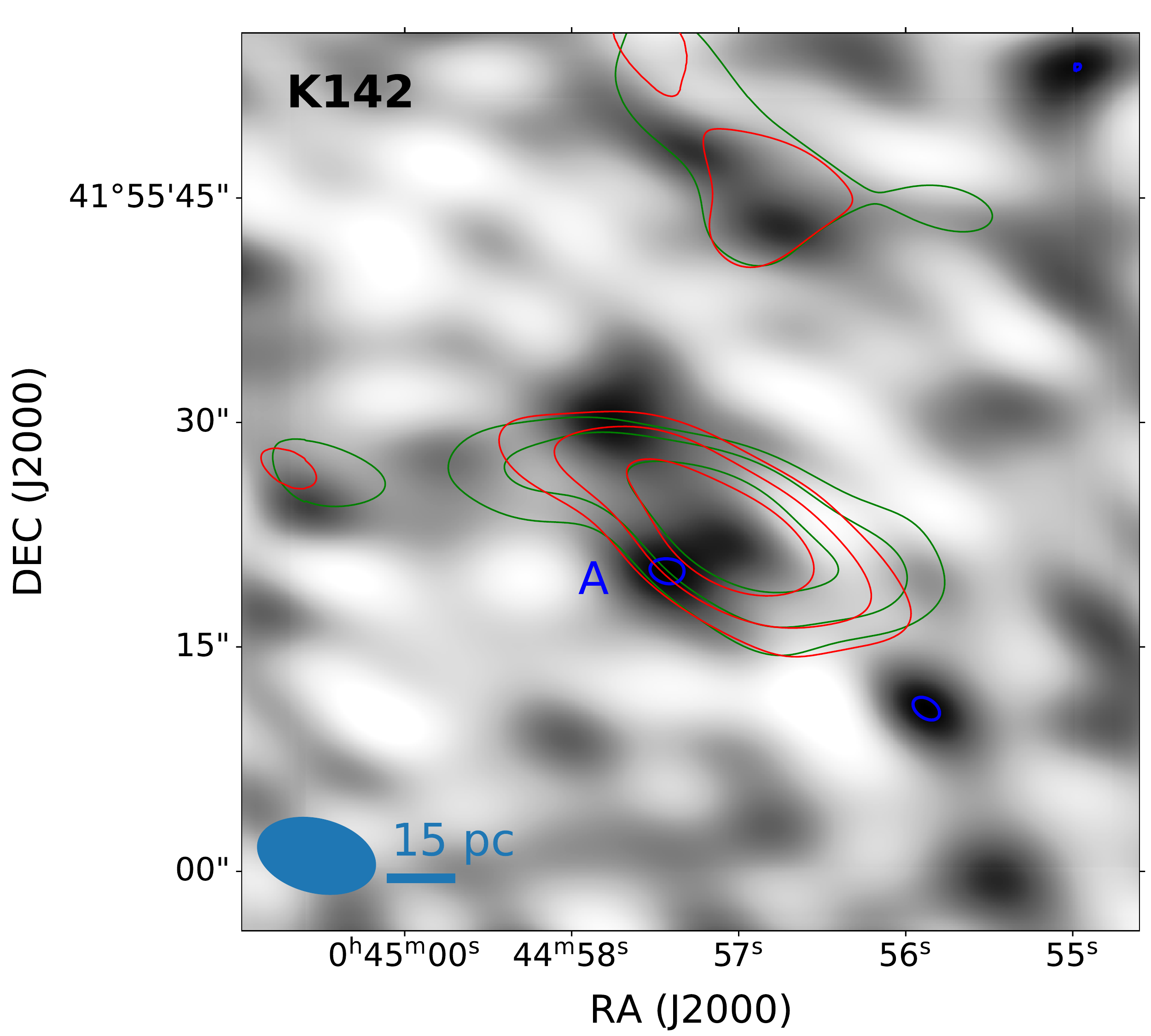}{0.33\textwidth}{}
          \fig{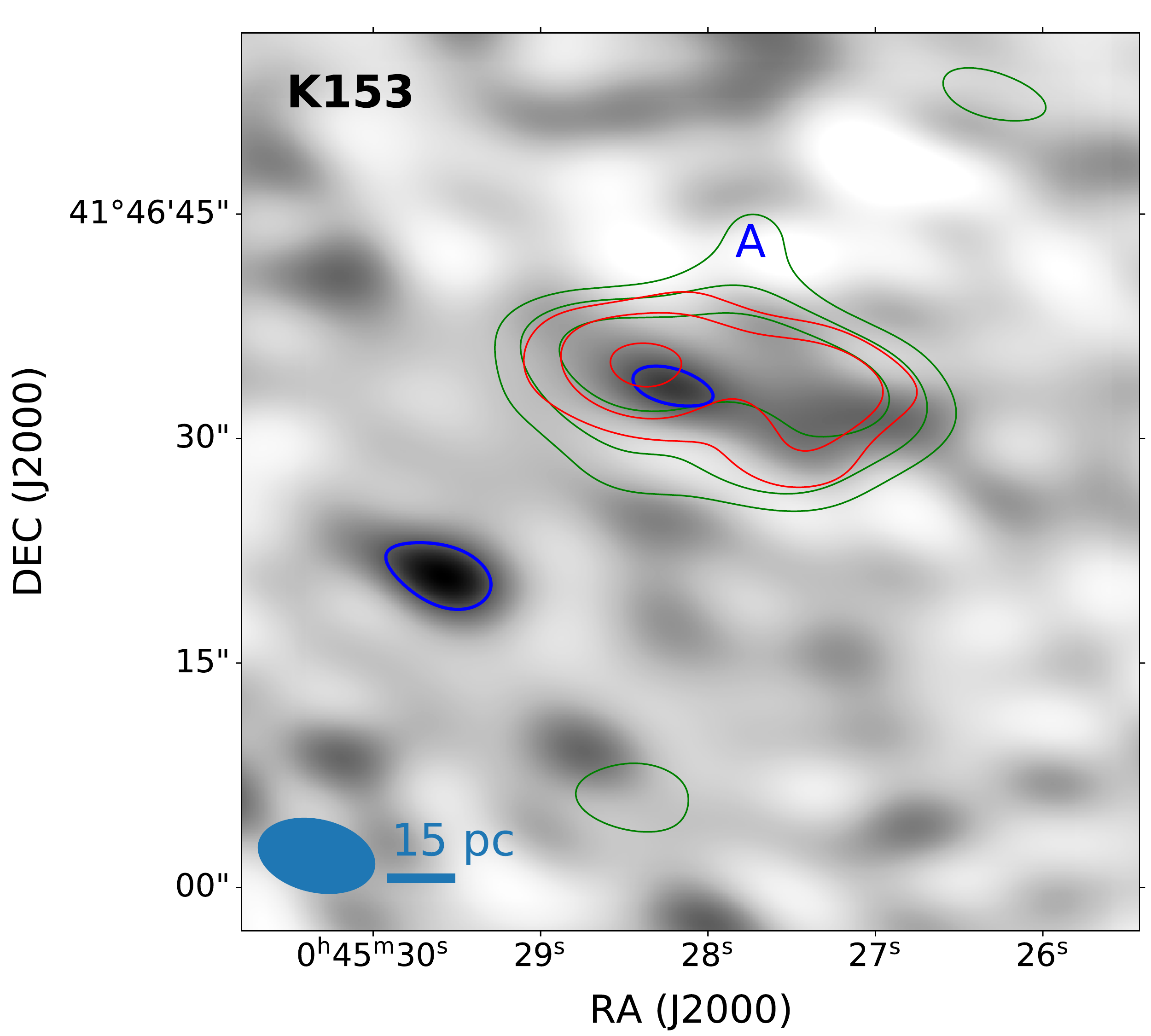}{0.33\textwidth}{}
          \fig{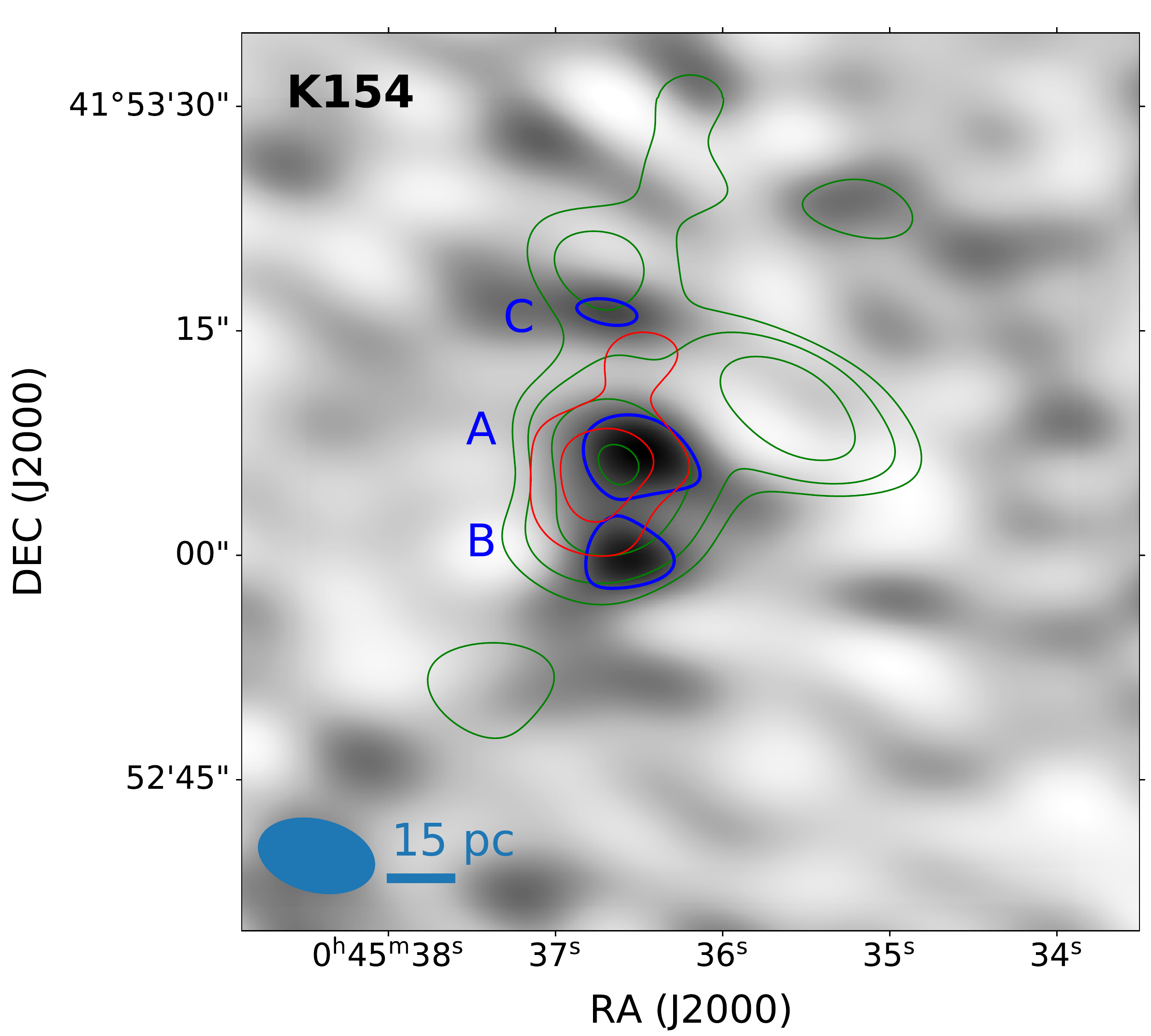}{0.33\textwidth}{}
          }
\figurenum{1a}
\caption{Combined maps of GMCs in Andromeda as seen by the SMA telescope, named according to the \citet{Kirk2015} catalog. The blue contours correspond to 3$\sigma$ and 6$\sigma$ levels of the dust continuum emission (greyscale background image). Overlaid are $^{12}$CO(2-1) contours in green ($3,6,12,24\sigma$ and $48\sigma$ where applicable) and $^{13}$CO(2-1) contours in red ($3-10\sigma$). The beam is shown in the bottom left corner. \label{fig:maps}}
\end{figure*}

\begin{figure*}
\gridline{\fig{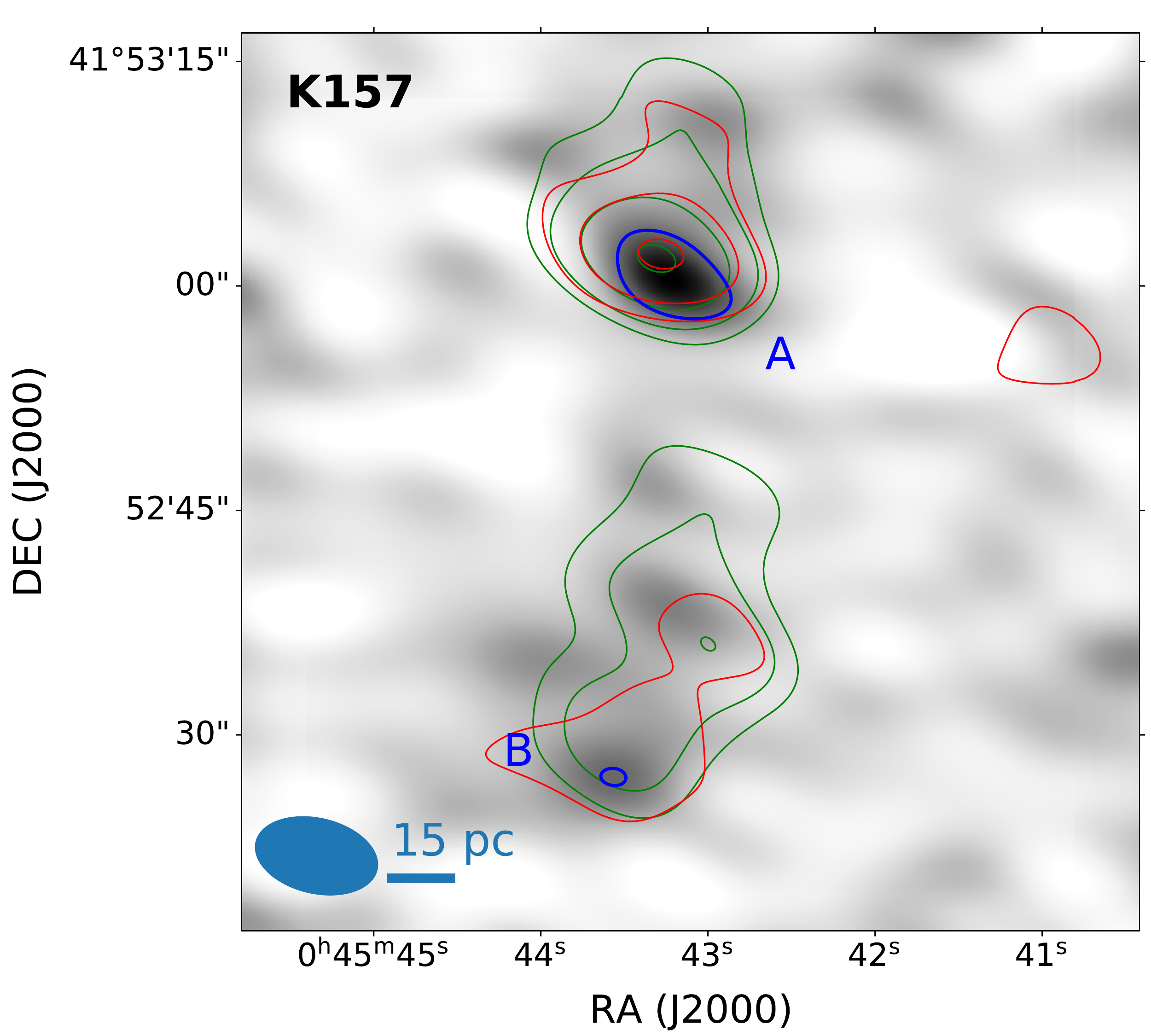}{0.33\textwidth}{}
          \fig{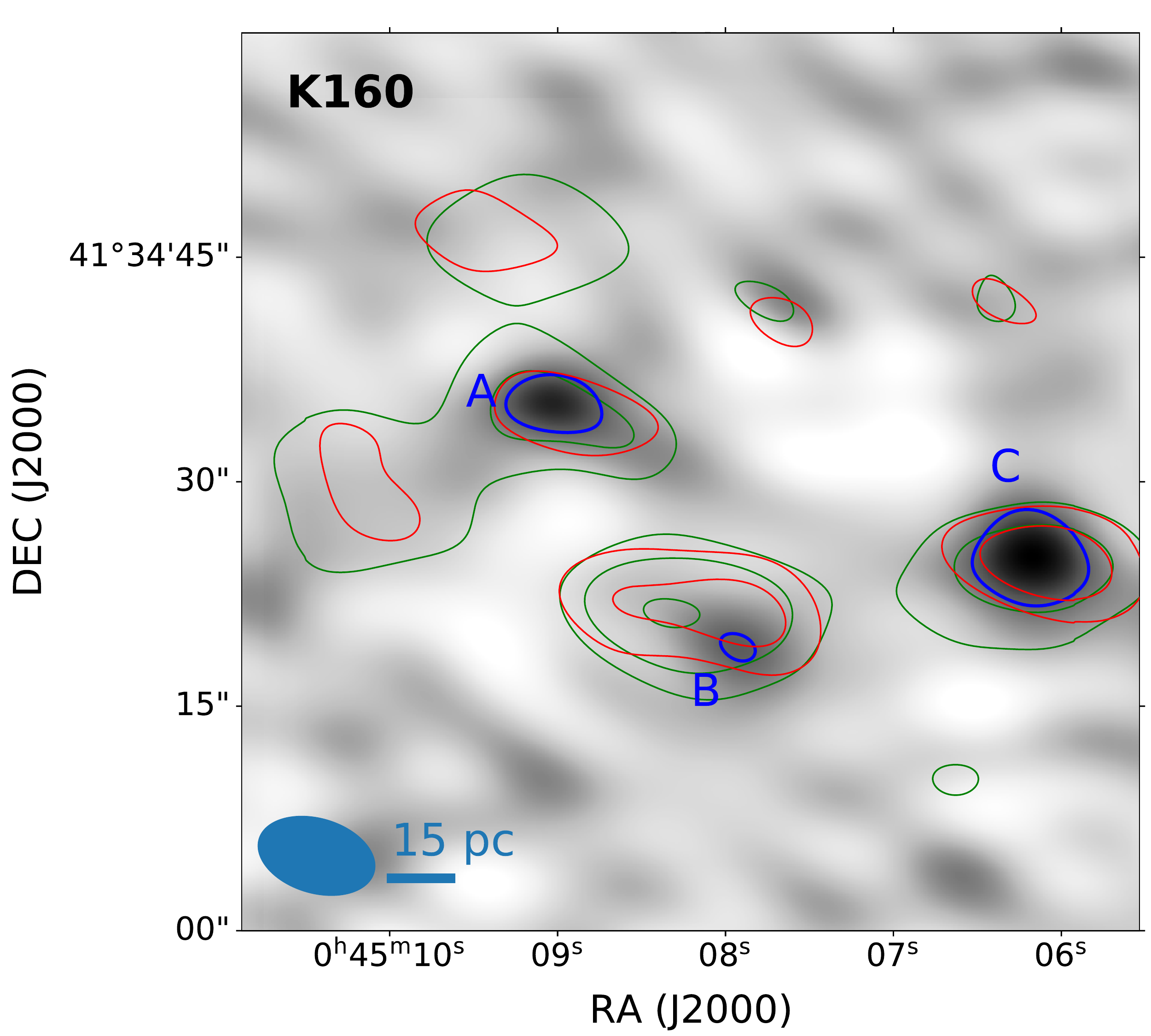}{0.33\textwidth}{}
          \fig{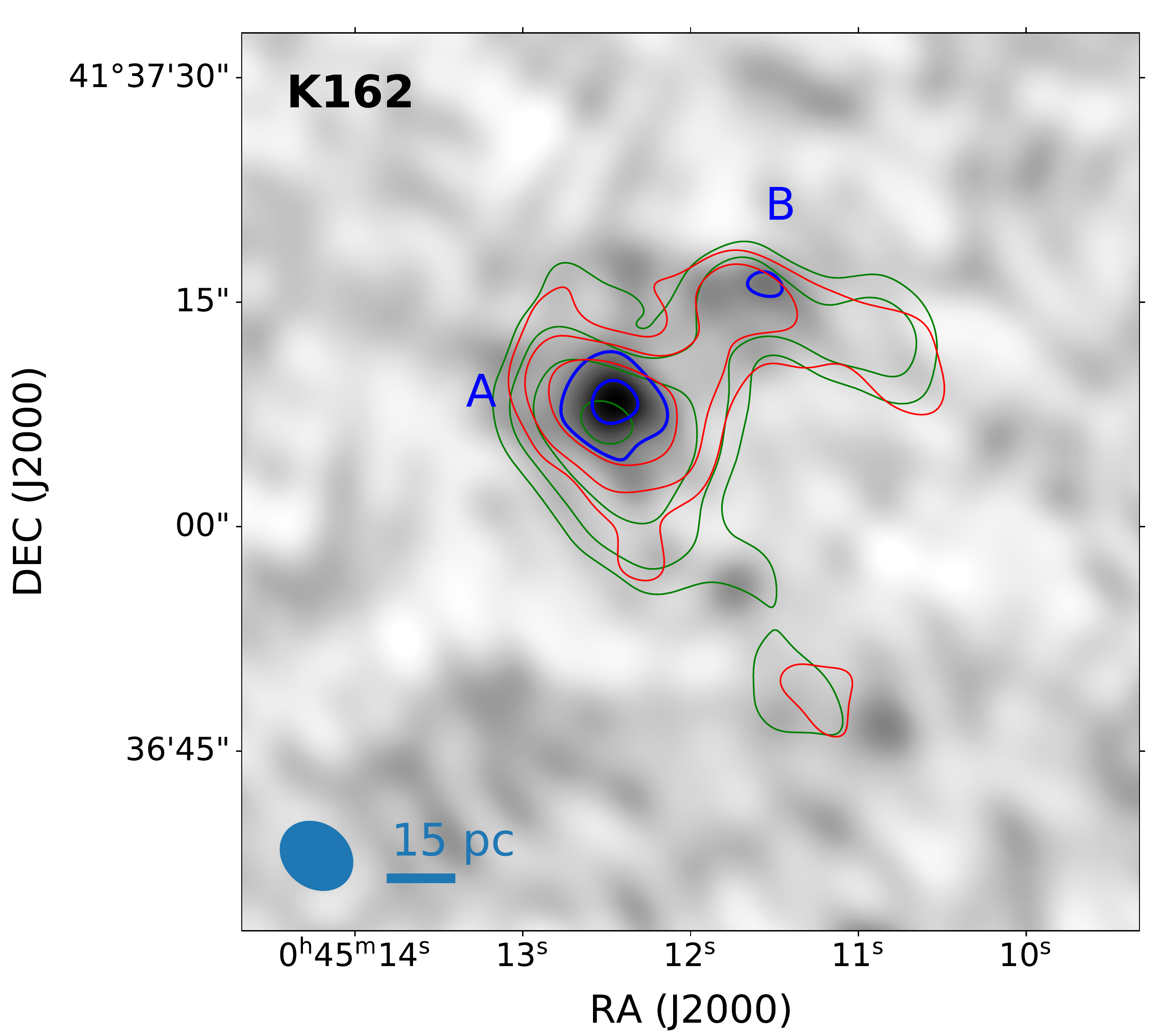}{0.33\textwidth}{}
          }
\gridline{\fig{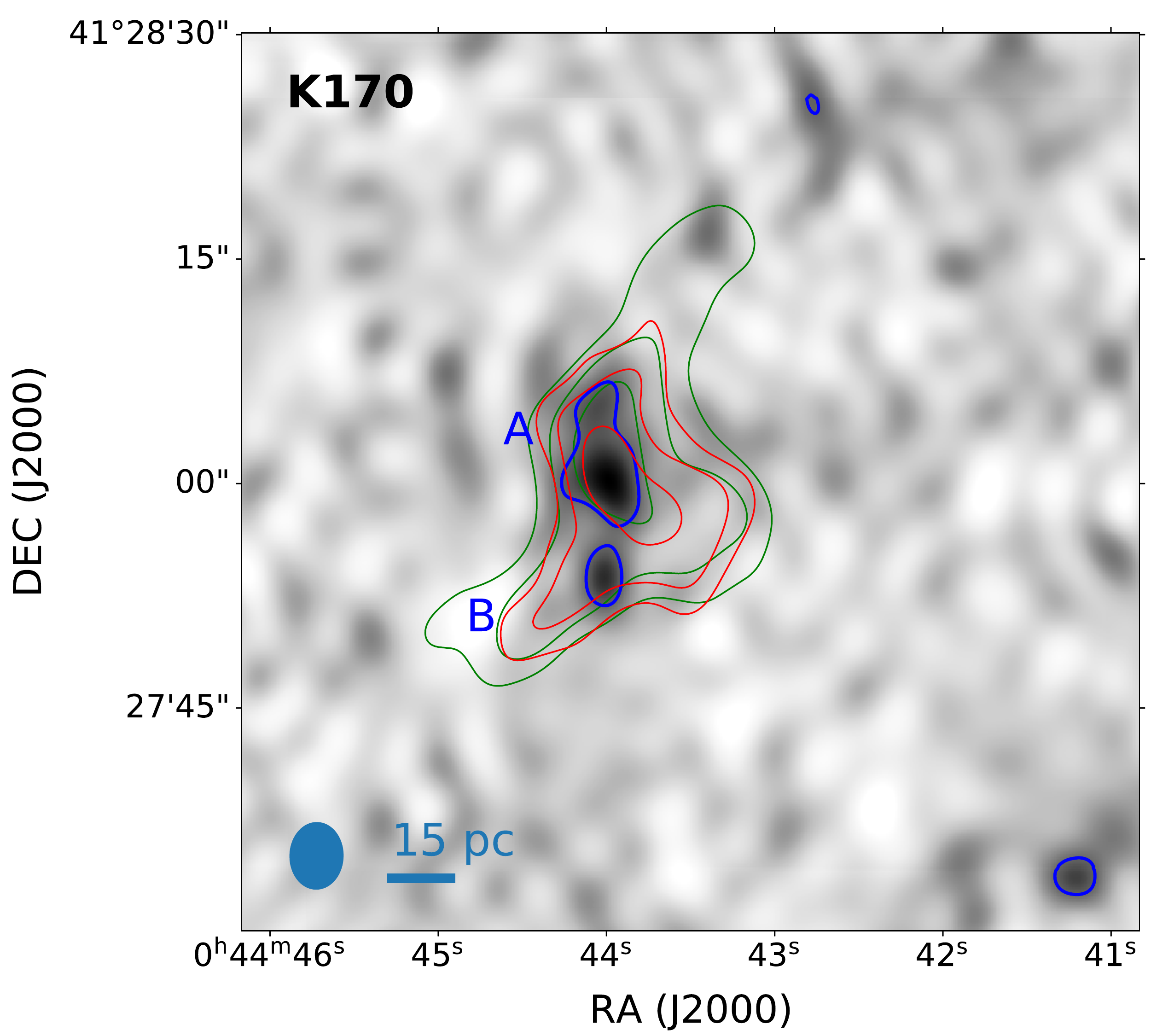}{0.33\textwidth}{}
          \fig{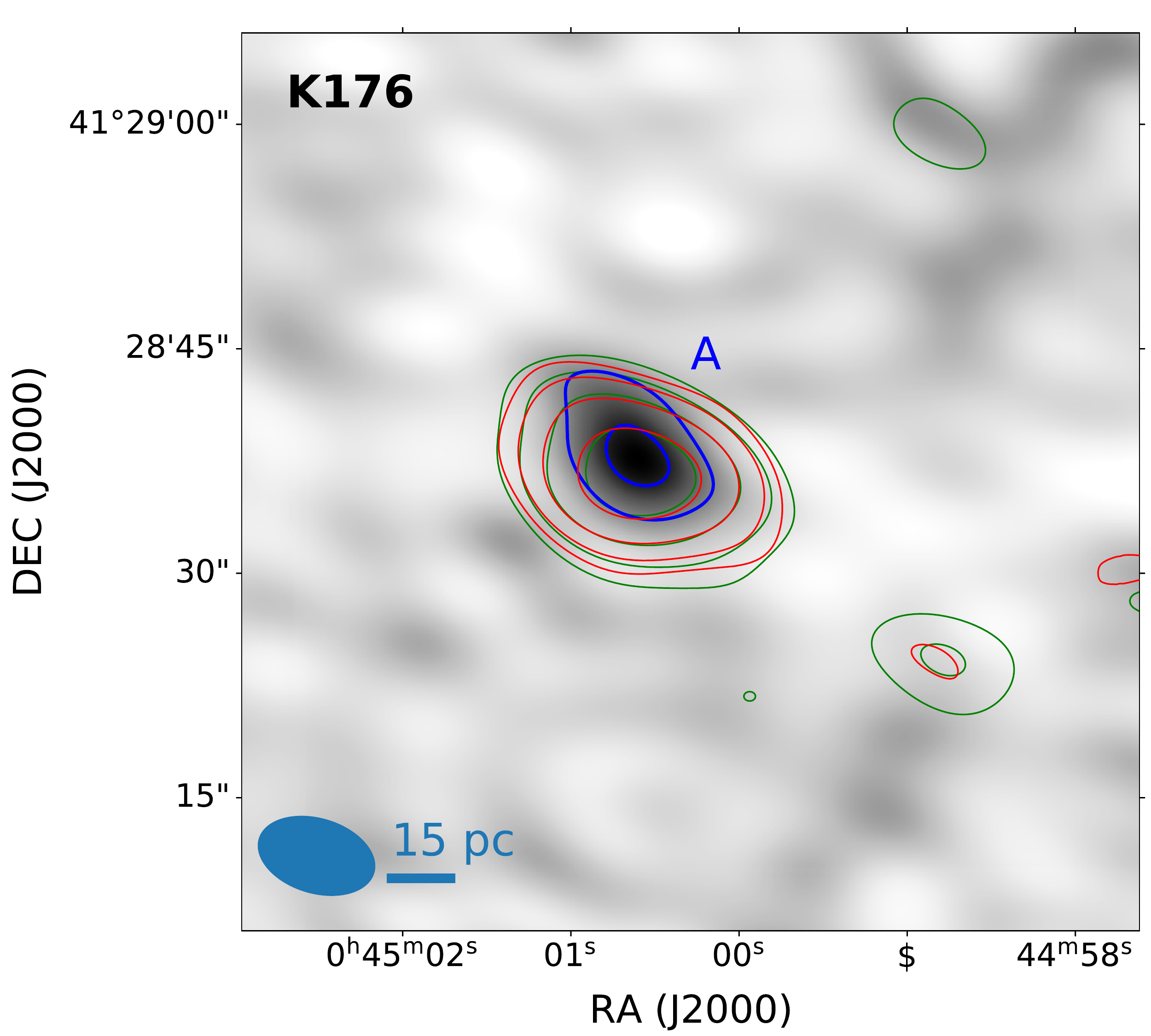}{0.33\textwidth}{}
          \fig{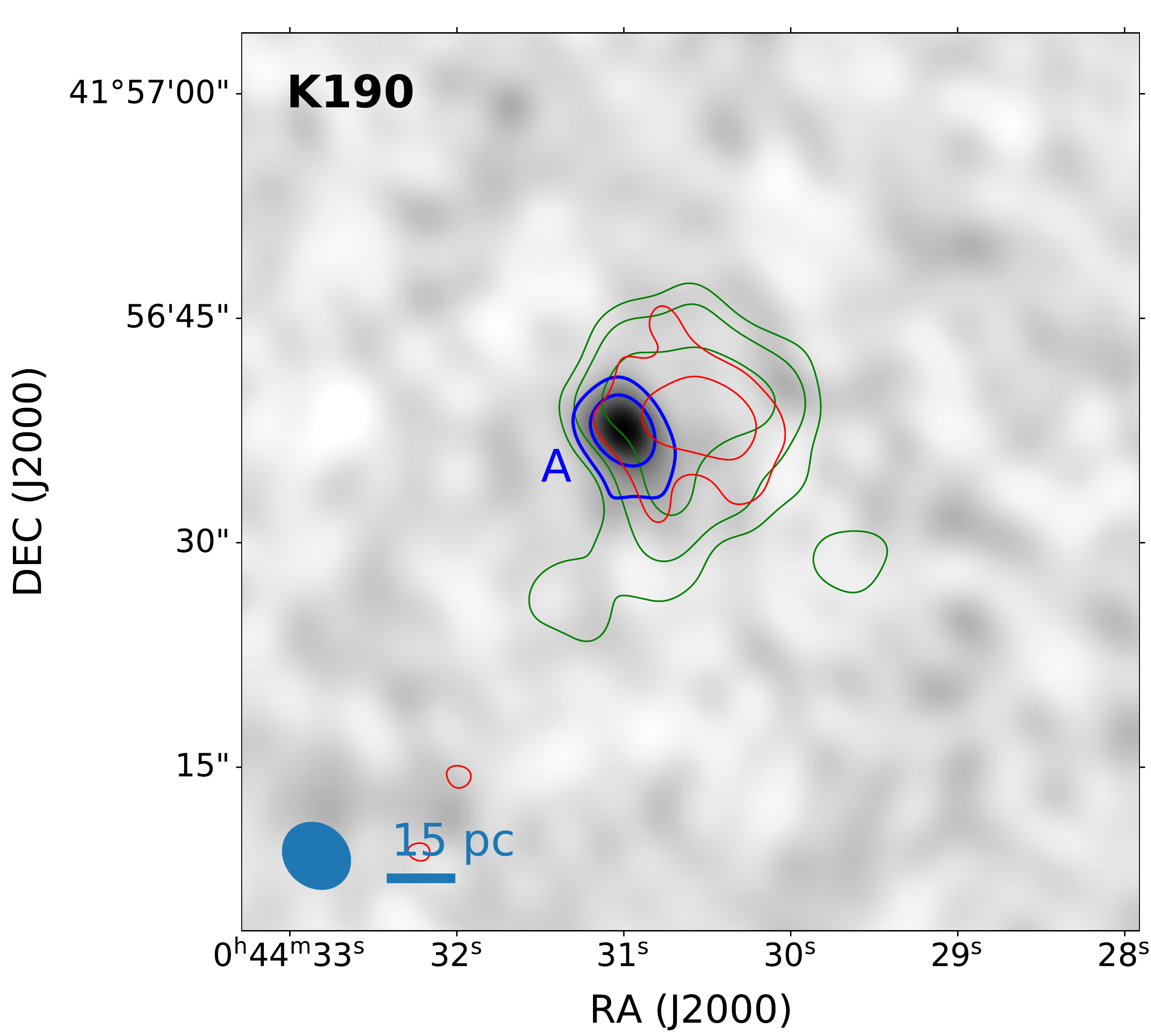}{0.33\textwidth}{}
          }
\gridline{\fig{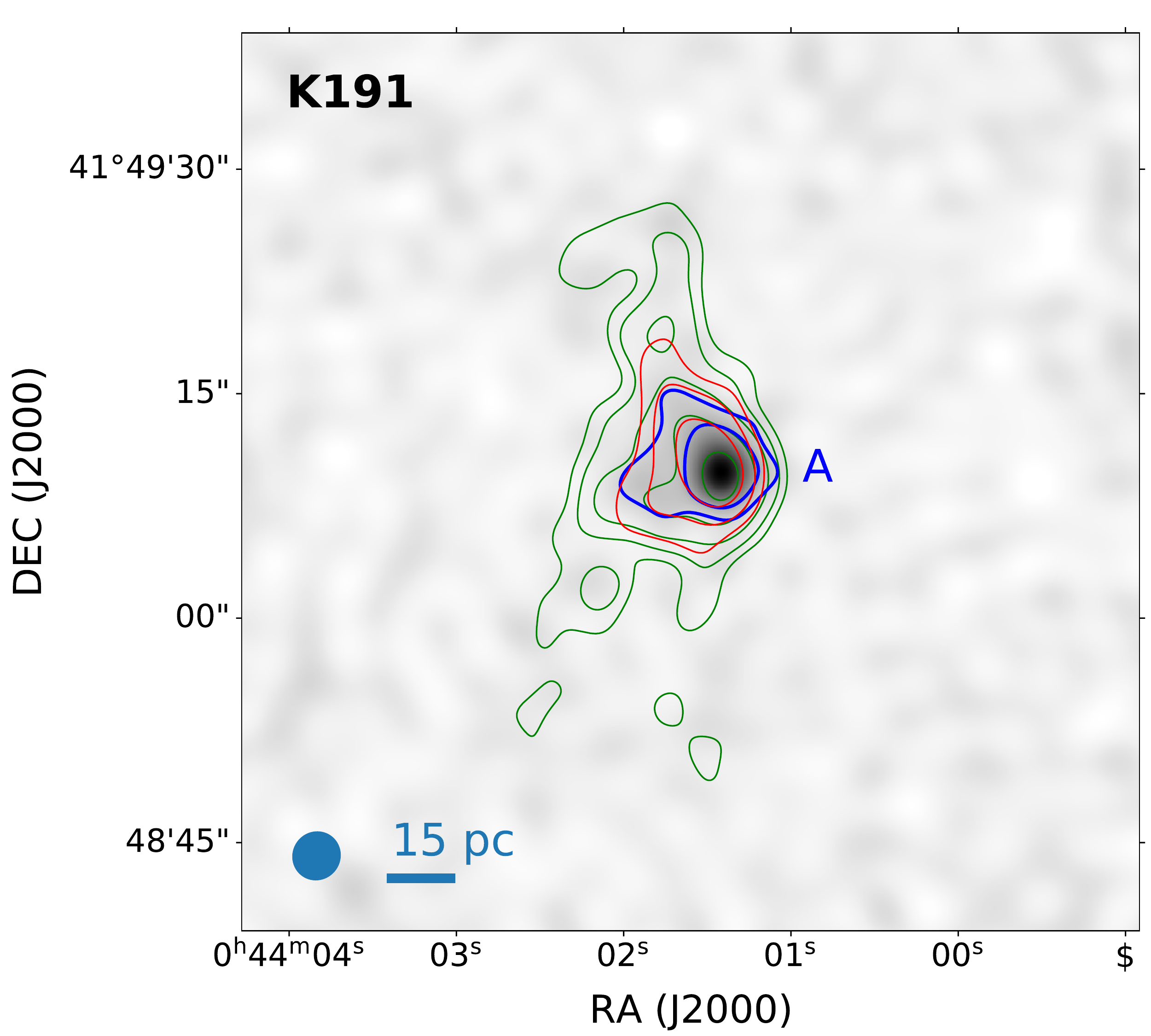}{0.33\textwidth}{}
          \fig{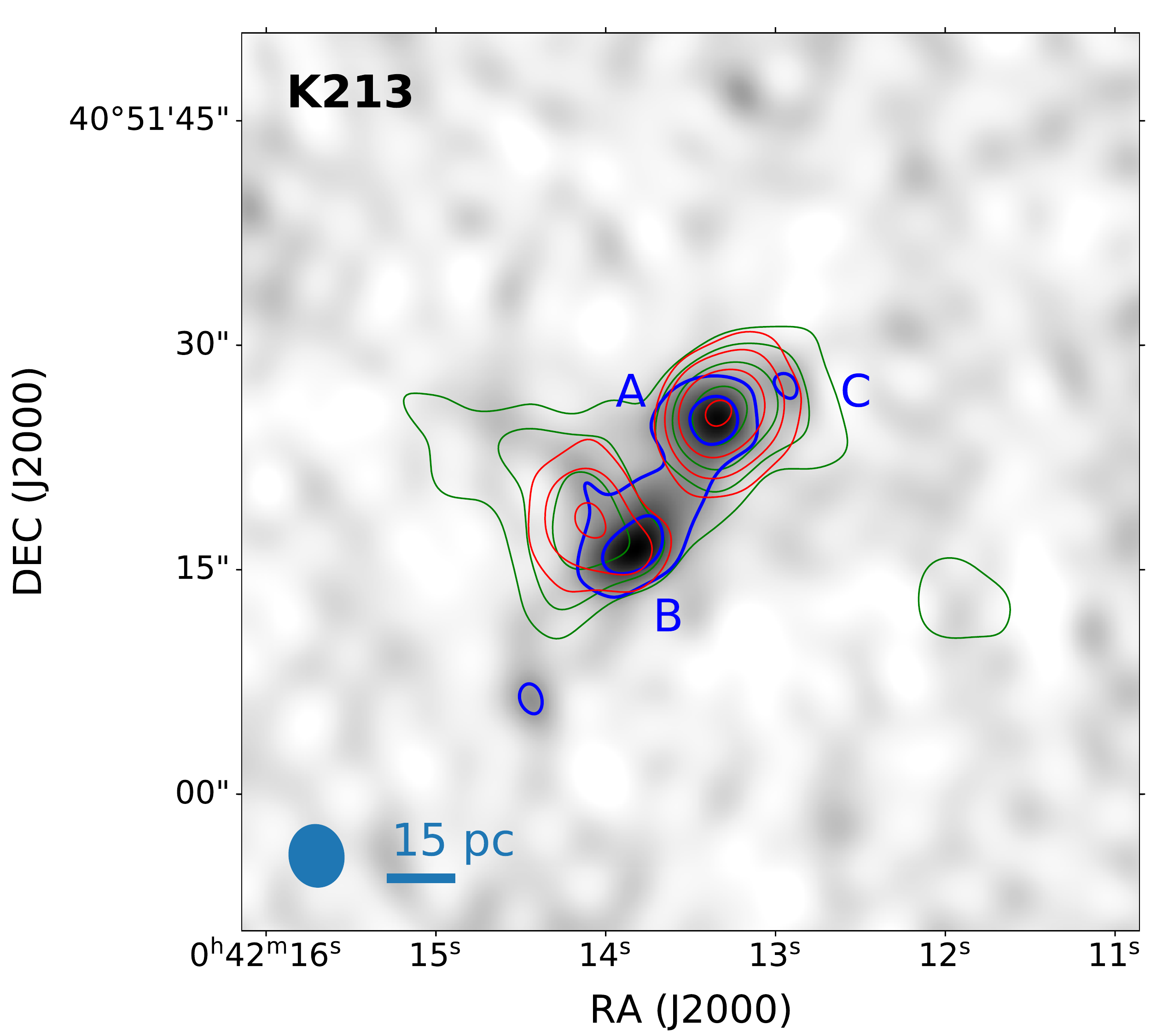}{0.33\textwidth}{}
          \fig{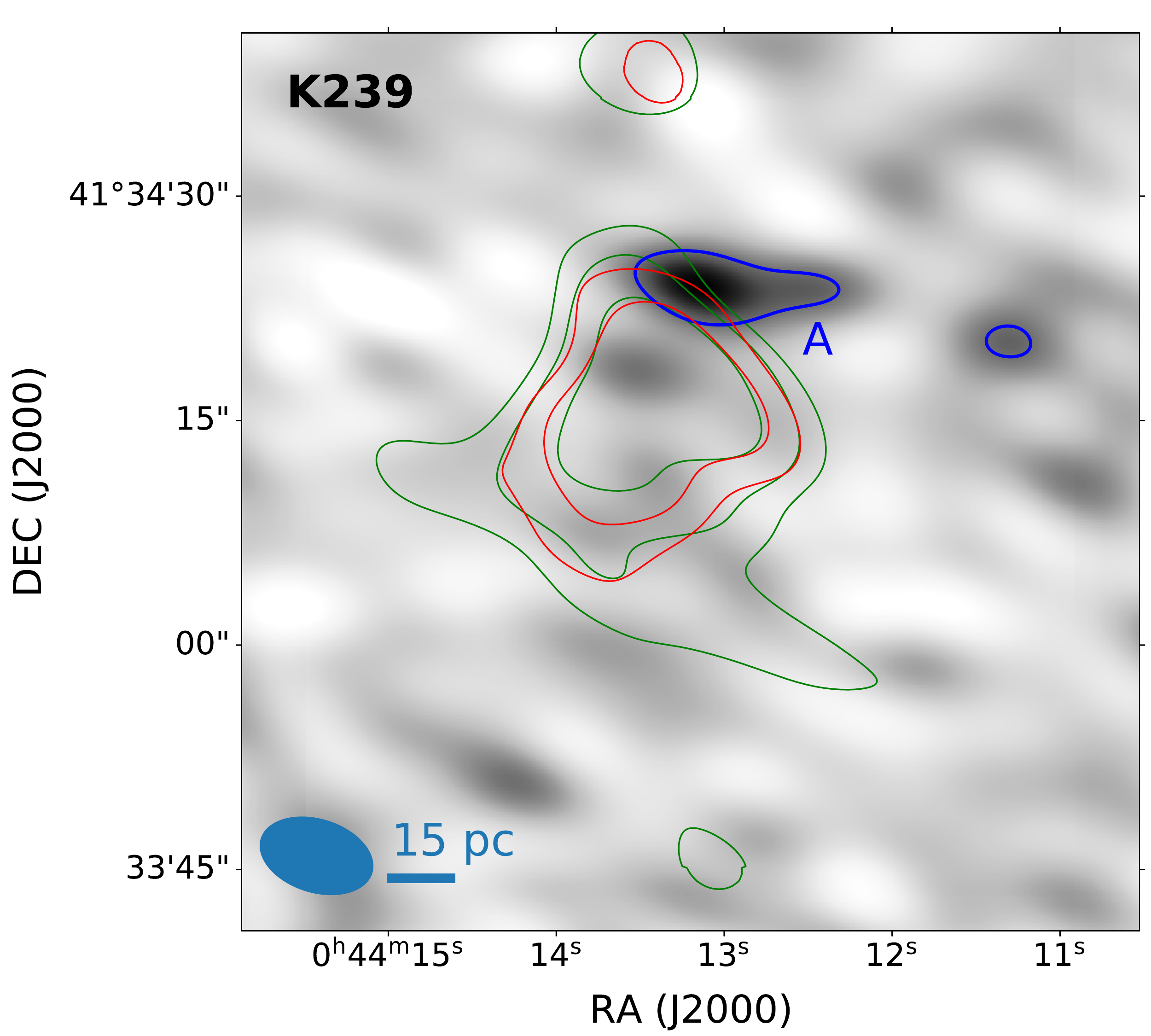}{0.33\textwidth}{}
          }
\figurenum{1a}
\caption{continued}
\end{figure*}

\begin{figure*}
\gridline{\fig{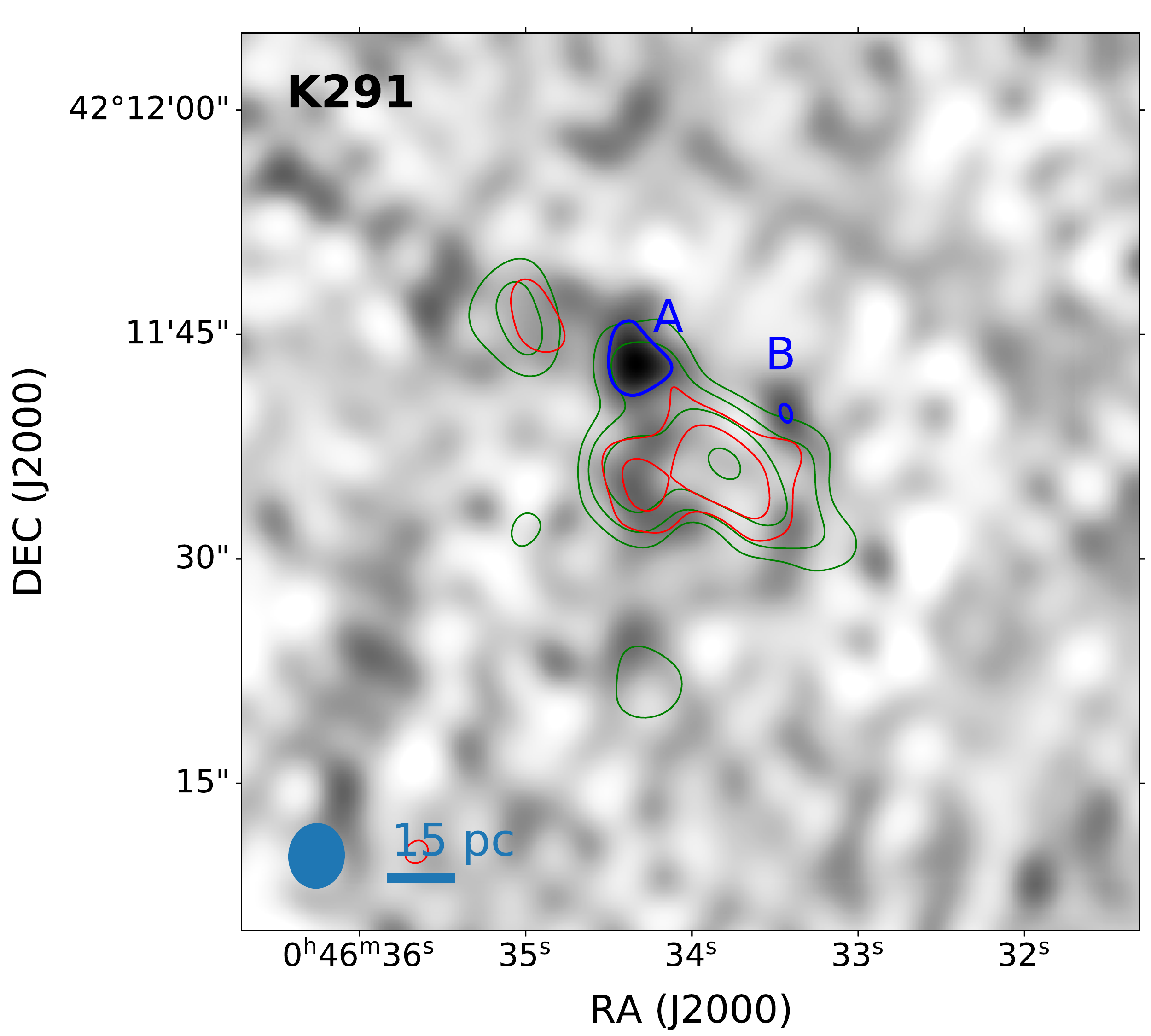}{0.33\textwidth}{}
          \fig{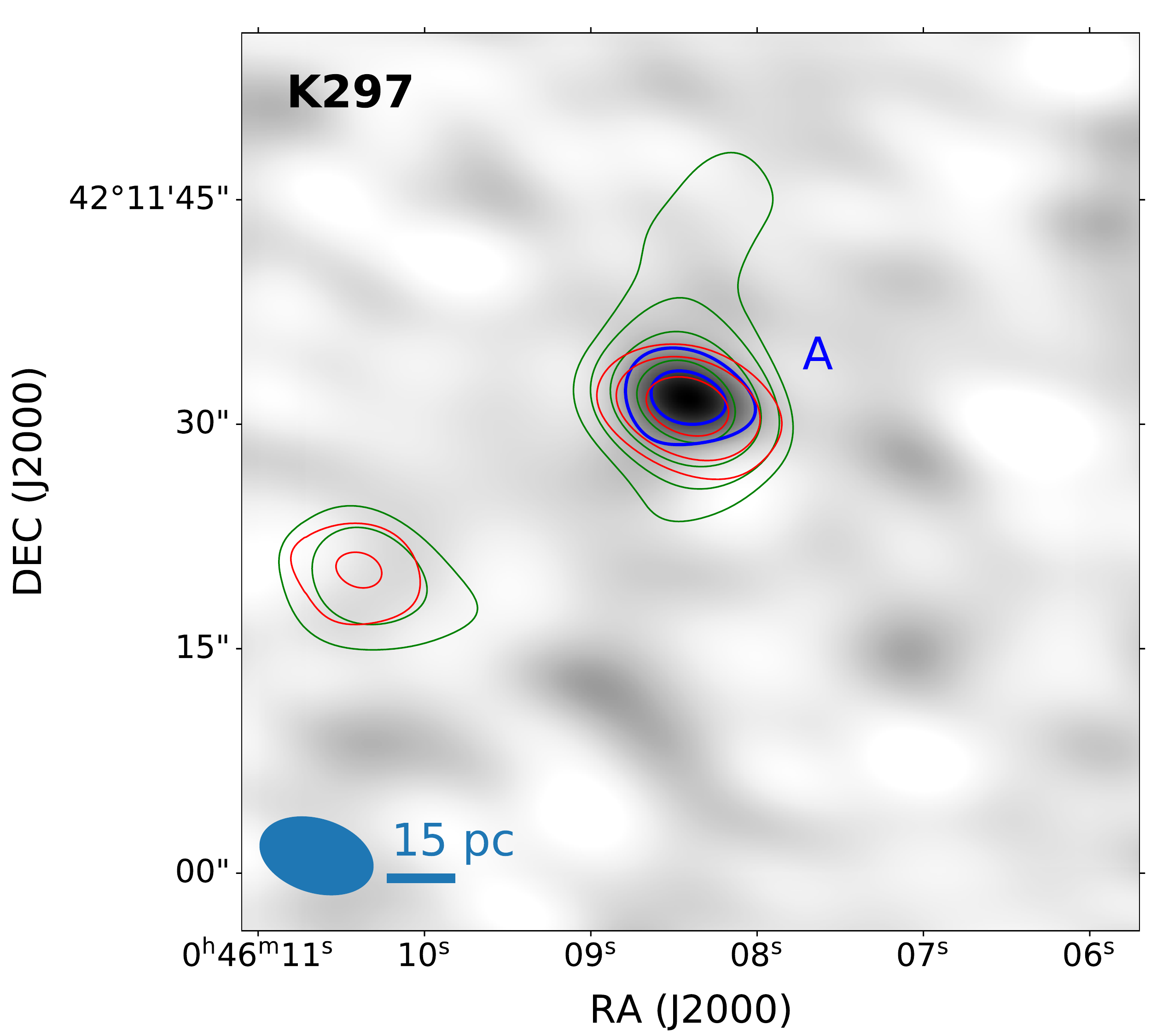}{0.33\textwidth}{}}
\figurenum{1a}
\caption{continued}
\end{figure*}

\begin{figure*}
\gridline{\fig{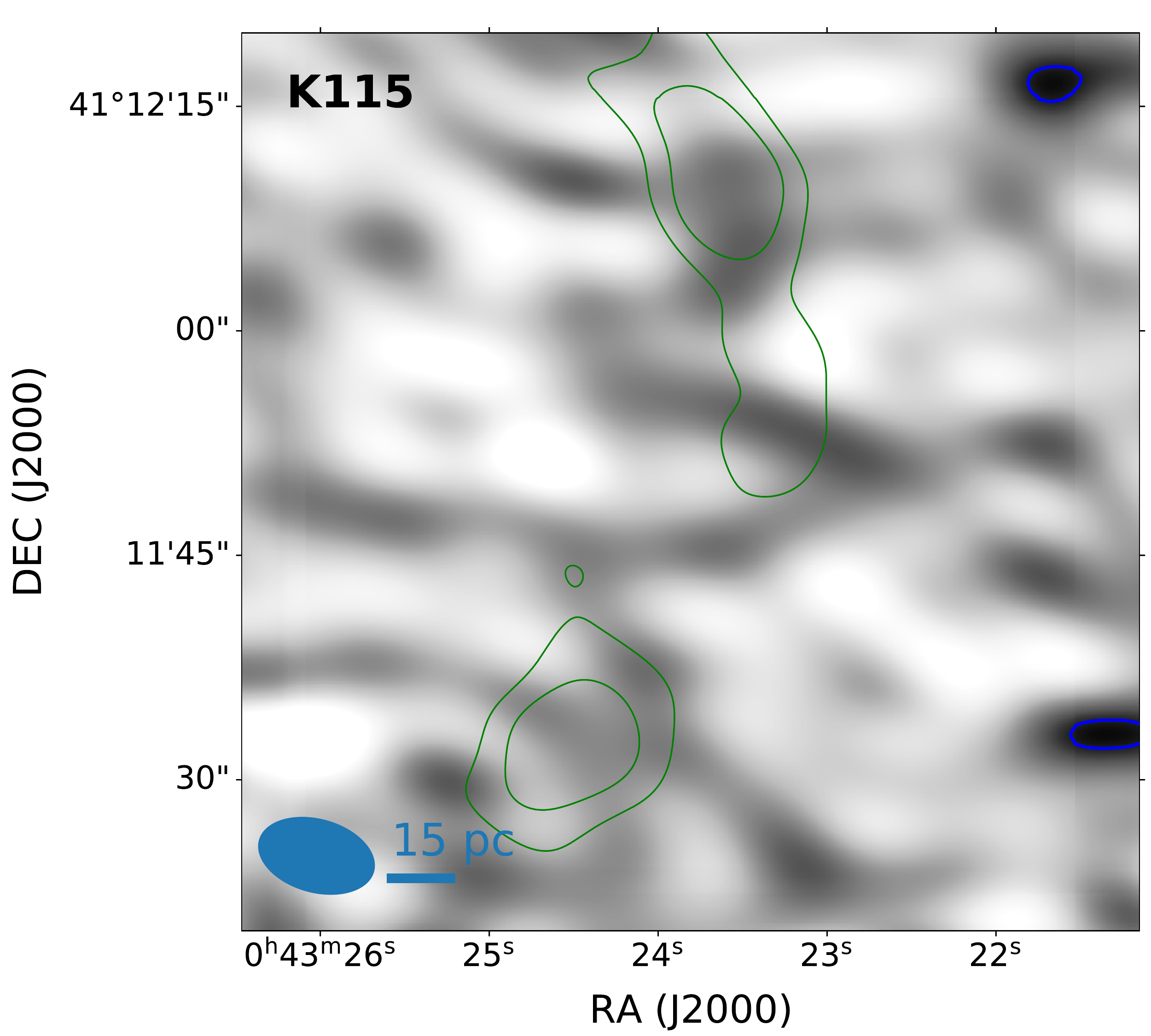}{0.33\textwidth}{}
          \fig{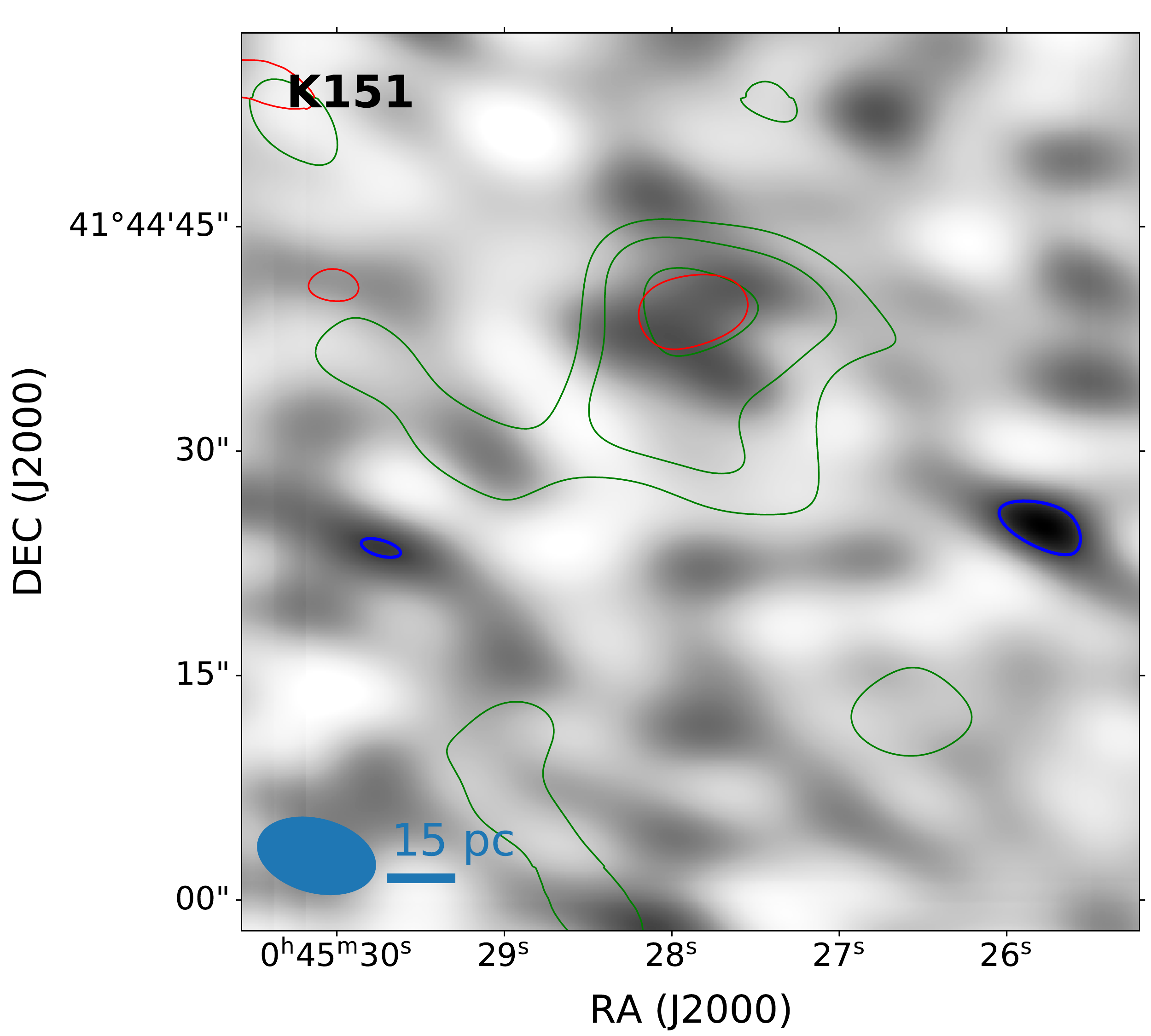}{0.33\textwidth}{}
          \fig{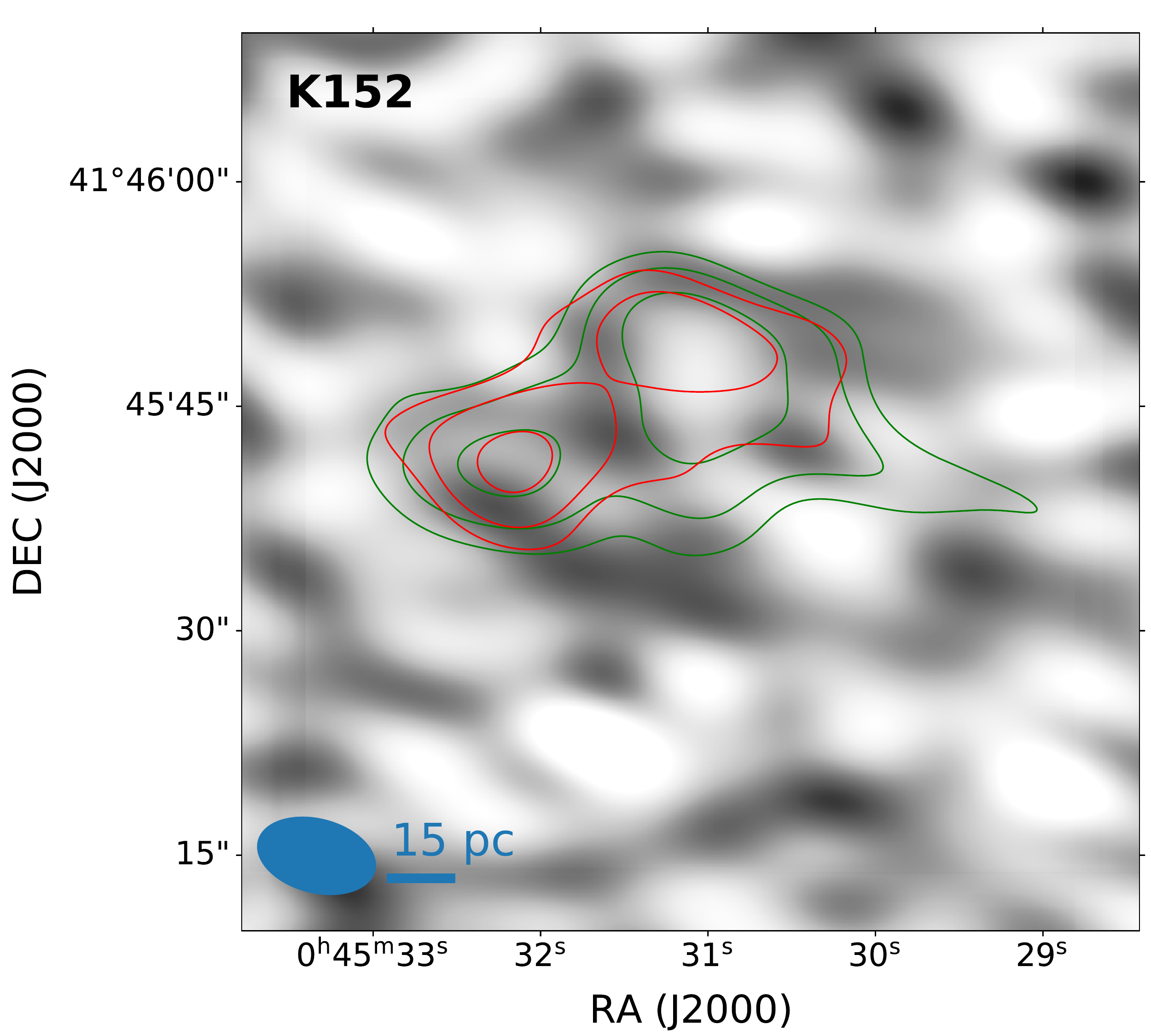}{0.33\textwidth}{}
          }
\gridline{\fig{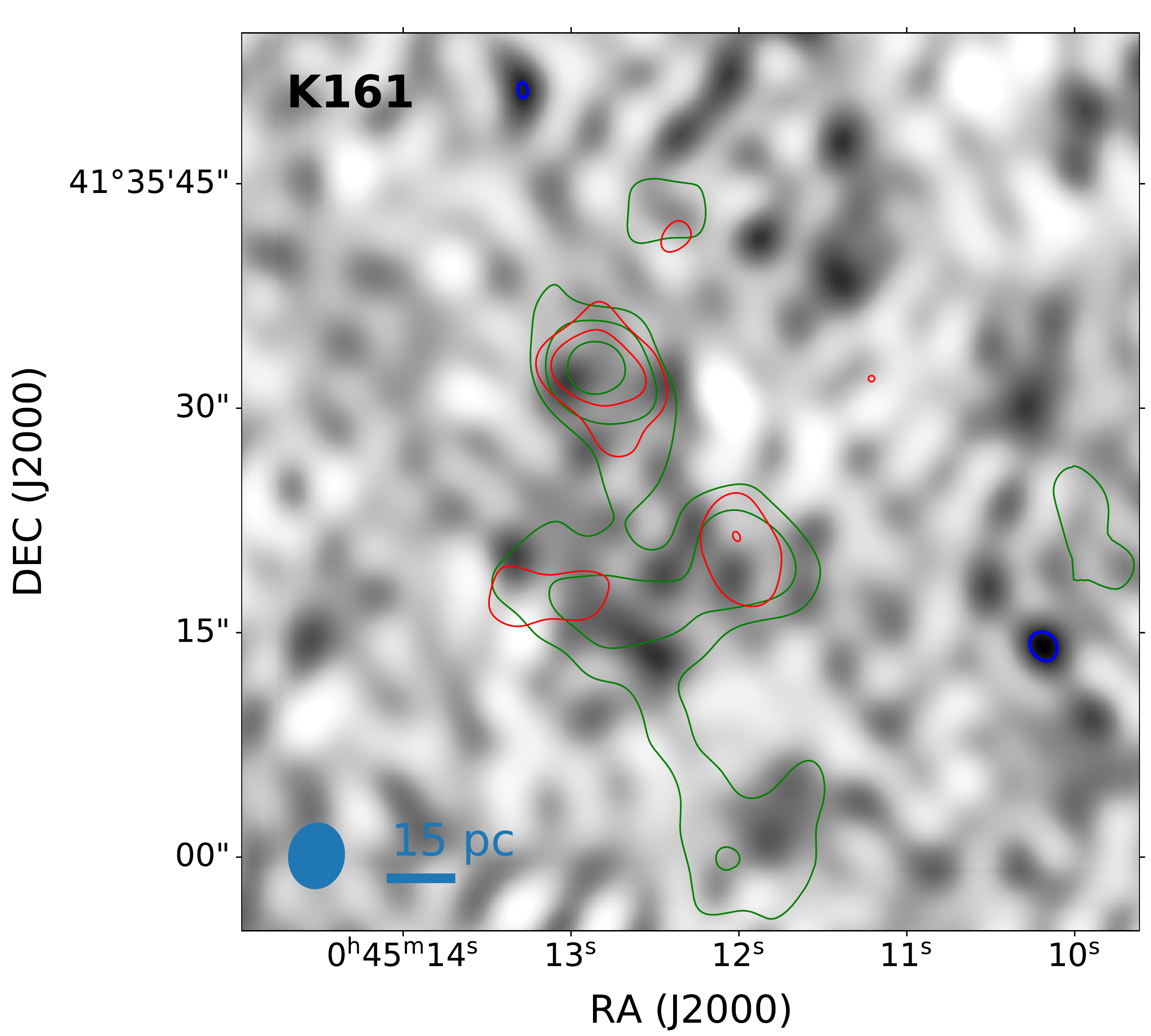}{0.33\textwidth}{}
		  \fig{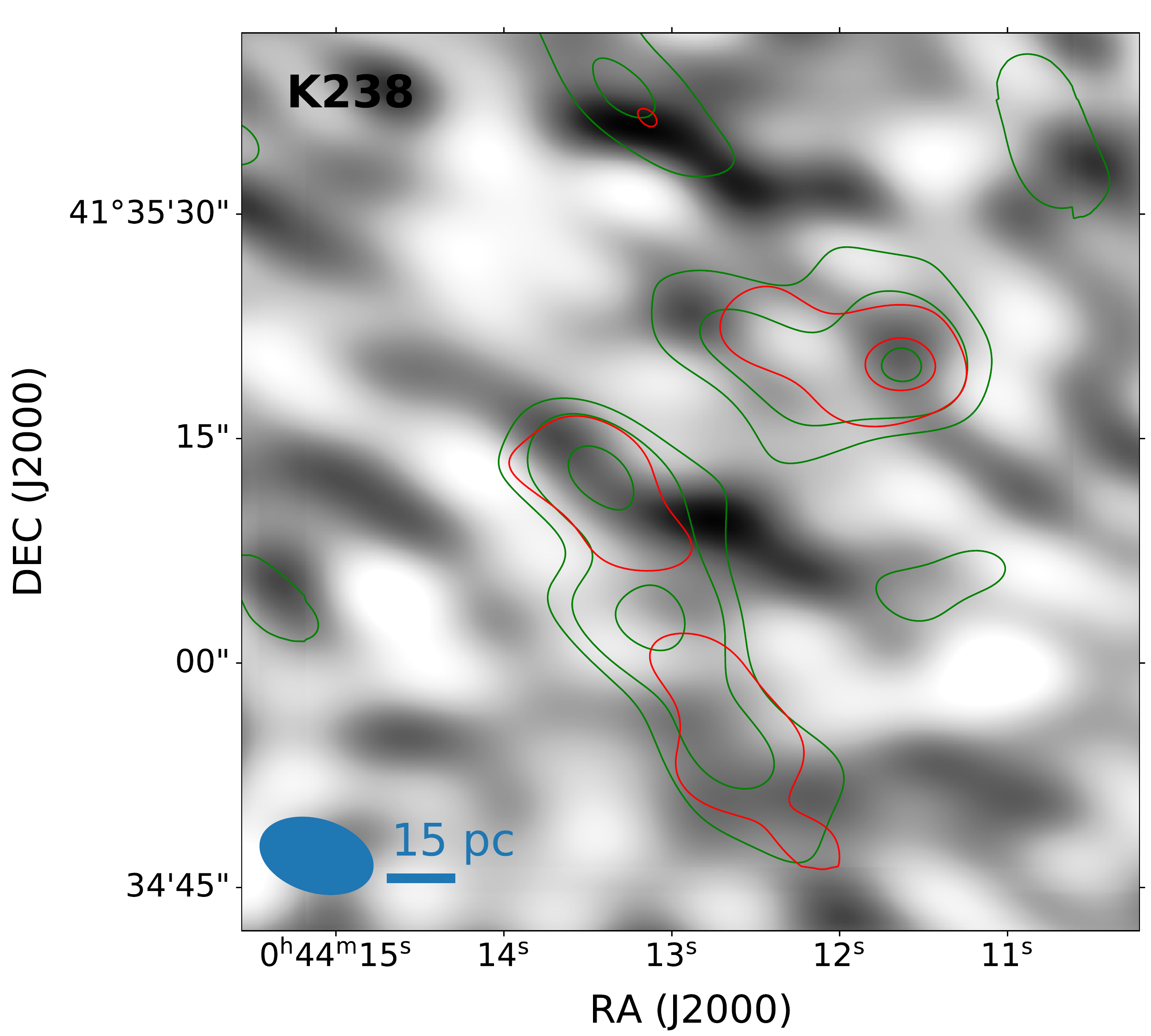}{0.33\textwidth}{}
		  }
\figurenum{1b}
\caption{continued, but now showing the pointings without continuum detection coinciding with \thco emission.}
\end{figure*}

\section{Results} \label{sec:results}

\subsection{Maps}  \label{sec:maps}

Fig.~\ref{fig:maps} shows the continuum maps of the individual pointings in grayscale. The blue contours correspond to 3 and 6 times the RMS level in the continuum. Although the dust emission can be resolved (10 out of 32 sources), in general the detections are rather compact, likely due to limits of our  sensitivity. This is also suggested by simulated observations of the Orion clouds (Paper I). On the panels in Fig.~\ref{fig:maps} we add contour levels for the integrated line emission from $^{12}$CO(2-1) and $^{13}$CO(2-1). 

The $^{12}$CO(2-1) and $^{13}$CO(2-1) emission is generally more extended than the dust continuum emission and the \twco emission is more extended than the \thco emission. Dust and CO peaks do not always spatially align, which we can evaluate due to the fact that both the continuum and line observations have been obtained simultaneously with the same calibration, imaging, and astrometry.  In a number of GMCs the peaks in dust emission are offset from peaks in CO emission even though they are still located within the \twco emitting region. Such circumstances may indicate a spatial separation of colder gas and hotter (brighter) dust heated by ongoing star formation at the GMC edges. It is also possible that the continuum emission occasionally originates within an associated HII region as our observations may be sufficiently sensitive to be contaminated by free-free emission as shown in Paper I.  K026, K093, K190, K213, K239 and K291 are all examples of bright continuum emission occurring at the edge of a CO peak. In many cases (9 of 25 fields)  there are  weak but significant continuum peaks that are found well outside the CO emitting regions. It remains unclear whether these sources are associated with any GMCs in M31. They are likely of  a different physical nature (e.g., background galaxies) (Paper I). They will not be considered further in the present paper. There is only one pointing where we do not detect any $^{13}$CO (or continuum) emission: K115. Here, even the $^{12}$CO emission is weak and offset from the center of the field. 

\begin{figure*}
\gridline{\fig{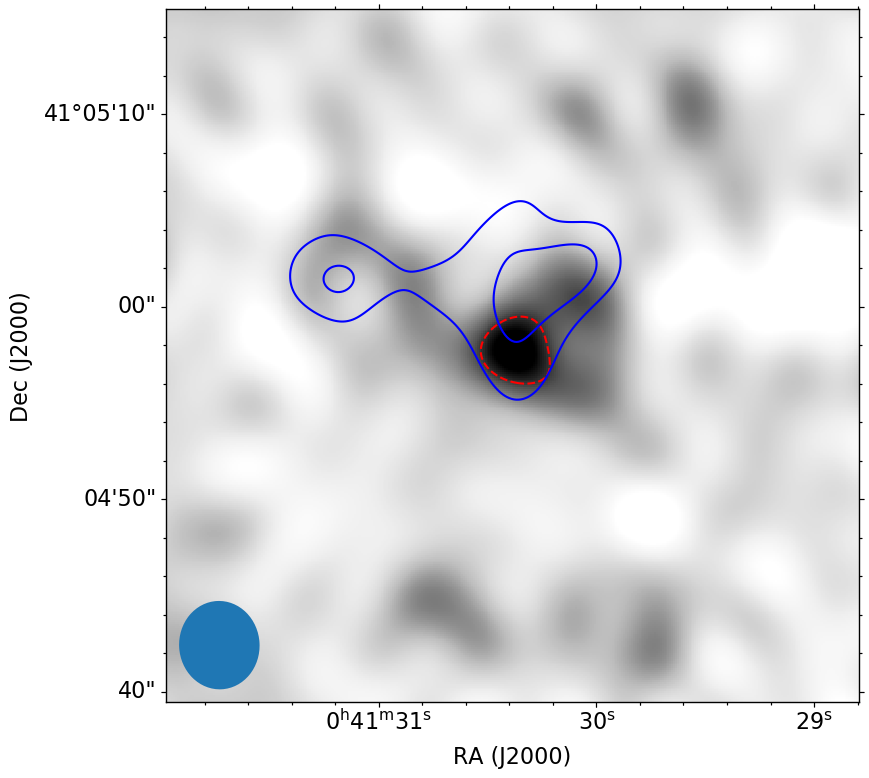}{0.40\textwidth}{}
          \fig{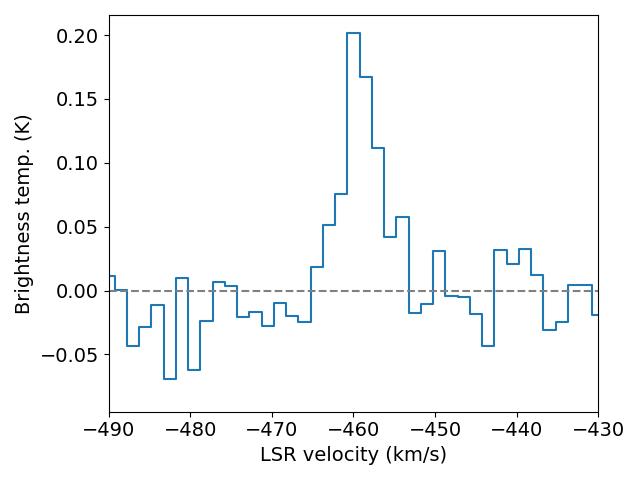}{0.48\textwidth}{}
          }
\gridline{\fig{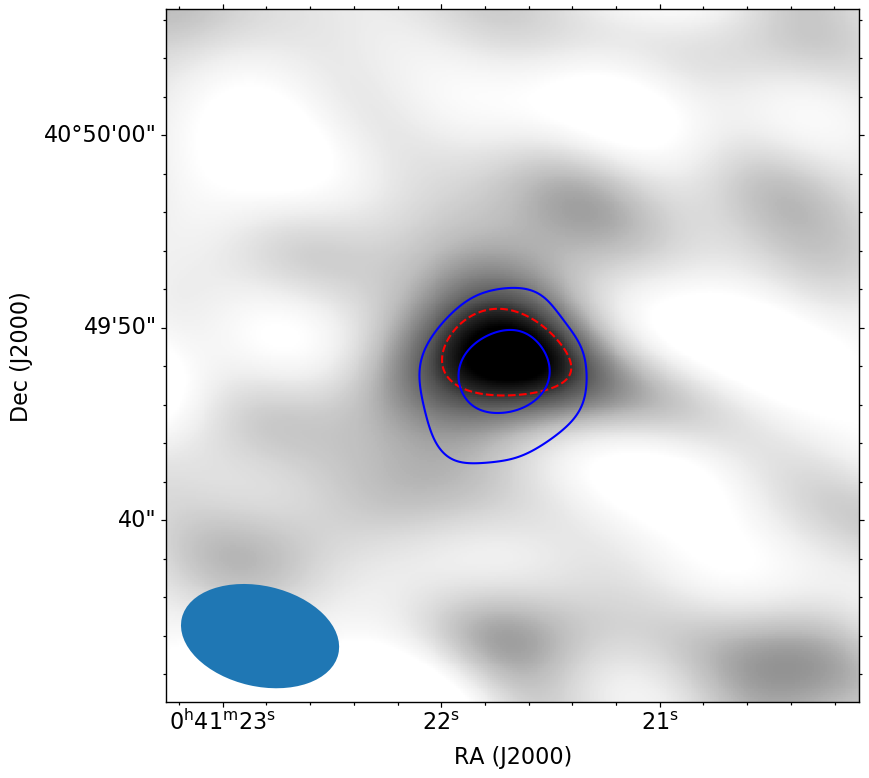}{0.40\textwidth}{}
          \fig{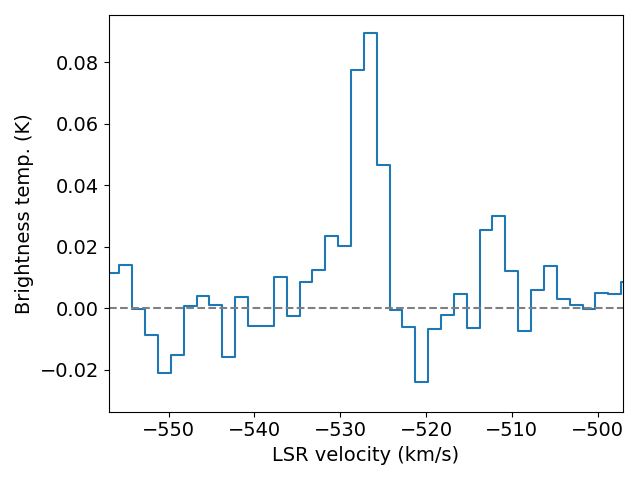}{0.48\textwidth}{}
          }
\gridline{\fig{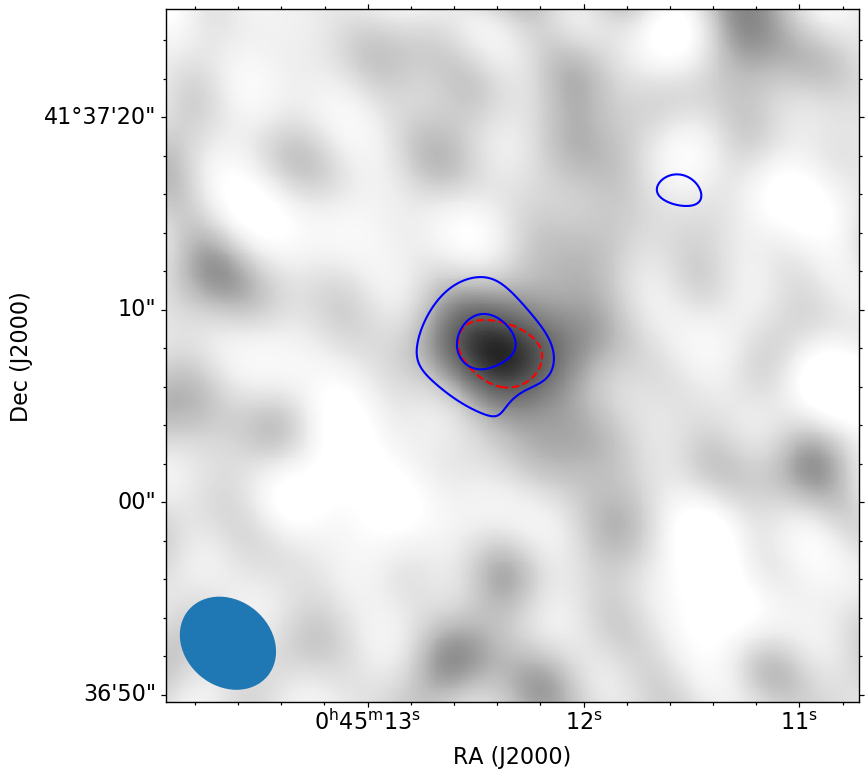}{0.40\textwidth}{}
          \fig{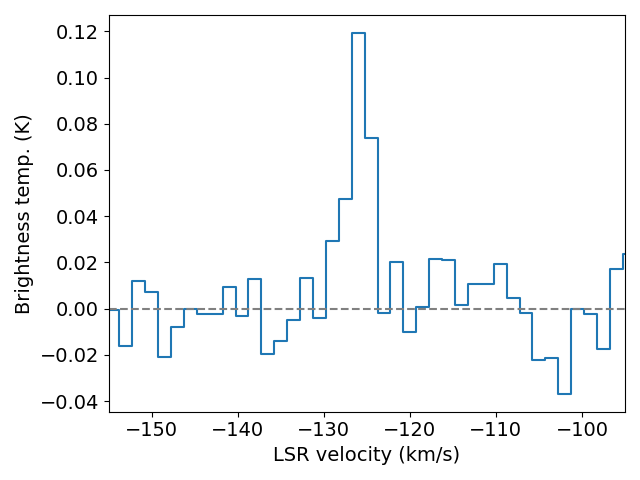}{0.48\textwidth}{}
          }
\figurenum{2}
\caption{The three highest-S/N detections of C$^{18}$O(2-1), in K026 (top panels), K092 (center panels) and K162 (bottom panels). In each case, the left panel shows the channel map for the peak spectral channel, and the corresponding peak spectrum in the right-hand panel. The dust continuum at 3$\sigma$ and 6$\sigma$ (blue) and C$^{18}$O(2-1) at 3$\sigma$ (red) are indicated with contours in the left-hand panels.}\label{fig:c18o}
\end{figure*}

\subsection{Detection of \ceighto}

Since both $^{12}$CO and $^{13}$CO emission is detected at very high signal-to-noise, it is worthwhile to also search for the weaker C$^{18}$O(2-1) isotopologue in these clouds. Using the $^{12}$CO peak velocity as a reference we searched each of the clouds for statistically significant emission lines at the expected  C$^{18}$O frequency. Figure~\ref{fig:c18o} shows \ceighto spectra in the clouds with the highest S/N line profiles.

These are, to our knowledge, among the first detections of C$^{18}$O emission lines from individual GMCs at this spatial resolution in an external galaxy (cf. \citealp{tok20}). The low abundance of this isotopologue ensures optically thin lines and thus provides a look into the deeper, presumably more dense parts of the GMCs. This experiment demonstrates that C$^{18}$O emission can be used as a potential probe of dense gas in individual extragalactic GMCs. It opens possibilities to test the proposed tight connection between dense gas and the star formation rate \citep[e.g.,][]{2010ApJ...724..687L, 2012ApJ...745..190L} in GMC populations beyond the Milky Way. For four clouds the quality and SNR of the \ceighto line within the dust emission mask allows for more detailed analysis (see also Table~\ref{tab:clouds_c18o}).

Although we clearly have detected the \ceighto emission line in four M31 dust clouds, for the rest of the clouds in our sample this signal may be hidden below the noise. However, an average C$^{18}$O profile with better signal-to-noise properties can be constructed by stacking the individual GMC spectra. This is possible because, given the high quality of the $^{12}$CO and $^{13}$CO spectra, we know exactly what velocity channels are likely to contain C$^{18}$O emission in each cloud. Figure 3 shows the result of median stacking 29 individual $^{12}$CO, \thco and \ceighto spectra\footnote{In 3 of the 32 cataloged clouds the \ceighto line was at the edge of the bandpass due to tuning and could not be measured.} after first spatially integrating the spectra over the spatial dust mask.  Doing so we are able to boost the signal-to-noise ratio of the \ceighto line to 17, producing a prominent and narrow spectral profile. 

\begin{figure*} 
\gridline{\fig{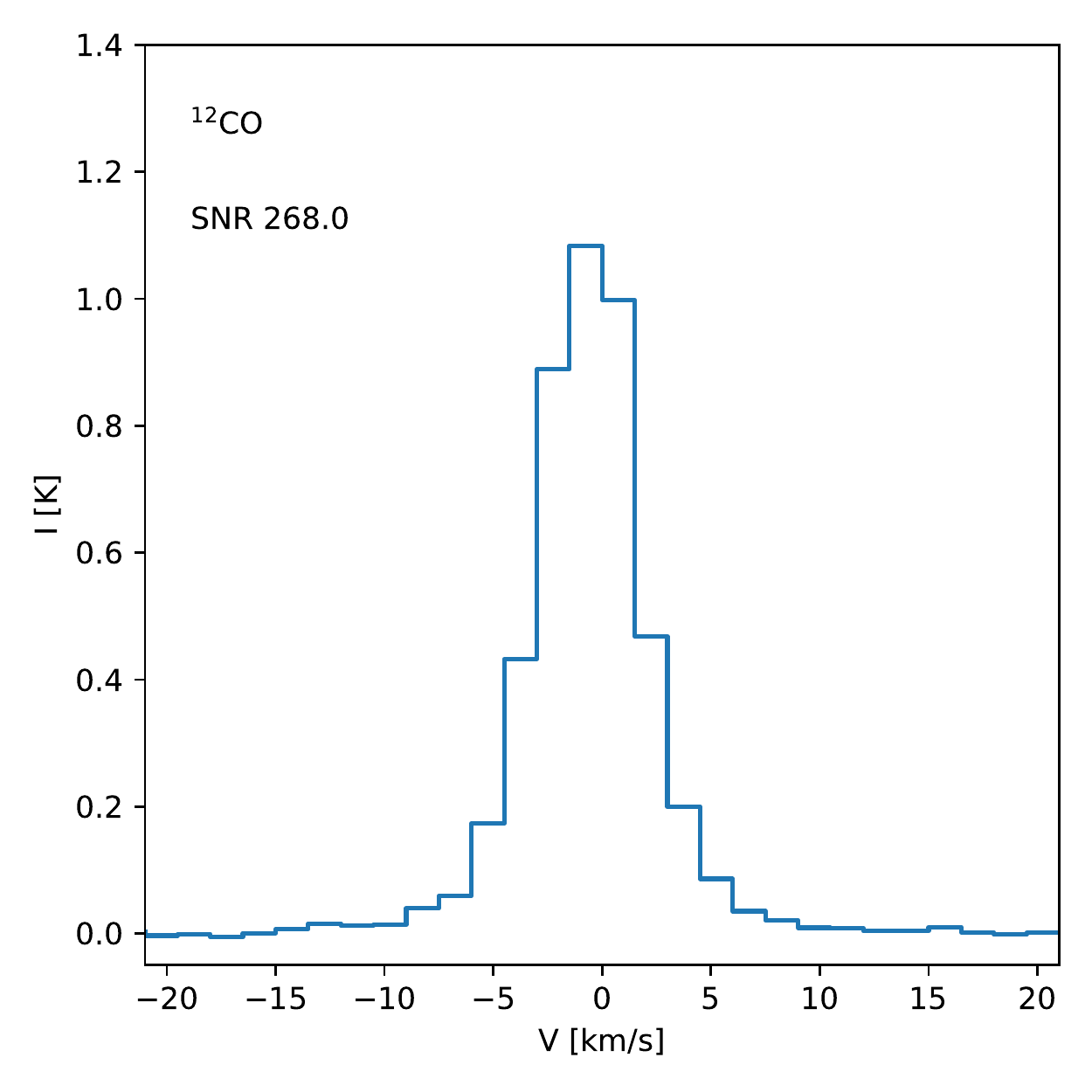}{0.33\textwidth}{}
          \fig{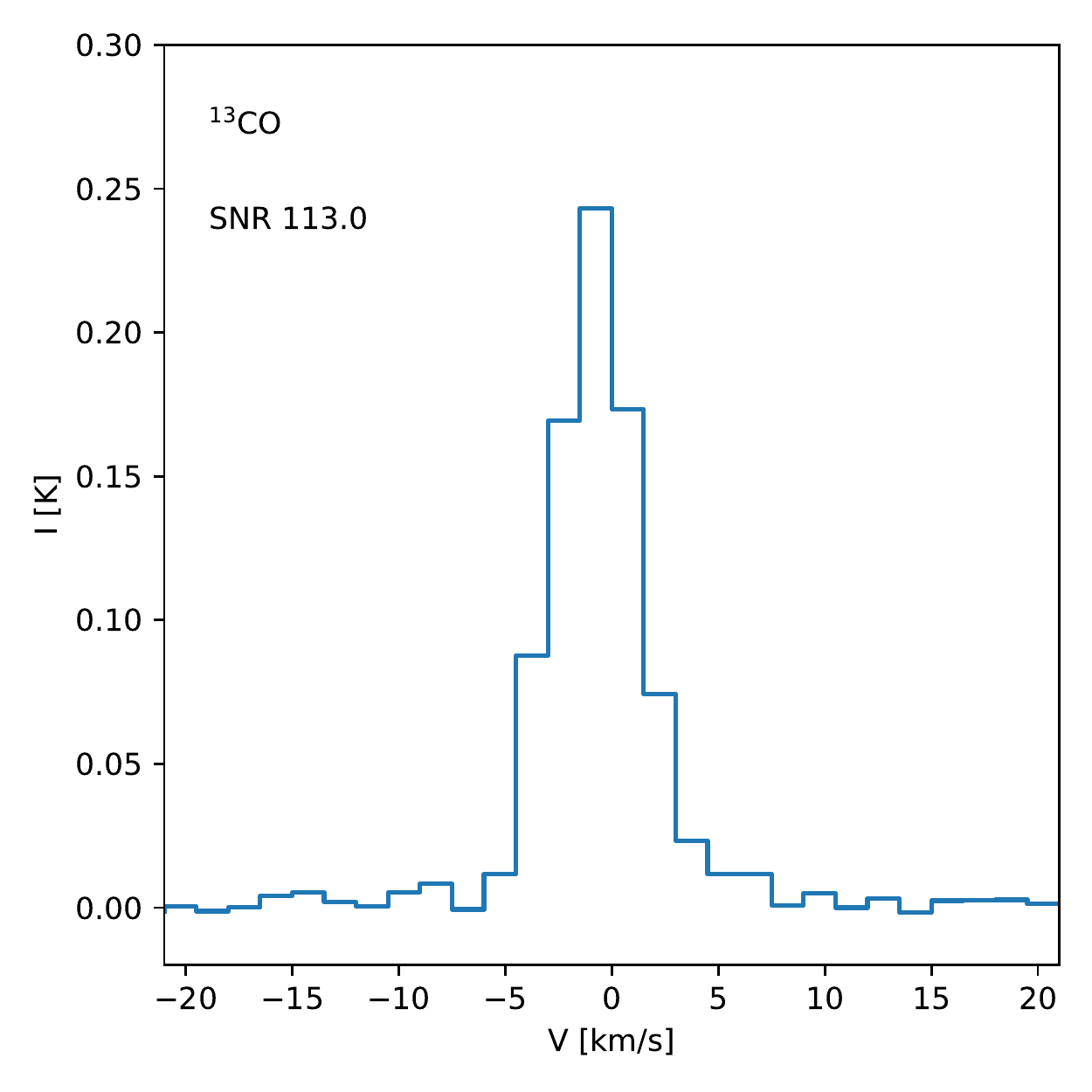}{0.33\textwidth}{}
          \fig{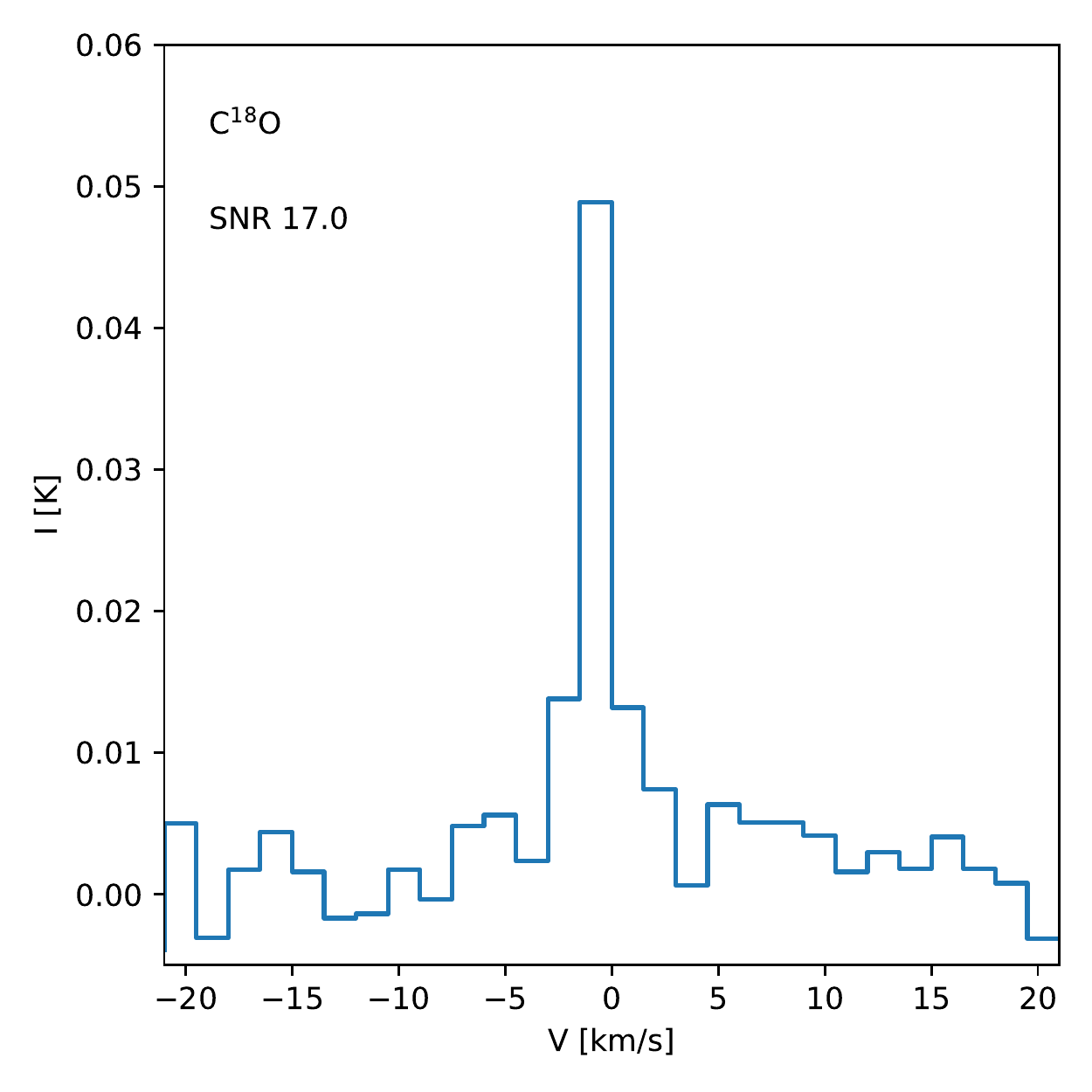}{0.33\textwidth}{}
          }
\figurenum{3}
\caption{Median stacked spectra of 29 clouds for the three CO isotopologues. The resulting signal-to-noise is indicated in the top left of each panel. Note that the y-axis has a different range in each panel.\label{fig:stack_C18O}}
\end{figure*}

\section{Physical Cloud Properties}

\subsection{Dust Masses}

Following Paper I we we define spatial continuum masks based on a 3$\sigma$ detection contour. The 3$\sigma$ threshold was computed as three times the RMS noise in the continuum image. One third of the continuum images harbour more than one distinct emission peak within or adjacent to a GMC.  We split the mask into multiple dust clouds when needed using the \texttt{AgglomerativeClustering} algorithm from \texttt{scikit-learn} \citep{scikit-learn}. This hierarchical clustering method successively merges pixels together until a set number of clusters remain. For the merge condition we found the best performance with average linkage. As a result, we obtain $32$ individual (dust) clouds to be used for further analysis. The sample contains $10$ resolved continuum sources, meaning their spatial extent is at least $20\%$ larger than the synthesized beam.

To convert the continuum emission within one mask into a dust mass $M_\mathrm{dust}$ we rely on a modified black body emission model \citep{Hildebrand1983}: 
\begin{equation} \label{eq:mdust}
    M_\mathrm{dust} = S_\nu D^2/(\kappa_\nu B_\nu(T_\mathrm{dust}))
\end{equation}
Where $S_\nu$ is the continuum flux and D the distance to the source. As outlined in \citet{Forbrich2020}, we take the opacity $\kappa_\nu$ from the dust model of \citet{Jones2017} for Milky Way dust. At 230 GHz this becomes $\kappa_\nu = 0.0425$ m$^2$kg$^{-1}$. We further assume a temperature $T_\mathrm{dust}$ of $20K$ for the cold dust component. The dust masses computed for each of the 32 clouds are listed in Table~\ref{tab:clouds}. Although 23 of the clouds were studied in Paper I we report them here as well, since some previously reported targets now have better data, and also for completeness and to facilitate comparison with the \thco and new \twco observations reported here for the first time.

\begin{longrotatetable}
\begin{deluxetable}{lccccccccccc}
\tablecaption{Cloud properties. \label{tab:clouds}}
\tablehead{
Cloud & $R$ & $F_\mathrm{cont}$ &  RMS & Beam size & $M_\mathrm{dust}$  &  $I_{12}$ &  $I_{13}$ &  $\alpha^\prime_\mathrm{12,dust}$ & $\alpha^\prime_\mathrm{13,dust}$ & $\frac{I_{12}}{I_{12}}$ & $N^\mathrm{tot}_{13}$ \\
  \\
 & kpc & mJy & mJy/beam & $a\times b$ (arcsec) & $M_\odot$ & $[\mathrm{K}\,\mathrm{km}\,\mathrm{s}^{-1}]$ & $[\mathrm{K}\,\mathrm{km}\,\mathrm{s}^{-1}]$ & $[\frac{M_\odot}{\mathrm{K}\,\mathrm{km}\,\mathrm{s}^{-1}\mathrm{pc}^{2}}]$ & $[\frac{M_\odot}{\mathrm{K}\,\mathrm{km}\,\mathrm{s}^{-1}\mathrm{pc}^{2}}]$ & & $10^{15}cm^{-2}$ \\
}
\decimalcolnumbers
\startdata
K026A\tablenotemark{a} &  5.8 &  $2.91 \pm 0.33$ &  0.19 & $4.5\times4.0$ & $ 828 \pm  93$ &  $ 15.17 \pm 0.13$ &  $3.37 \pm 0.13$ & $0.05 \pm 0.01$ & $0.26 \pm 0.05$ &  $ 4.50 \pm 0.06$ & $47.3 \pm 0.8$ \\
K026B &                   5.8 &  $1.21 \pm 0.19$ &  0.19 & $4.5\times4.0$ & $ 344 \pm  53$ &  $  4.66 \pm 0.14$ &  $0.80 \pm 0.14$ & $0.12 \pm 0.03$ & $0.86 \pm 0.20$ &  $ 5.82 \pm 0.33$ & $26.9 \pm 1.8$ \\
K048A &                   5.7 &  $0.86 \pm 0.27$ &  0.27 & $4.4\times3.5$ & $ 244 \pm  77$ &  $ 15.65 \pm 0.13$ &  $2.72 \pm 0.13$ & $0.06 \pm 0.02$ & $0.28 \pm 0.09$ &  $ 5.76 \pm 0.09$ & $34.2 \pm 0.7$ \\
K092A &                   8.0 &  $1.45 \pm 0.23$ &  0.19 & $6.9\times5.0$ & $ 411 \pm  66$ &  $ 10.28 \pm 0.14$ &  $2.39 \pm 0.14$ & $0.04 \pm 0.01$ & $0.18 \pm 0.05$ &  $ 4.30 \pm 0.10$ & $50.2 \pm 1.5$ \\
K093A &                   8.1 &  $1.13 \pm 0.20$ &  0.20 & $4.2\times3.6$ & $ 321 \pm  58$ &  $  7.75 \pm 0.14$ &  $1.91 \pm 0.14$ & $0.12 \pm 0.03$ & $0.43 \pm 0.11$ &  $ 4.06 \pm 0.11$ & $42.1 \pm 1.5$ \\
K093B &                   8.1 &  $0.64 \pm 0.20$ &  0.20 & $4.2\times3.6$ & $ 181 \pm  58$ &  $ 10.52 \pm 0.14$ &  $2.85 \pm 0.14$ & $0.06 \pm 0.02$ & $0.20 \pm 0.07$ &  $ 3.69 \pm 0.07$ & $64.6 \pm 1.6$ \\
K134B\tablenotemark{b} & 10.5 &  $0.92 \pm 0.27$ &  0.27 & $5.5\times4.7$ & $ 260 \pm  76$ &  $  6.55 \pm 0.13$ &  $1.34 \pm 0.13$ & $0.04 \pm 0.01$ & $0.20 \pm 0.06$ &  $ 4.89 \pm 0.17$ & $30.0 \pm 1.3$ \\
K136A &                  11.3 &  $2.26 \pm 0.31$ &  0.20 & $5.6\times4.7$ & $ 644 \pm  90$ &  $  7.55 \pm 0.13$ &  $1.59 \pm 0.13$ & $0.05 \pm 0.01$ & $0.30 \pm 0.10$ &  $ 4.73 \pm 0.14$ & $32.8 \pm 0.9$ \\
K142A &                  11.6 &  $0.79 \pm 0.25$ &  0.25 & $8.1\times5.0$ & $ 225 \pm  72$ &  $  4.56 \pm 0.13$ &  $0.85 \pm 0.13$ & $0.04 \pm 0.01$ & $0.21 \pm 0.07$ &  $ 5.37 \pm 0.31$ & $22.2 \pm 1.6$ \\
K153A &                  11.2 &  $0.77 \pm 0.21$ &  0.21 & $8.1\times5.0$ & $ 220 \pm  59$ &  $  8.23 \pm 0.14$ &  $1.01 \pm 0.14$ & $0.03 \pm 0.01$ & $0.22 \pm 0.07$ &  $ 8.18 \pm 0.35$ & $22.9 \pm 1.1$ \\
K154A &                  11.4 &  $1.02 \pm 0.20$ &  0.20 & $8.0\times4.9$ & $ 291 \pm  58$ &  $  4.28 \pm 0.14$ &  $0.62 \pm 0.14$ & $0.10 \pm 0.03$ & $0.44 \pm 0.15$ &  $ 6.92 \pm 0.49$ & $23.4 \pm 1.8$ \\
K154C &                  11.4 &  $0.69 \pm 0.20$ &  0.20 & $8.0\times4.9$ & $ 196 \pm  58$ &  $  1.84 \pm 0.14$ &  $0.08 \pm 0.14$ & $0.27 \pm 0.09$ & $1.11 \pm 0.42$ &  $22.58\pm 11.29$ & $4.2  \pm 3.9$ \\
K154B &                  11.4 &  $0.92 \pm 0.20$ &  0.20 & $8.0\times4.9$ & $ 261 \pm  58$ &  $  2.28 \pm 0.14$ &  $0.24 \pm 0.14$ & $0.12 \pm 0.03$ & $0.56 \pm 0.20$ &  $ 9.37 \pm 1.54$ & $12.0 \pm 2.5$ \\
K157A &                  11.9 &  $1.69 \pm 0.31$ &  0.31 & $8.4\times5.1$ & $ 482 \pm  89$ &  $  7.11 \pm 0.14$ &  $1.02 \pm 0.14$ & $0.06 \pm 0.01$ & $0.39 \pm 0.11$ &  $ 7.00 \pm 0.30$ & $23.8 \pm 1.2$ \\
K157B &                  11.9 &  $0.97 \pm 0.31$ &  0.31 & $8.4\times5.1$ & $ 274 \pm  89$ &  $  3.22 \pm 0.14$ &  $0.39 \pm 0.14$ & $0.10 \pm 0.03$ & $0.58 \pm 0.19$ &  $ 8.24 \pm 0.93$ & $23.2 \pm 2.3$ \\
K160A &                  12.3 &  $0.99 \pm 0.22$ &  0.22 & $8.1\times5.1$ & $ 282 \pm  62$ &  $  3.33 \pm 0.13$ &  $0.53 \pm 0.13$ & $0.07 \pm 0.02$ & $0.54 \pm 0.18$ &  $ 6.25 \pm 0.54$ & $26.6 \pm 3.0$ \\
K160B &                  12.3 &  $0.70 \pm 0.22$ &  0.22 & $8.1\times5.1$ & $ 199 \pm  62$ &  $  4.96 \pm 0.13$ &  $0.55 \pm 0.13$ & $0.04 \pm 0.01$ & $0.32 \pm 0.11$ &  $ 8.99 \pm 0.68$ & $29.2 \pm 2.6$ \\
K160C\tablenotemark{c} & 12.3 &  $1.19 \pm 0.22$ &  0.22 & $8.1\times5.1$ & $ 338 \pm  62$ &  $  4.28 \pm 0.13$ &  $0.79 \pm 0.13$ & $0.07 \pm 0.02$ & $0.40 \pm 0.13$ &  $ 5.43 \pm 0.32$ & $33.5 \pm 2.4$ \\
K162A &                  11.8 &  $1.20 \pm 0.22$ &  0.19 & $5.2\times4.3$ & $ 342 \pm  62$ &  $ 21.38 \pm 0.14$ &  $3.56 \pm 0.14$ & $0.02 \pm 0.01$ & $0.17 \pm 0.04$ &  $ 6.01 \pm 0.07$ & $37.8 \pm 0.5$ \\
K162B &                  11.8 &  $0.64 \pm 0.19$ &  0.19 & $5.2\times4.3$ & $ 181 \pm  53$ &  $ 10.60 \pm 0.13$ &  $1.62 \pm 0.13$ & $0.03 \pm 0.01$ & $0.22 \pm 0.07$ &  $ 6.56 \pm 0.18$ & $34.8 \pm 1.4$ \\
K170A &                  11.8 &  $1.46 \pm 0.29$ &  0.23 & $4.5\times3.6$ & $ 417 \pm  84$ &  $ 21.13 \pm 0.13$ &  $3.12 \pm 0.13$ & $0.03 \pm 0.01$ & $0.24 \pm 0.07$ &  $ 6.77 \pm 0.10$ & $55.7 \pm 0.9$ \\
K170B &                  11.8 &  $1.00 \pm 0.23$ &  0.23 & $4.5\times3.6$ & $ 285 \pm  65$ &  $ 12.78 \pm 0.13$ &  $2.01 \pm 0.13$ & $0.06 \pm 0.02$ & $0.35 \pm 0.09$ &  $ 6.36 \pm 0.14$ & $24.6 \pm 0.7$ \\
K176A &                  13.8 &  $1.71 \pm 0.29$ &  0.24 & $8.1\times5.1$ & $ 486 \pm  84$ &  $ 10.14 \pm 0.13$ &  $1.86 \pm 0.13$ & $0.04 \pm 0.01$ & $0.24 \pm 0.06$ &  $ 5.45 \pm 0.13$ & $31.2 \pm 1.0$ \\
K190A &                  13.9 &  $1.63 \pm 0.22$ &  0.17 & $4.9\times4.2$ & $ 463 \pm  63$ &  $  4.28 \pm 0.14$ &  $0.45 \pm 0.14$ & $0.16 \pm 0.04$ & $1.68 \pm 0.73$ &  $ 9.59 \pm 0.92$ & $11.5 \pm 1.5$ \\
K191A &                  12.1 &  $6.33 \pm 0.46$ &  0.21 & $3.3\times3.2$ & $1804 \pm 130$ &  $ 35.29 \pm 0.14$ &  $4.12 \pm 0.14$ & $0.07 \pm 0.01$ & $0.34 \pm 0.07$ &  $ 8.57 \pm 0.09$ & $38.3 \pm 0.5$ \\
K213A &                  12.0 &  $3.91 \pm 0.52$ &  0.31 & $4.3\times3.7$ & $1115 \pm 147$ &  $ 19.68 \pm 0.13$ &  $4.02 \pm 0.13$ & $0.07 \pm 0.01$ & $0.35 \pm 0.08$ &  $ 4.90 \pm 0.06$ & $59.3 \pm 1.1$ \\
K213B &                  12.0 &  $3.37 \pm 0.46$ &  0.31 & $4.3\times3.7$ & $ 959 \pm 131$ &  $ 12.82 \pm 0.13$ &  $2.25 \pm 0.13$ & $0.12 \pm 0.03$ & $0.67 \pm 0.19$ &  $ 5.69 \pm 0.11$ & $30.2 \pm 0.7$ \\
K213C &                  12.0 &  $1.03 \pm 0.31$ &  0.31 & $4.3\times3.7$ & $ 294 \pm  89$ &  $ 11.63 \pm 0.13$ &  $1.70 \pm 0.13$ & $0.07 \pm 0.02$ & $0.38 \pm 0.12$ &  $ 6.83 \pm 0.18$ & $33.0 \pm 1.1$ \\
K239A &                   5.9 &  $1.34 \pm 0.23$ &  0.23 & $7.8\times4.9$ & $ 382 \pm  64$ &  $  3.37 \pm 0.14$ &  $0.45 \pm 0.14$ & $0.09 \pm 0.02$ & $0.68 \pm 0.19$ &  $ 7.50 \pm 0.74$ & $14.9 \pm 2.5$ \\
K291A &                  16.0 &  $1.09 \pm 0.22$ &  0.22 & $4.4\times3.8$ & $ 311 \pm  62$ &  $  3.43 \pm 0.14$ &  $0.24 \pm 0.14$ & $0.23 \pm 0.06$ & $1.86 \pm 0.72$ &  $14.20 \pm 2.42$ & $14.2 \pm 2.4$ \\
K291B &                  16.0 &  $0.67 \pm 0.22$ &  0.22 & $4.4\times3.8$ & $ 191 \pm  62$ &  $  1.51 \pm 0.13$ &  $0.26 \pm 0.13$ & $0.40 \pm 0.13$ & $1.68 \pm 0.55$ &  $ 5.84 \pm 1.03$ & $23.6 \pm 4.5$ \\
K297A &                  15.9 &  $2.06 \pm 0.25$ &  0.25 & $7.8\times5.0$ & $ 587 \pm  71$ &  $  7.58 \pm 0.12$ &  $1.05 \pm 0.12$ & $0.07 \pm 0.01$ & $0.48 \pm 0.10$ &  $ 7.24 \pm 0.30$ & $24.6 \pm 1.3$ \\
\enddata
\tablenotetext{a}{The dust mass of K026A is $36\%$ higher than in paper~I because the addition of new data drove down the RMS, picking up more continuum flux above the $3\sigma$ threshold.}
\tablenotetext{b}{K134A from paper~I was found to have a SNR of $2.8\sigma$ and did not make significance cut for this study.}
\tablenotetext{c}{K160C was not included in paper~I due to a stricter radial cut from the phase center.}
\tablecomments{(1) cloud name, (2) Galactocentric radius \citep{Kirk2015} (3) Continuum flux, (4) continuum RMS, (5) Continuum beam FWHM of major and minor axis (6) Dust mass, (7-8) Integrated CO intensities, (9-10) CO line conversion factors, (11) line intensity ratio, (12) isotopologue column density.}
\end{deluxetable}
\end{longrotatetable}

\begin{deluxetable}{lcccc}
\tablecaption{\ceighto cloud properties \label{tab:clouds_c18o}}
\tablehead{
Cloud & $R$ & $I_{18}$ & $\frac{I_{13}}{I_{18}}$ & $N^\mathrm{tot}_{18}$ \\
  \\
 & kpc & $[\mathrm{K}\,\mathrm{km}\,\mathrm{s}^{-1}]$ & & $10^{15}cm^{-2}$ \\
}
\decimalcolnumbers
\startdata
K026A &  5.8 & $  0.36 \pm 0.07$ &  $ 9.41 \pm 1.83$ & $4.3 \pm 1.1$  \\
K092A &  8.0 & $  0.20 \pm 0.07$ &  $11.90 \pm 3.93$ & $4.3 \pm 1.6$  \\
K136A & 11.3 & $  0.20 \pm 0.06$ &  $ 7.82 \pm 2.46$ & $3.7 \pm 1.4$  \\
K162A & 11.8 & $  0.27 \pm 0.07$ &  $12.99 \pm 3.32$ & $2.3 \pm 0.8$  \\
\enddata
\tablecomments{(1) cloud name (2) galactocentric radius \citep{Kirk2015} (3) integrated \ceighto intensity (4) line intensity ratio, (5) isotopologue column density}
\end{deluxetable}

\subsection{CO Conversion Factors} \label{sec:COfactor}

Molecular hydrogen, H$_2$, is the primary constituent of GMCs, but it is not observable in emission in the cold physical conditions that characterize these GMCs. Consequently, CO, as the next most abundant molecular species and a bright emitter, has become the most utilized surrogate for tracing both the extent and the mass of the cold gas in GMCs.

The conversion from CO light to cloud mass involves a conversion factor $\alpha_\mathrm{CO}$ or $X_\mathrm{CO}$, which is traditionally determined for the $^{12}$CO(1-0) transition (but which we here measure for the $^{12}$CO(2-1) transition). The empirical calibration of the CO conversion factor has proven difficult because it requires an independent estimate of H${_2}$ mass \citep[see e.g.][for a review]{Bolatto2013}. Usually this has been done by comparison of CO emission with measurements of dust column density or mass because dust is considered a more direct tracer of H$_2$ by virtue of its constant and well established abundance relative to hydrogen in the Milky Way \citep[e.g.,][]{1978ApJ...224..132B,2002ApJ...577..221R} 

The CO conversion factor is typically defined as $\alpha_\mathrm{CO} = M_\mathrm{tot}/L_\mathrm{CO}$, where $M_\mathrm{tot}$ is the total mass of the cloud in units of solar mass and $L_\mathrm{CO}$ is the luminosity of the CO line, in units of $\mathrm{K}\,\mathrm{km}\,\mathrm{s}^{-1}\mathrm{pc}^{2}$. Since we are interested in using dust to calibrate cloud masses for the $\alpha_\mathrm{CO}$ measurement, both $M_\mathrm{tot}$ and $L_\mathrm{CO}$ are computed only within the continuum mask for each cloud. Following Paper I we first derive the more direct observational quantity $\alpha^\prime_{CO-dust} = M_\mathrm{dust}/L_\mathrm{CO}$, where $M_\mathrm{dust}$ is the dust mass within the continuum mask of each cloud that is derived from eq.~\ref{eq:mdust}. The value of $\alpha_\mathrm{CO}$ follows from knowledge of the dust-to-gas ratio, $R_{g2d}$, i.e., 
\begin{equation}
\alpha_\mathrm{CO} = R_{g2d} \times \alpha^\prime_{CO-dust}
\end{equation}

\begin{figure*} 
    \fig{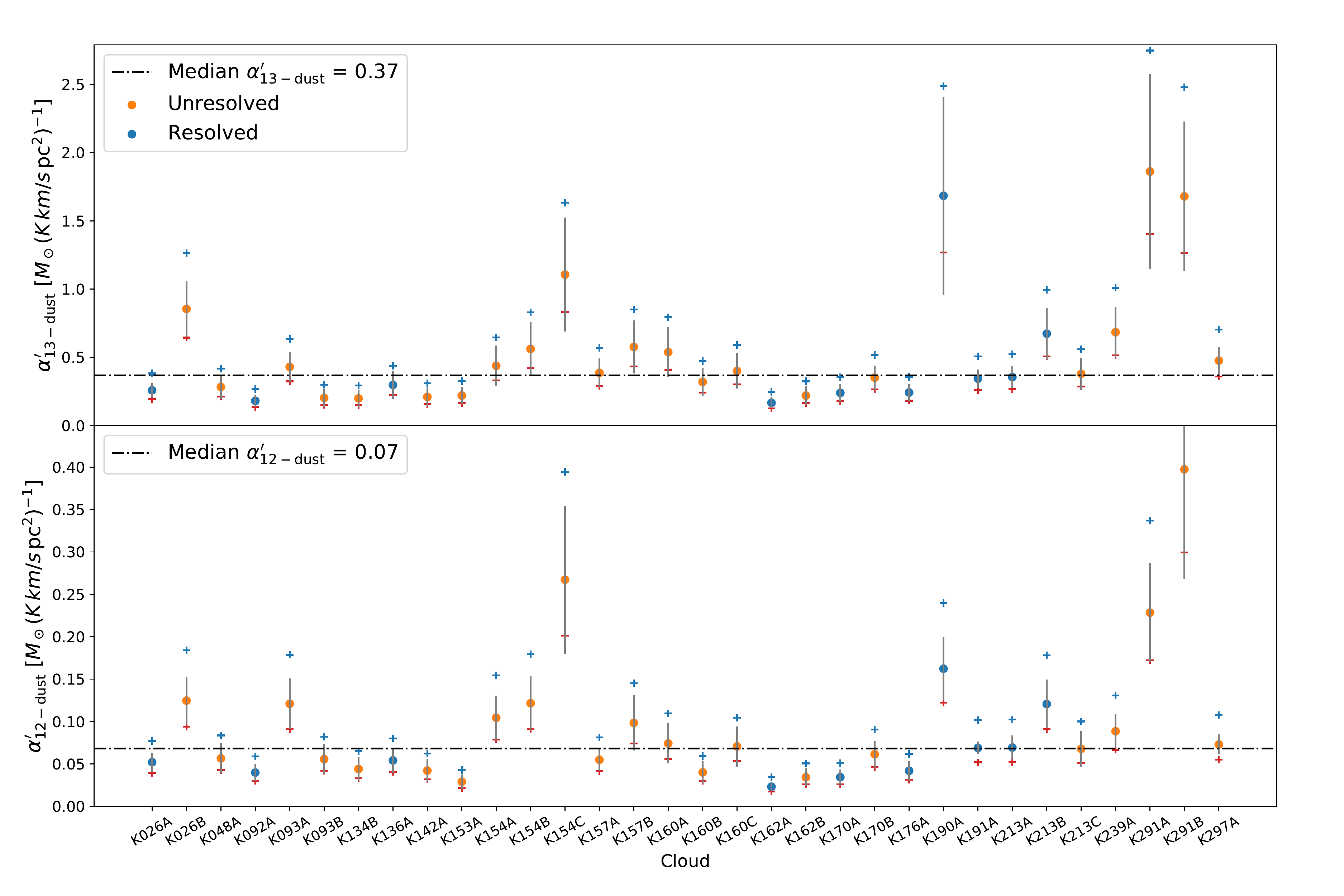}{\textwidth}{}
    \figurenum{4}
    \caption{CO-dust conversion factors $\alpha^\prime_{12-\mathrm{dust}}$ and $\alpha^\prime_{13-\mathrm{dust}}$ for $^{12}$CO and $^{13}$CO, respectively, for individual dust clouds. Blue dots correspond to resolved clouds. The range indicated by the blue and red crosses corresponds to changes in the assumed dust temperature between 15 and 25K, respectively. \label{fig:alpha}}
\end{figure*}

\subsubsection{\twco and \thco}

Figure \ref{fig:alpha} shows the conversion factors ($\alpha'$) we derived for both the $^{12}$CO and $^{13}$CO lines toward individual GMCs in M31. There is considerable scatter around the median values, with no distinction between resolved or unresolved dust clouds. There are a few clouds with significantly higher $\alpha^\prime_{\mathrm{CO}-\mathrm{dust}}$ and uncertainties (see also Table~\ref{tab:clouds}). Most of these (K026B, K190A and K291A) were already identified by \citet{Forbrich2020} as candidate sources for non-dust continuum emission. For the full dust cloud sample we find a median  $\alpha^\prime_{12}$= 0.068\  $\mathrm{M}_\odot\,(\mathrm{K}\,\mathrm{km}\,\mathrm{s}^{-1}\mathrm{pc}^{2})^{-1}$ and $\alpha^\prime_{13} = 0.37$ $\mathrm{M}_\odot\,(\mathrm{K}\,\mathrm{km}\,\mathrm{s}^{-1}\mathrm{pc}^{2})^{-1}$. 

Following Paper I we also calculate the mean value for $\alpha^\prime_{12}$ after eliminating the four most extreme outliers in Figure 4, as these are likely to be contaminants, as well as  the source K026B which was shown in Paper I to likely constitute an unrelated extragalactic object. We find $<\alpha^\prime_{12}> =$ 0.064 $\pm$ 0.029 $\mathrm{M}_\odot\,(\mathrm{K}\,\mathrm{km}\,\mathrm{s}^{-1}\mathrm{pc}^{2})^{-1}$. This is in excellent agreement with the median value and also with our previous estimate (0.06) for the J $=$ 2-1 transition (i.e., Paper I and  \citealp{for20err}). It is somewhat higher than, but consistent within the uncertainties, with the value of 0.045 $\mathrm{M}_\odot\,(\mathrm{K}\,\mathrm{km}\,\mathrm{s}^{-1}\mathrm{pc}^{2})^{-1}$, estimated for Milky Way GMCs in Paper I. 

Similarly for the same clouds we find $<\alpha^\prime_{13}> =$ 0.36 $\pm$ 0.15 $\mathrm{M}_\odot\,(\mathrm{K}\,\mathrm{km}\,\mathrm{s}^{-1}\mathrm{pc}^{2})^{-1}$, essentially the same as the median value listed above. There is as yet no agreed upon empirical measurement or estimate of $\alpha^\prime_{13}$ for Milky Way GMCs. Our measurements of $\alpha^\prime_{13}$ in M31 are likely unique in terms of the size and uniformity of the cloud sample for which they were determined.

By assuming a gas to dust ratio, R$_{g2d}$, one can derive total gas masses and the corresponding mean values of $\alpha_{CO}$ for the clouds in the sample:
$\alpha_{12} = (8.7 \pm 3.9) \times [R_{g2d}/136]$~${M}_\odot\,(\mathrm{K}\,\mathrm{km}\,\mathrm{s}^{-1}\mathrm{pc}^{2})^{-1}$ and $\alpha_{13} = (48.9 \pm 20.4) \times [R_{g2d}/136]$~${M}_\odot\,(\mathrm{K}\,\mathrm{km}\,\mathrm{s}^{-1}\mathrm{pc}^{2})^{-1}$. Here $R_{g2d}$ is defined to include the presence of primordial helium ($36\%$). This also assumes the Milky Way value of $R_{g2d} = 136$ which is within the range of gas-to-dust ratios derived for M31 \citep{Leroy2011,Smith2012, Draine2014}. For comparison we note that for Galactic GMCs, $\alpha_{CO(MW)} = 6.1\  {M}_\odot\,(\mathrm{K}\,\mathrm{km}\,\mathrm{s}^{-1}\mathrm{pc}^{2})^{-1}$ for the $J=2-1$ transition of CO which, though a  lower value, is within the uncertainties of the M31 value. Our value is also within the range (3--14) inferred for selected regions in M31 on much larger spatial scales (170 pc) by Leroy et al. (2011) from {\it Spitzer} infrared, CO and H\,{\sc i} observations.  To illustrate the magnitude of a possible systematic uncertainty due to our assumption of a constant dust temperature of 20 K for all clouds, we additionally calculated the dust masses and corresponding $\alpha^\prime$s assuming dust temperatures of $15$K and $25$K. The range in possible values of $\alpha^\prime_{12}$ and $\alpha^\prime_{13}$ are indicated in Figure 4. We furthermore observe no significant difference between resolved and unresolved sources, which indicates that possible differences in the CO and dust beam filling factors are not affecting these measurements. Our method effectively probes the conversion factor in the GMCs by measuring $\alpha^\prime$ in the dust emitting areas of the GMCs. Differences between clouds inside the same GMC do not exceed the general scatter in the sample. We thus expect these probes to be representative of the individual GMCs in our sample. Finally we emphasize here that the above conversion factors were derived for the J(2-1) transitions of CO. To convert these values to ones appropriate for another transition of CO  one needs to apply a correction factor. For example,  the factor typically used in the literature to convert J=2-1 values to  J=1-0 values is  0.7 \citep[][and references therein]{Bolatto2013, Nishimura2015}.

\subsubsection{C$^{18}$O }

For the stacked spectra of 29 clouds the median dust mass surface density is also obtained. This again allows the computation of a conversion factor. We thus find a $\alpha^\prime_\mathrm{18,dust}$ conversion factor of $2.54^{+0.18}_{-0.23}$  \,\msun $(\mathrm{K}\,\mathrm{km}\,\mathrm{s}^{-1}\mathrm{pc}^{2})^{-1}$ corresponding to $\alpha_{18} = 345^{+25}_{-31}$\, \msun $(\mathrm{K}\,\mathrm{km}\,\mathrm{s}^{-1}\mathrm{pc}^{2})^{-1}$ with $R_{g2d} = 136$. This can be seen as a typical and representative conversion factor for the C$^{18}$O line for molecular clouds. It is the first time this number was derived at this spatial scale for extragalactic sources. It allows users to estimate the total molecular mass using an optically thin tracer of dense material in GMCs where \ceighto is detected. 

As a reference and sanity check we also derive the conversion factor for the stacked spectra of the two brighter isotopologues, for the same 29 clouds. We find  $\alpha^\prime_\mathrm{12,dust} = 0.072^{+0.006}_{-0.007}  \,{M}_\odot\, (\mathrm{K}\,\mathrm{km}\,\mathrm{s}^{-1}\mathrm{pc}^{2})^{-1}$ and $\alpha_{13} = 0.36^{+0.04}_{-0.03}  \,{M}_\odot\, (\mathrm{K}\,\mathrm{km}\,\mathrm{s}^{-1}\mathrm{pc}^{2})^{-1}$. These numbers are perfectly consistent with the values derived from individual clouds. We are thus confident that the derived value for C$^{18}$O is representative for the GMC population in Andromeda.

The uncertainties on the stacked products was estimated from Monte Carlo simulations. Clouds were removed from the stack at random, not least to evaluate the inadvertent inclusion of outlier contaminant sources. Between one to nine clouds were removed (i.e. up to one third of the sample) with 1000 iterations per number of clouds removed. The effect on the resulting spectra and the derived properties was subsequently analysed. The stacking procedure proved to be stable for the removal of clouds, with typical uncertainties of around $7-11\%$ on the conversion factors.

\subsection{Line Ratios and Column Densities}

In Sect.~\ref{sec:COfactor} the spectrally integrated moment zero maps were used to estimate the conversion between dust and gas emission within the dust emitting regions of the clouds. In this section we preserve the spectral information but spatially integrate the spectra of each cloud within the mask set by the dust emission. The integrated spectra of $^{12}$CO and $^{13}$CO are shown in Appendix~\ref{app:spectra}. The strength of the spectra varies significantly from cloud to cloud, with K191A having the highest peak flux. It is reassuring that for all clouds in Fig.~\ref{fig:spectra}, the $^{13}$CO line mimics the $^{12}$CO line closely in shape, with no offsets between the peaks. The FWHM of both species are also comparable, with a median value of $5.9$ km/s for $^{12}$CO and $4.9$ km/s for $^{13}$CO. 
This allows us to further compare the spectra, based on the assumption that they come from the same clouds. We note that at least $6 \, (19\%)$ sources exhibit a strong secondary peak at an offset velocity, suggesting further substructure or cloud blending along the line of sight. These secondary peaks are typically simultaneously present in both isotopologues and will not likely have a significant impact on our assumption that the \twco and \thco originate in the same clouds even in the presence of blending along the line of sight. 

\begin{figure} 
    \fig{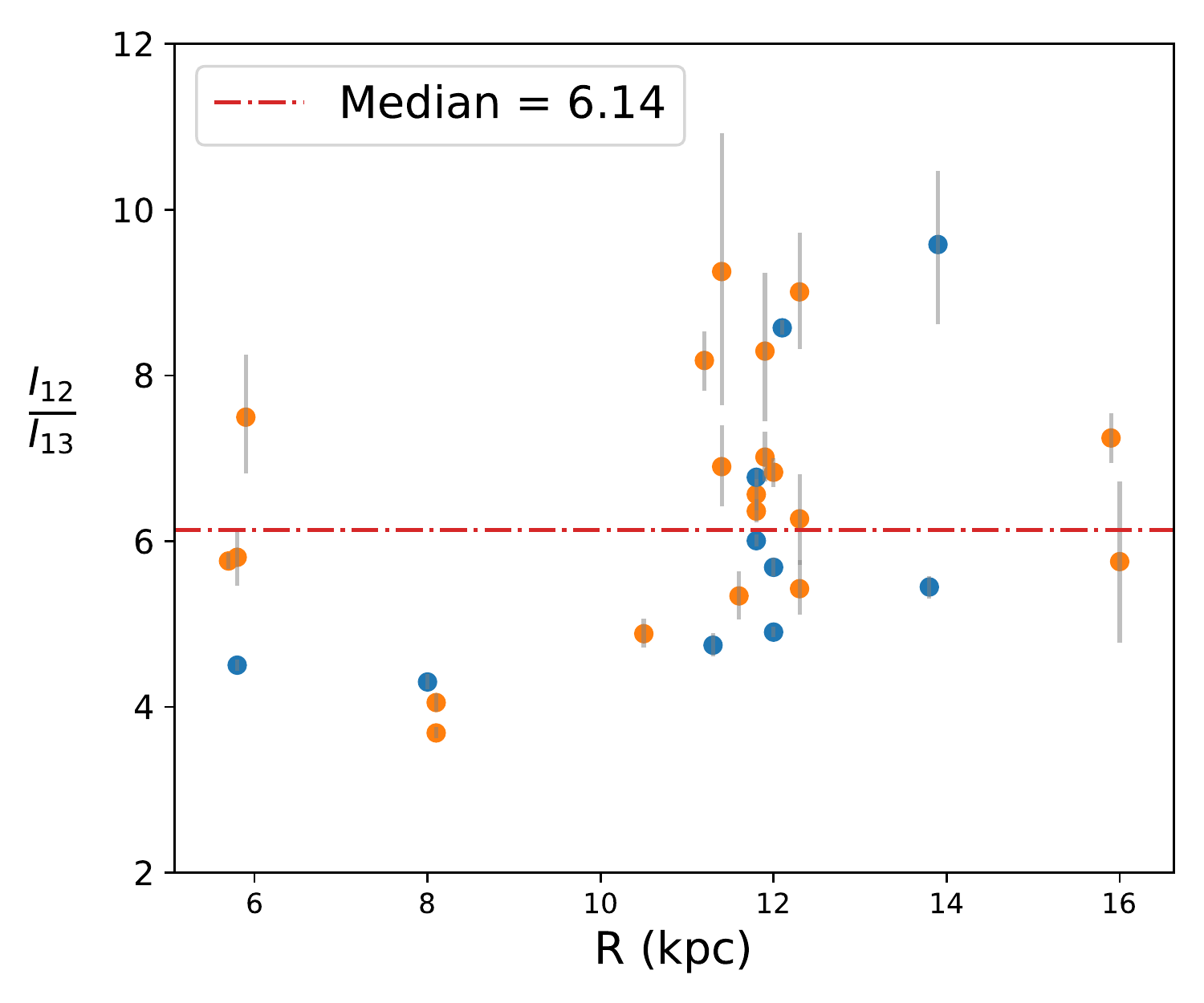}{0.5\textwidth}{}
    \figurenum{5}
    \caption{Intensity ratio of the $^{12}$CO and $^{13}$CO line for the clouds in our sample, shown according to galactocentric radius in M31. Blue dots correspond to resolved clouds, orange dots are unresolved by the telescope. \label{fig:ratio}}
\end{figure}

Figure~\ref{fig:ratio} shows the spatially integrated line intensity ratios $I_{12}/I_{13}$ of the cloud sample as a function of galactocentric radius in M31. K154C and K291A were excluded from this plot because their very weak \thco emission prevented a reliable derivation of the intensity ratio. No significant trend is seen on this scale. Despite the large scatter, $I_{12}/I_{13}$ is remarkably stable between 4-10, with a median value of $6.14$. We find no apparent dependencies of this ratio with dust mass surface density, integrated line intensity or resolvedness of the source. There is a notable clustering of sources between galactocentric radii of 11-13 kpc. This coincides with the main star-forming ring of Andromeda, where the majority of GMCs reside.

Our results are also generally consistent with previous measurements of $I_{12}/I_{13}$ in nearby galaxies obtained at lower (50 -- 200 pc) resolution. Our median value of 6.14 is similar to that found in the central parts of a GMA in a southern arm of M31 by \citet{tos07} and in the spiral arms of M51 by \citet{sch10}, , both using the (1-0) transitions. However, \citet{mel16} note a much higher $I_{12}/I_{13}$ of $\sim21$ in GMCs near the center of M31 in the (2-1) transitions, a result consistent with other measurements of CO isotopologues in galaxy centers (e.g., \citealp{kri10}, and references therein). This suggests that galaxy centers can show significant differences with respect to disks in gas physical conditions or isotope ratios.

The median value of 6.14 that we find for $I_{12}/I_{13}$ in M31 is comparable to the values (i.e., 5.5-6.7) measured for Milky Way clouds \citep[e.g.][]{Solomon1979,Polk1988}. The $I_{13}/I_{18}$ ratio is also relatively stable for the four clouds with a significant detection (see Table~\ref{tab:clouds}). It ranges between $8$ and $13$ with a median of 11. This stability suggests that the conditions that regulate these emission lines are stable for these isotopologues which likely probe the denser parts of individual GMCs. Again, our results show agreement with lower resolution studies of nearby galaxies: \citet{jim17} found a 13/18 ratio of 7.9 when averaged over 9 nearby galaxy disks.

The primary factors that can influence the $I_{12}/I_{13}$ ratio are isotopic abundance, excitation temperature and opacity. This can be seen from the consideration that the ratio of intensities is given by:
\begin{equation}
I_{12}/I_{13} = R = \frac{(J_\nu (T_{ex}^{12})-(J_\nu (T_{cmb}))(1 - e^{-\tau_{12}})}{(J_\nu (T_{ex}^{13})-(J_\nu (T_{cmb}))(1 - e^{-\tau_{13}})}
\end{equation}
Here, $J_\nu(T) = h\nu/k (e^{h\nu/kT} - 1)^{-1}$ and $T_{ex}$ is the excitation temperature of the appropriate isotopologue and $T_{cmb}$ is the temperature of the cosmic microwave background.
If we assume that the excitation temperatures are the same for both isotopologues (i.e., $T_{ex}^{12} = T_{ex}^{13}$) then in the optically thin limit for both lines $R \approx \tau_{12}/\tau_{13}$ which is just the isotopic abundance ratio since $\tau$ is directly proportional to column density.
For the dust clouds we detect here, the \twco line is likely very optically thick and saturated with an intensity near its excitation temperature. Therefore, the intensity ratio should depend primarily on the opacity of the \thco line with the ratio decreasing with increasing \thco opacity and column density. Since we expect that, similar to the Milky Way, \thco is less abundant than \twco by a factor likely between 50-90, the \thco emission is likely optically thin and thus $R \approx \tau_{13}^{-1}$ The intensity ratios we measured in M31 therefore suggest that $\tau_{13} = 0.1 -0.25$ with an median value of 0.16. 

The relatively stable intensity ratios seen in Figure~\ref{fig:ratio} may result from a combination of a stable isotopic abundance ratio coupled with the fact that  our observations are presently only sensitive to a limited range of \thco column densities. This is because as a starting point these particular measurements are confined to the dust emitting regions of the GMCs that are already at relatively high column density levels. Moreover,  our spatial resolution is relatively poor compared to the scale of these regions. As a result the measurements are heavily smoothed and spatially averaged, obscuring any internal gradients in GMC column density that might exist in these regions.

The above considerations suggest that the \thco emission we detect is likely optically thin. We can use this fact to derive reasonable optical depths and column densities for both \thco and \ceighto. To calculate these column densities we use the standard LTE formalism as outlined in detail in Appendix A. Briefly this method assumes that 1) the excitation temperatures of the isotopologues are all the same and can be obtained directly from the \twco profile which is assumed to be saturated as mentioned above and 2) the \thco and\ceighto lines are in the optically thin regime. We derive total column densities for both species: $N_{13}({\rm CO})$ and $N_{18}({\rm CO})$. These are listed in Table~\ref{tab:clouds}. The median $N_{13}({\rm CO}) = 29.6\times 10^{15} \mathrm{cm}^{-2}$ and $N_{18}({\rm CO}) = 4.0\times 10^{15} \mathrm{cm}^{-2}$. 

We can use the stacked spectra to perform the same LTE calculations as for the individual spectra. We find optical depths of 0.24 and 0.04 for \thco and \ceighto , respectively. These numbers are again consistent with the values found for individual sources. Translated into column densities, that becomes $N_{13}({\rm CO}) = 29.0\times 10^{15} \mathrm{cm}^{-2}$ and $N_{18}({\rm CO}) = 4.3\times 10^{15} \mathrm{cm}^{-2}$ in excellent agreement with the median values derived for the individual GMCs. 

\subsection{Isotopic Abundances}

Comparing with the dust continuum measurements, we can now determine the relative abundance of the CO isotopologues with respect to H$_2$. We use the CO column densities derived from the stacked spectra to determine the isotopic abundances for both \thco and C$^{18}$O for the observed GMC population. From a similar stacking of the dust measurements we derive the corresponding molecular hydrogen density to be:
\begin{equation}
N(H_2) = 9.84 \times 10^{21}\left(\frac{R_{g2d}^\prime}{100}\right) \mathrm{cm}^{-2}
\end{equation}
where $R_{g2d}^\prime$ is the hydrogen gas-to-dust ratio without correction for Helium and heavier elements (i.e.,  $R_{g2d}^\prime$ = $R_{g2d}/1.36$ = 100 for our adopted value of $R_{g2d}=$136).
We find the following abundance ratios:
\begin{equation} \label{eq:X13}
[^{13}\mathrm{CO}] = \frac{(N_{13}({\rm CO}))}{(N(H_2))} = 2.95 \times 10^{-6} \left(\frac{100}{R_{g2d}^\prime}\right)
\end{equation}
\begin{equation} \label{eq:X18}
[\mathrm{C}^{18}\mathrm{O}] = \frac{(N_{18}({\rm CO}))}{(N(H_2))} = 
0.44 \times 10^{-6}  \left(\frac{100}{R_{g2d}^\prime}\right)
\end{equation}
We estimate the uncertainties to be 19\% and 20\% for $[^{13}\mathrm{CO}]$ and  [C$^{18}$O] respectively. Both these uncertainties are dominated by the estimated error in the dust column density. 

For comparison, direct measurements for the isotopic abundance ratio of \thco  in the Milky Way range between $1.3 - 2.8 \times 10^{-6}$ with a mean $\sim$ 2 $\times 10^{-6}$ \citep{Dickman1978,Frerking1982,Bachiller1986,Duvert1986,Pineda2008,Roueff2020,Lewis2020}. Our M31 estimate for the \thco abundance is thus  similar to the values reported for  Milky Way clouds.  There are few direct estimates of [C$^{18}$O] for Milky Way clouds. \citet{Frerking1982} found $[C^{18}O] = 1.7 \times 10^{-7}$ for the Taurus and Rho Ophiuchi clouds  and more recently \citet{Lewis2020} report a value of 1.3 $\times 10^{-7}$ for the California cloud and \citet{Roueff2020} find a value of 2.0 $\times 10^{-7}$ for the Orion B cloud. These values are about 2 times lower than the value we find for M31.  The isotopic abundance ratio we find (6.7$\pm$ 2.9) from equations 4 and 5 is somewhat lower than, but within the uncertainties of, the Milky Way value (8.1; \citealt{Wilson1999}). 
 
\subsubsection{$^{13}$CO mask} \label{subsec:13COmask}

In most of this work we have focused on the area of the GMCs marked by the 3$\sigma$ dust mask. This allowed us to compute conversion factors and abundance ratios by directly linking continuum and line emission to the same physical area of the GMCs under study. We can expand the LTE analysis to a larger part of the GMC since the $^{13}$CO emission extends beyond the dust mask (see Fig.~\ref{fig:maps}). In this experiment, we create spatially integrated spectra over the entire field of view, summing all pixels above the 3$\sigma$ threshold in the $^{13}$CO cube. We no longer split the GMC emission in clouds. 

In 12 GMCs, emission was detected above $3\sigma$ at the expected frequency  of \ceighto. The SNR was determined by comparing the peak channel emission to the RMS in the line-free channels on both sides of the emission line. For further analysis we adopt a more conservative threshold of $4\sigma$ as we visually assessed that below this value the emission line measurements are not reliable. This leaves 5 GMCs with \ceighto emission. For \thco the sample consists of 24 GMCs. We perform the LTE calculations for this subsample and again found typical optical depths of 0.18 and 0.02 for $^{13}$CO and C$^{18}$O, respectively. The median $N_{13}({\rm CO}) = 34.7\times 10^{15} \mathrm{cm}^{-2}$ and $N_{18}({\rm CO}) = 3.1\times 10^{15} \mathrm{cm}^{-2}$. The median ratio $\frac{N_{13}({\rm CO})}{N_{18}({\rm CO})}$ is $8.04$. These numbers are similar to those derived within the dust mask indicating that the properties we derive for material within the dust masks is representative of the overall properties of the GMCs.

\subsection{GMC sizes}

\begin{figure} 
    \fig{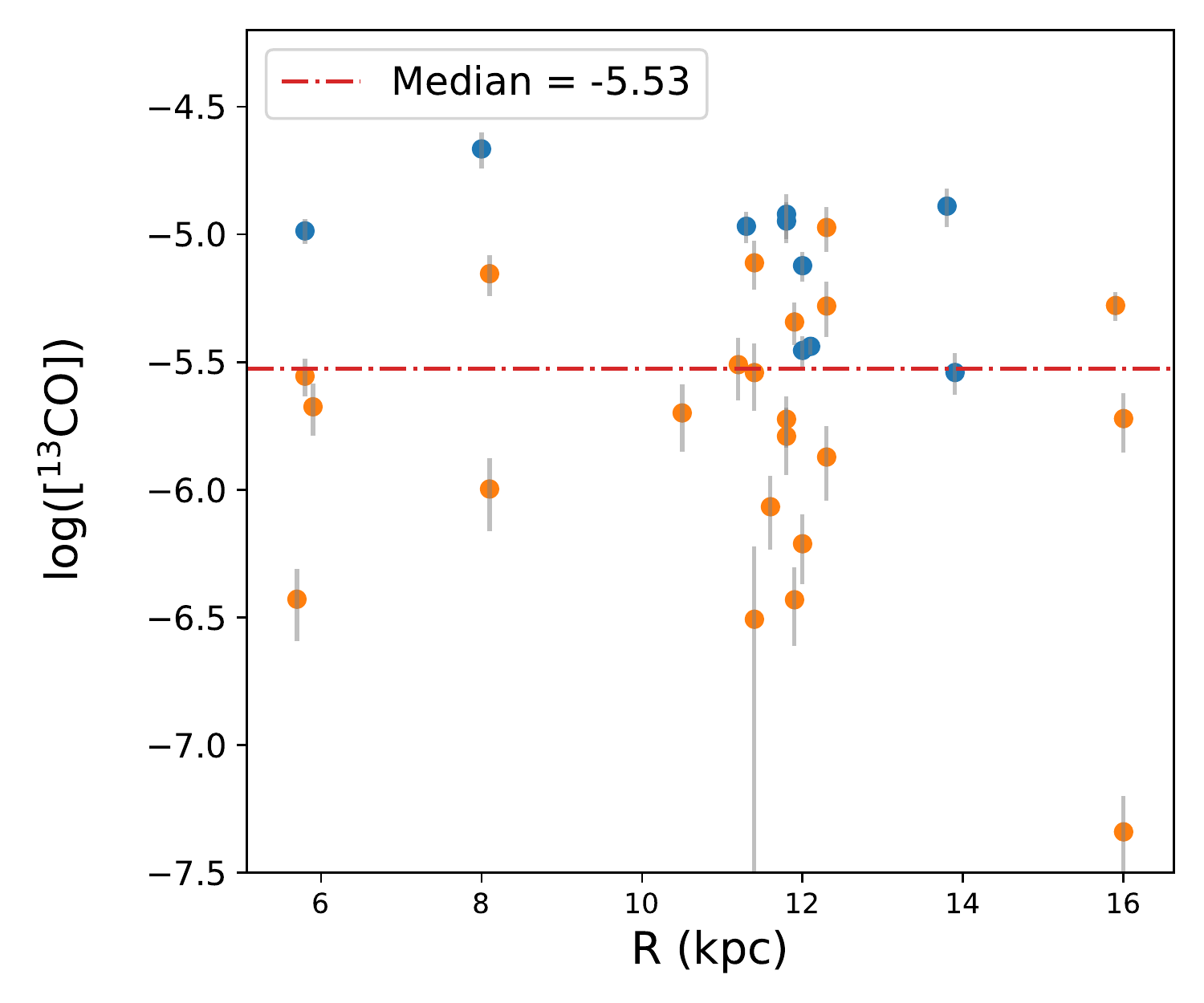}{0.5\textwidth}{}
    \figurenum{6}
    \caption{$[^{13}\mathrm{CO}]$, the \thco abundance (relative to $H_2$) for the GMCs in our sample, shown according to galactocentric radius in M31. Blue dots correspond to resolved dust clouds, orange dots are unresolved by the telescope. \label{fig:abundance_radius}}
\end{figure}

\begin{figure} 
    \fig{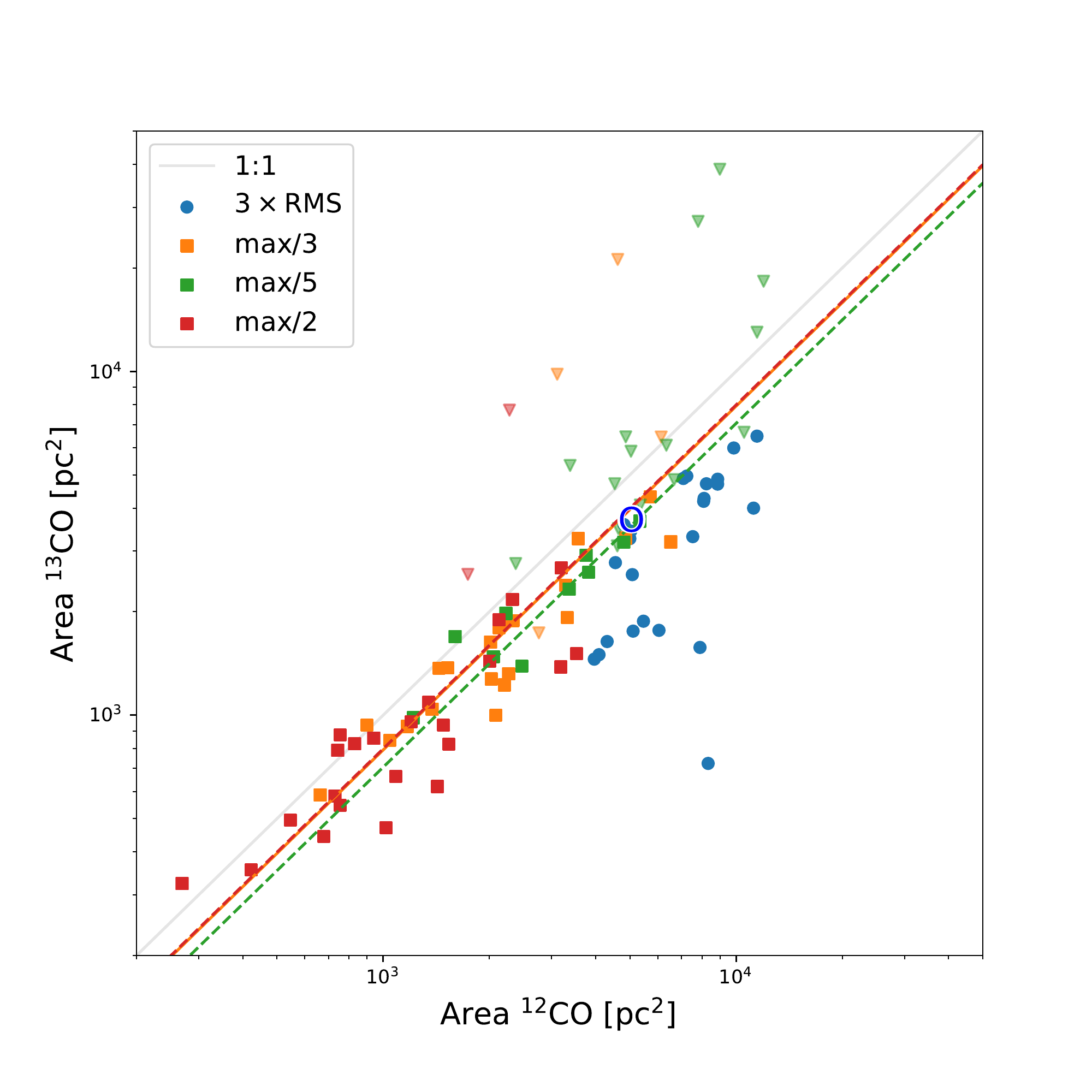}{0.5\textwidth}{}
    \figurenum{7}
    \caption{GMC sizes in $^{12}$CO and $^{13}$CO when described as the area within the contour enclosing a certain emission threshold. Blue dots show a direct comparison of the $3\sigma$ contours for each GMC. The blue 'O' indicates the $3\sigma$ areas for the simulated Orion observations. Squares correspond to thresholds that are a set percentage of the peak emission, as indicated in the figure key. Triangles are upper limits for GMC where the threshold lies below the $3\sigma$ significance level. The lines correspond to median area ratios.  \label{fig:sizes}}
\end{figure}

The obvious difference in spatial extent between CO and dust emission signifies that we only detect the brightest peaks of continuum emission. The deep integrations do facilitate detection of CO lines across a wider area of the GMC. Indeed, as we will show in the next section, we likely detect \twco emission from the entire CO emitting area of an M31 GMC. The GMC sizes as they appear in the full extent of the $^{12}$CO and $^{13}$CO moment zero maps (beyond the dust masks) are compared in Fig.~\ref{fig:sizes}. We quantify the size of a GMC as the area of the contour above a certain emission threshold. Determining the threshold in the moment zero maps is not trivial because the image RMS noise in the moment zero maps of $^{12}$CO is higher than for $^{13}$CO. Given the significantly stronger and more extended $^{12}$CO emission, there are fewer truly emission-free areas in the map, and the nominal noise is also more affected by imaging artifacts due to spatial filtering. We therefore compare the GMC sizes for several different thresholds. 

First, we compare the nominal 3$\sigma$ contour areas (corresponding to the outer contours in Fig.~\ref{fig:maps}). These are indicated as blue points in Fig.~\ref{fig:sizes} and exhibit a moderate, non-linear relation. As we will show below in Section~\ref{sec_sim_obs}, these size ratios are compatible with what we would expect from simulated observations of the Orion molecular clouds, as if they were located in M\,31.

Second, we determine the threshold as half the peak emission value. The area is then a 2D analogue of the full-width-half-maximum (FWHM) metric and shown as red squares in Fig.~\ref{fig:sizes}. In a few cases, this threshold lies below the 3$\sigma$ level. In these cases the sizes are upper limits because several noise peaks are added to the total area. They are indicated as triangles in the figure. There is a clear linear trend in the measured areas for both emission lines. The $^{12}$CO emission is consistently more more extended than the $^{13}$CO, by a constant $25\%$. To estimate the dependency of this result on the 2D FWHM threshold, we also plot the peak/3 and peak/5 sizes in Fig.~\ref{fig:sizes}. These of course yield larger sizes as the threshold is lowered, and consequently also spawn more upper limits. The linear trend does remain unaffected, with  $^{12}$CO sizes $26\%$ larger than $^{13}$CO sizes for the peak/3 threshold. For the peak/5 threshold, the increase is roughly constant at $41\%$, which is still consistent with the other trends. In summary, we find that $^{13}$CO emission, despite its lower abundance, is well resolved with an extent typically 75\% of that of the \twco emission.

\subsection{Simulated observations}
\label{sec_sim_obs}

We demonstrate the plausibility of our findings and provide Galactic context by simulating observations of the Orion molecular clouds as if observed, with the SMA, in M\,31. For this purpose, we use the large-scale moment 0 maps in $^{12}$CO(2-1), $^{13}$CO(2-1), and C$^{18}$O(2-1) obtained by \citet{Nishimura2015}. As in Paper I, we first convert the main beam temperatures to flux densities before scaling them and the pixel and beam sizes to the distance of M\,31. We also modify the position on the sky of Orion to that of M\,31 to ensure a realistic simulated $(u,v)$ coverage and thus viewing geometry. These images were then used as input for simulating full tracks with the SMA in its subcompact configuration in CASA~5.3, using the tasks \texttt{simobserve} and \texttt{simanalyze}. 

To estimate the resulting S/N ratio, we then compare the predicted integrated flux densities with the sensitivity reached by the SMA, assuming very good observing conditions at 1~mm pwv, over a bandwidth corresponding to a typical linewidth of 5 km\,s$^{-1}$. The choice of a typical linewidth is necessary in this estimate to obtain a realistic S/N over a specified bandwidth from a moment~0 map. For simplicity, we use a typical linewidth of 5~km\,s$^{-1}$ for all three isotopologues. This does represent the $^{12}$CO width in the brightest parts of the Orion clouds \citep{Nishimura2015}, and it is also compatible with typical $^{12}$CO linewidths found in our observations of M\,31 clouds. 

\begin{figure} 
    \fig{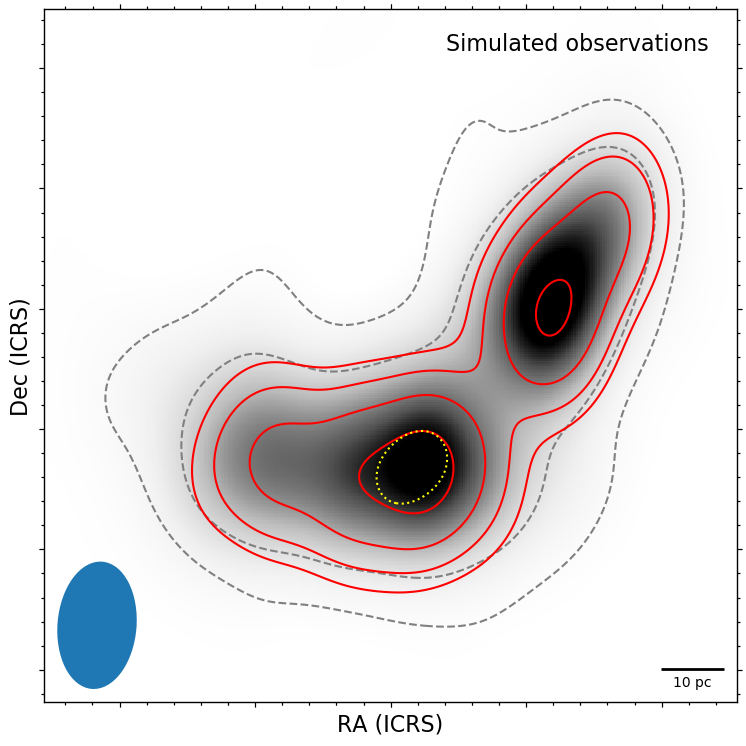}{0.45\textwidth}{}
    \figurenum{8}
    \caption{Simulated CO isotopologue observations of the Orion A and B clouds, as if located in M\,31 and conducted with the SMA in its subcompact configuration, based on single-dish observations of the Orion clouds \citep{Nishimura2015}. The greyscale image displays the simulated $^{12}$CO(2-1) observation, with the dashed grey contour line indicating contours at S/N$\sim$10 and S/N$\sim$100. The red contour lines show simulated $^{13}$CO(2-1) in steps of S/N$\sim$10, 20, 40, and 80, while the yellow dotted contour indicates the simulated detection of C$^{18}$O(2-1) at S/N$\sim$3 
    \label{fig_sim}}
\end{figure}

Figure~\ref{fig_sim} shows the simulated observations, highlighting three insights: 1) As discussed in Paper~I, the simulated Orion A and B clouds in M\,31 are easily detected and resolved in $^{12}$CO(2-1), 2) the sensitivity of these long integrations is sufficient to comfortably resolve the GMCs even in $^{13}$CO(2-1), and 3) even C$^{18}$O(2-1) is detected, but marginally at S/N$\sim$3 and only toward the simulated Orion A cloud. This experiment shows that in M31 the detection of resolved GMCs in $^{13}$CO(2-1) emission is clearly plausible, and even the detection of C$^{18}$O(2-1) on these scales is not unexpected. The apparent size ratio of the simulated $^{13}$CO(2-1) and $^{12}$CO(2-1) observations, if each observation is evaluated at its S/N=3 contour, is 72\%.

\section{Conclusions} \label{sec:conclusion}

As part of an ongoing survey with the SMA \citep{Forbrich2020} we have obtained resolved observations on $\sim$ 15 pc spatial scales of both dust continuum and high fidelity J=2-1  $^{12}$CO, $^{13}$CO, and \ceighto line emission within individual GMCs of the Andromeda galaxy. Simultaneous collection of these data was enabled by the wide-band frequency coverage of the SMA spectrometer. In this paper we primarily analyze emission from the \thco and \ceighto isotopologues. We compare these observations with simultaneous measurements of dust continuum emission at 230 GHz. In the GMCs, we isolate individual dust peaks that we designate as clouds and derive the column densities, abundances and conversion factors for the rare CO isotopologues observed within them. Similar results for the $^{12}$CO line were reported earlier \citep{Forbrich2020}. Here we also report updated measurements of dust masses and the $^{12}$CO conversion factors from an expanded and growing sample of GMCs in our survey of M31.  The primary conclusions of this present effort are as follows:

\begin{itemize}
    \item Our continuing survey confirms that dust continuum emission can routinely be detected on $\sim$ 15 pc scales in individual GMCs in M31. Across 25 pointings we so far have identified 32 dust emitting clouds contained within 20 GMCs. Ten of these dust clouds   are spatially resolved in dust emission.
    \item The survey observations are sufficiently deep to permit high-fidelity measurements of $^{12}$CO emission that essentially span the entire extent of the CO gas contained within the GMC, enabling accurate measurements of GMC areas and sizes.
    \item $^{13}$CO emission is detected with high signal-to-noise and resolved in all but one pointing. The extent of its emission is nearly comparable to that of $^{12}$CO, typically covering 75\% of the area of the $^{12}$CO emission.  
    \item We report the first detections of C$^{18}$O on these spatial scales, in M31 GMCs. C$^{18}$O is clearly detected within the dust emitting regions of four clouds. This number rises to 12 formal detections when looking beyond the regions of dust emission.
    \item We use the dust continuum to derive dust masses and the J = 2-1 CO conversion factors for the isotopologues. We find a mean 
    $\alpha_{12} = 8.7 \pm 3.9 \times \left(\frac{R_{g2d}}{136}\right)$ M$_\odot$\,$(\mathrm{K}\,\mathrm{km}\,\mathrm{s}^{-1}\mathrm{pc}^{2})^{-1}$ and $\alpha_{13} = 48.9 \pm 20.4 \times \left(\frac{R_{g2d}}{136}\right)$ M$_\odot$\,$(\mathrm{K}\,\mathrm{km}\,\mathrm{s}^{-1}\mathrm{pc}^{2})^{-1}$ where $R_{g2d}$ is the dust-to-gas ratio. A stacking analysis returns very similar values for $\alpha_{12}$ and $\alpha_{13}$  and from this analysis we are also able to derive a conversion factor of $\alpha_{18}= 345^{+25}_{-31}$  \, \msunsp $ \mathrm{K}\,\mathrm{km}\,\mathrm{s}^{-1}\mathrm{pc}^{2}$ for C$^{18}$O with an assumed dust-to-gas ratio of 136.
    \item We perform an LTE analysis on the spectra to derive column densities and (gas phase) abundances for  $^{13}$CO and C$^{18}$O. Assuming a Milky Way gas-to-dust ratio for the GMCs in our current M~31 sample, we find abundances of $[^{13}\mathrm{CO}] = 2.95 \times 10^{-6}$ and [C$^{18}$O] $= 0.44 \times 10^{-6}$, relative to H$_2$. These values are similar to those of Milky Way GMCs.  Moreover, isotopic abundance ratio,  [$^{13}$CO]/[C$^{18}$O] is found to be 6.7 $\pm$ 2.9 somewhat lower but consistent with that (8.1) measured in Milky Way clouds.
\end{itemize}
Overall we conclude that the fundamental properties of CO gas contained within the individual GMCs we observed in M31 are essentially in the same range as those that characterize individual GMCs in the Milky Way. 

\acknowledgments
The Submillimeter Array is a joint facility of the Smithsonian Astrophysical Observatory and the Academia Sinica Institute of Astronomy and Astrophysics. It is operated by the Smithsonian Astrophysical Observatory on Mauna Kea, Hawaii with funding by the Smithsonian Institution and the Academia Sinica. \\
SV acknowledges the Flemisch Research Fund (FWO) for supporting this research. SV would also like to thank the University of Hertfordshire for their support during these observations. We thank Jonathan Toomey for helpful discussions.

%


 
\vfill\eject
\bibliography{M31_13CO}{}
\bibliographystyle{aasjournal}
\facilities{SMA}
\software{
Astropy \citep{2013A&A...558A..33A},
MIRIAD \citep{miriad}, 
CASA v5.3 \citep{mcm07},
Scikit-Learn \citep{scikit-learn}}


\appendix

\section{LTE analysis}

To compute the optical depth for $^{13}$CO, we assume the commonly used formalism of local thermal equilibrium (LTE, \citealt{Dickman1978,Rohlfs2004,Pineda2008}). For the J(2-1) transition, this becomes \citep{Indebetouw2013,Nishimura2015}:

\begin{equation}
T_{\rm ex}^{J=2-1} = 11.06 \left\{ \ln \left[ 1+ \frac{11.06}{T_{\rm peak}^{12,J=2-1} + 0.19} \right] \right\}^{-1}.
\end{equation}

The excitation temperature $T_{\rm ex}^{J=2-1}$ is derived from the peak temperature of the $^{12}$CO line: $T_{\rm peak}^{12,J=2-1}$ under the assumption that this line is optically thick. We further assume that $^{13}$CO and $^{12}$CO are well mixed in the area under study and that they will both have the same excitation temperature. The optical depth for each velocity bin, $\tau_{J=2}^{13}(v)$, can then be calculated as
\begin{equation}
\tau_{J=2-1}^{^{13}CO}(v) =
\end{equation}
\begin{equation*}
- \ln \left\{ 1 - \frac{T_{\rm mb}^{13,J=2-1}(v)}{10.58} \left[ \frac{1}{\exp(10.58/T_{\rm ex}) - 1} - 0.02 \right]^{-1} \right\},
\end{equation*}
with $T_{\rm mb}^{13,J=2-1}(v)$ the brightness temperature of $^{13}$CO in each velocity bin. The LTE derived \thco optical depths are found to in the range 0.1 - 0.4 similar to that estimated earlier from the \twco to \thco line intensity ratios indicating optically thin emission for \thco. Comparison of the two opacity estimations is shown in Figure A1. The mean LTE opacity is 0.2 $\pm$ 0.07.

The optical depth can be integrated to provide  optical depth for the cloud in the continuum mask:
\begin{equation}
\label{eq:opdepth13CO}
\tau_{^{^{13}CO}}= \int \tau_{J=2-1}^{^{13}CO}(v) dv.
\end{equation}

\begin{figure} 
    \fig{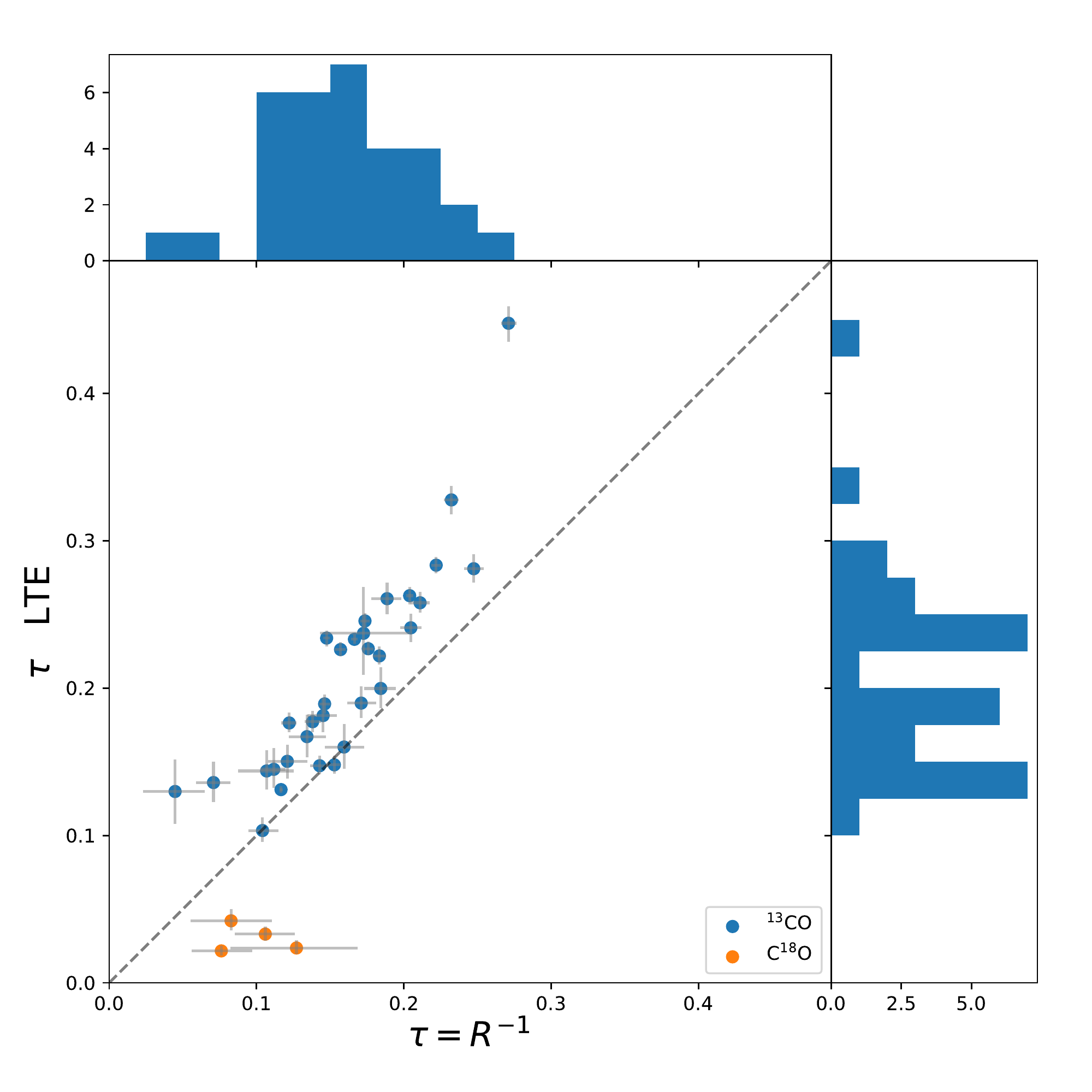}{0.45\textwidth}{}
    \figurenum{A.1}
    \caption{Comparison of the \thco and \ceighto optical depths estimated using both the inverse line ratio proxy and the LTE approximation. The dashed line in the main panel shows the 1:1 line. Top and right panel are axis projection histograms of the individual quantities for \thco.\label{fig:odepth}}
\end{figure}

As a final step in the LTE calculation, the integrated optical depths can be converted into column densities. We further follow the formulas outlined in \citet{Indebetouw2013} and \citet{Nishimura2015}. The column density of $^{13}$CO molecules occupying the $J=2$ rotational level is obtained as:
\begin{equation}
N_{J=2}^{13} = 1.65 \times 10^{16} \left[ \exp \left( \frac{10.58}{T_{\rm ex}} \right) - 1 \right]^{-1} \int \tau_{J=2-1}^{13}(v) dv
\end{equation}

To convert this into a total column density of the isotopologue, a correction factor $Z$ is required to include all other rotational transitions. This can be computed using the partition function:

\begin{equation}
Z = \sum_{J=0}^{\infty} (2J + 1) \exp \left[ - \frac{h B_0 J (J+1)}{k T_{\rm ex}} \right],
\end{equation}
where $B_0 = 5.51 \times 10^{10} \mathrm{s}^{-1}$ is the rotational constant for $^{13}$CO. The total column density of an isotopologue can then be computed as
\begin{equation}
N({\rm CO}) = N_{J} \frac{Z}{2J+1} \exp \left[ \frac{h B_0 J (J+1)}{k T_{\rm ex}} \right]
\end{equation}

A similar calculation can be done for the four clouds with a significant C$^{18}$O detection. The relevant equations are analogous to the ones for $^{13}$CO and are described in \citet{Nishimura2015}. As expected, \ceighto is even more optically thin than \thco (see also fig.~\ref{fig:odepth}.)

\vskip 4.0in
\section{Cloud spectra} \label{app:spectra}

Plots of the spectra for individual clouds.
\vfill
\begin{figure*} 
\gridline{\fig{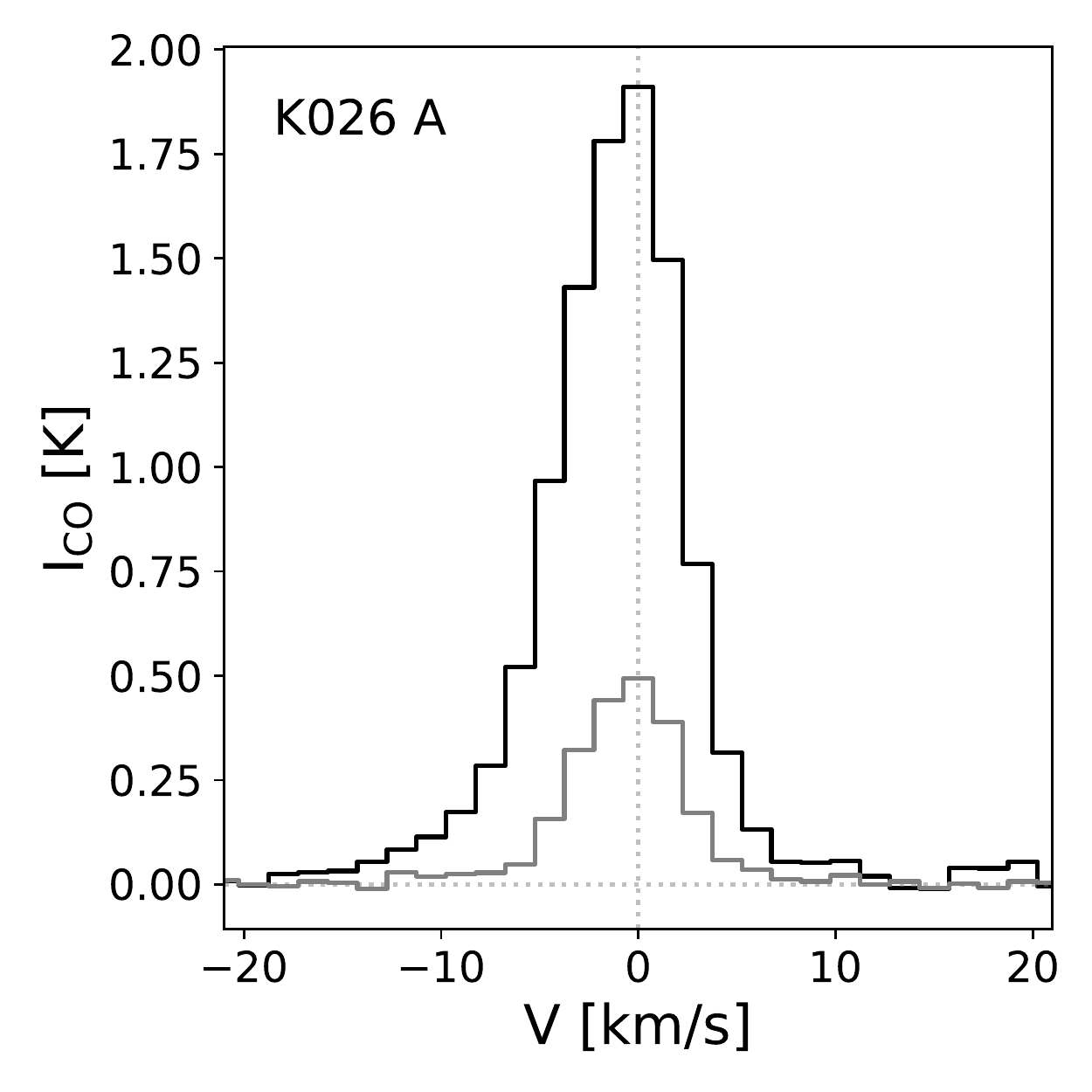}{0.25\textwidth}{}
          \fig{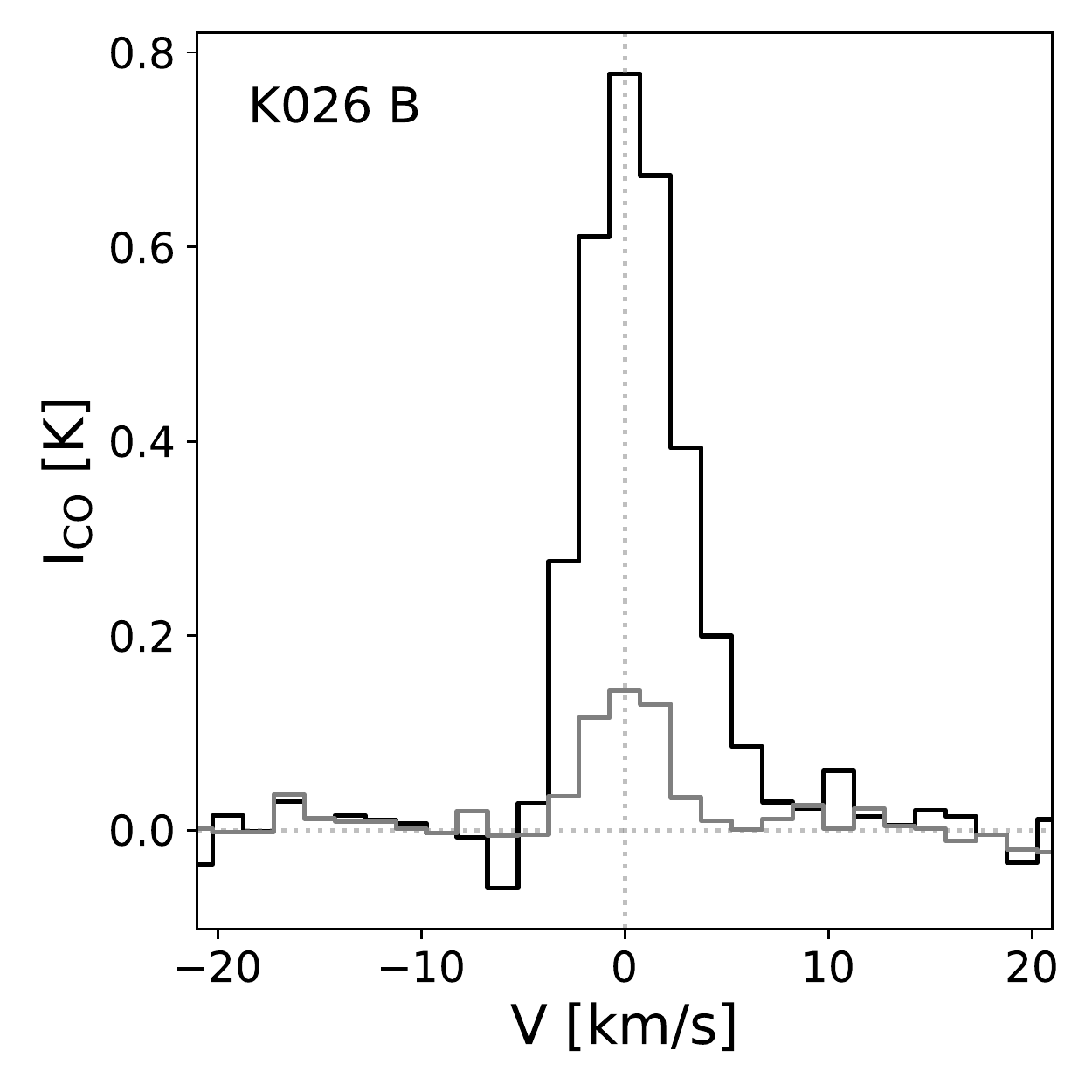}{0.25\textwidth}{}
          \fig{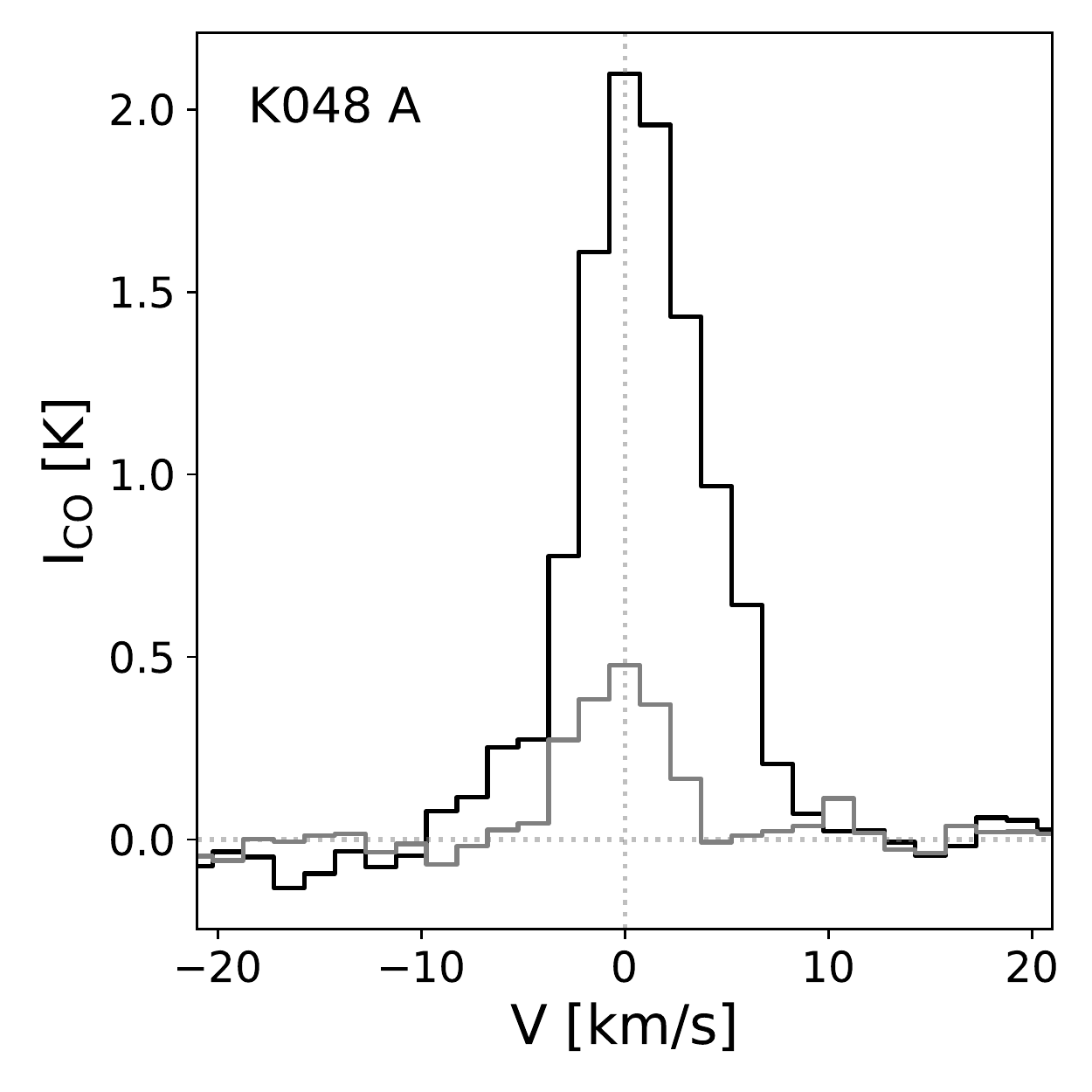}{0.25\textwidth}{}
          \fig{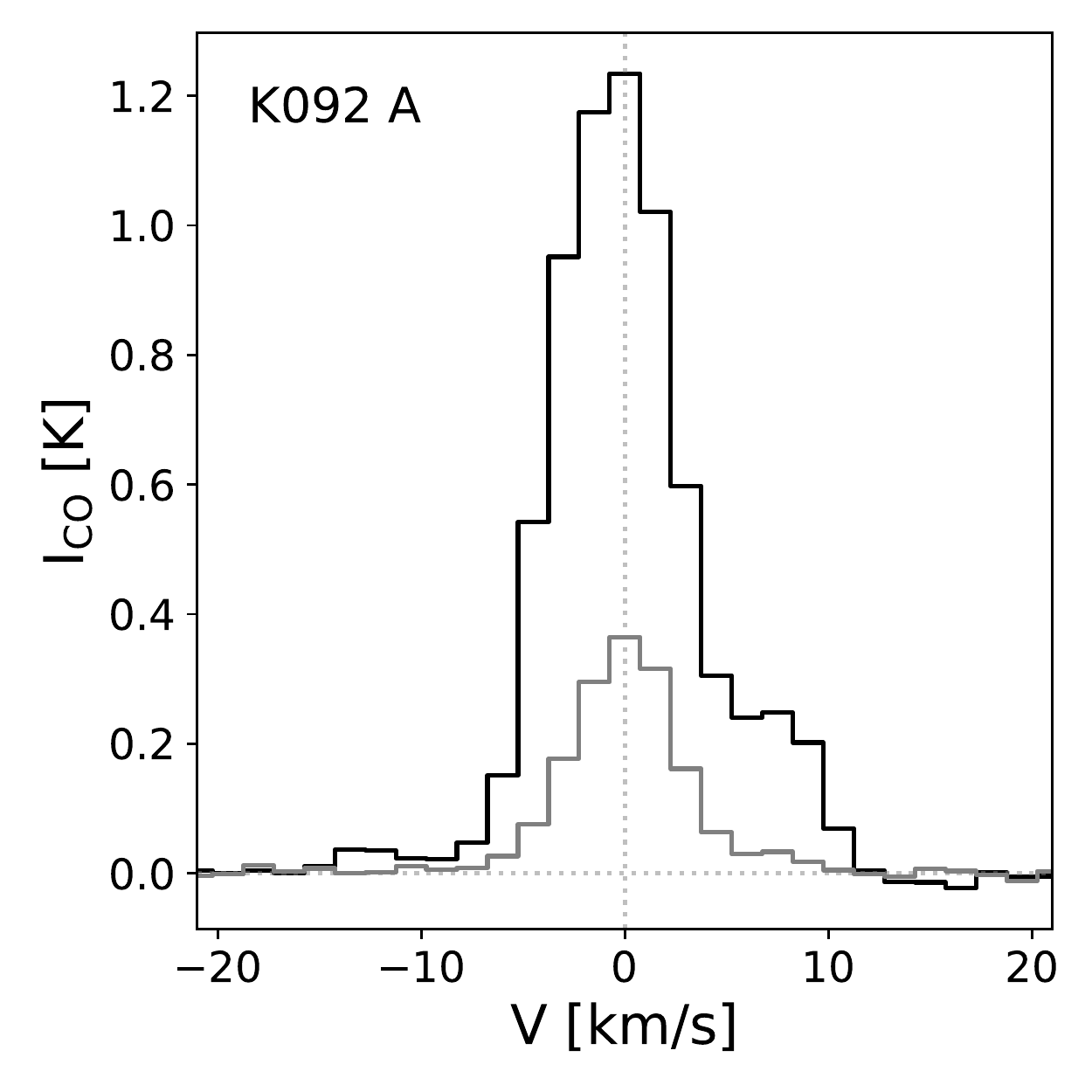}{0.25\textwidth}{}
          }
\gridline{\fig{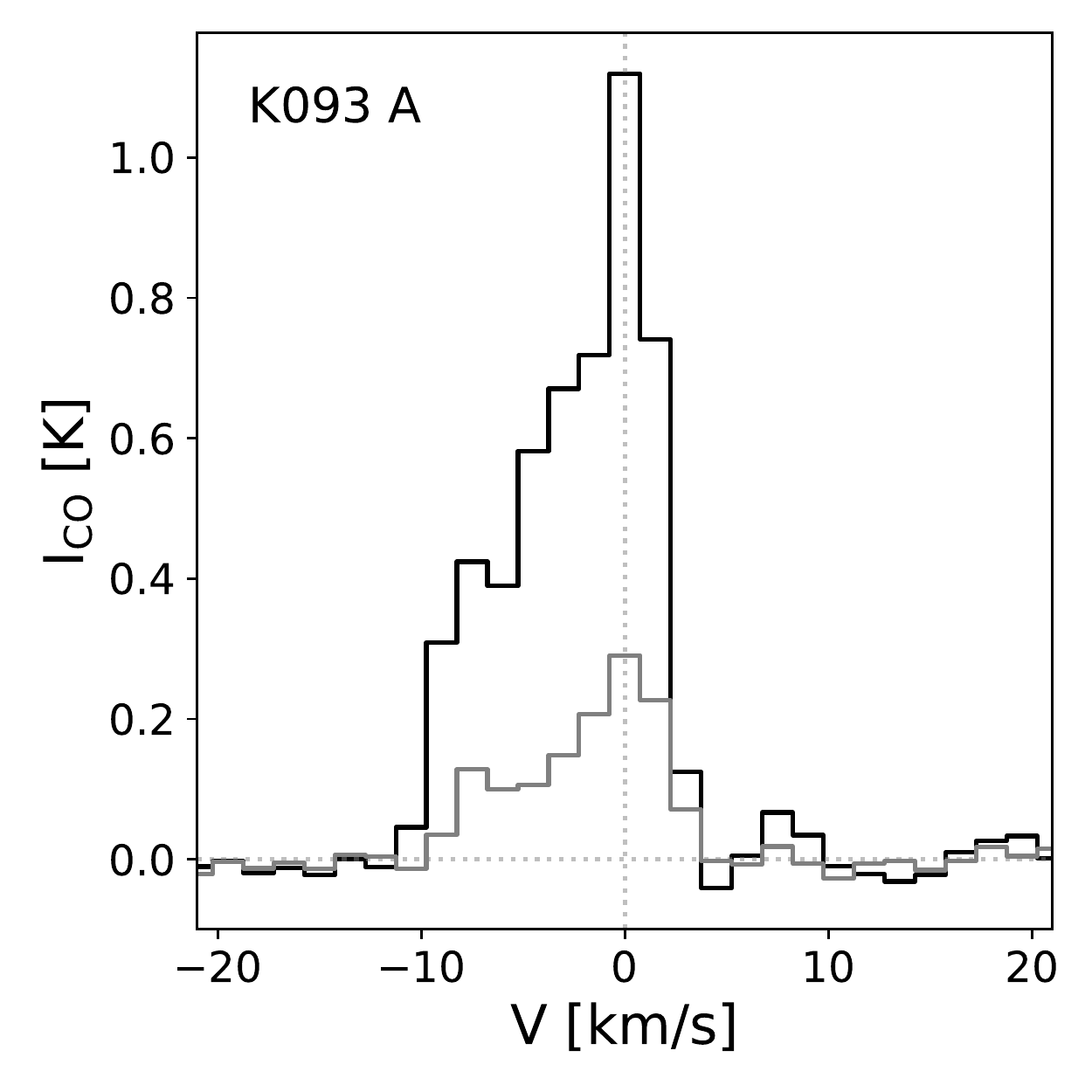}{0.25\textwidth}{}
          \fig{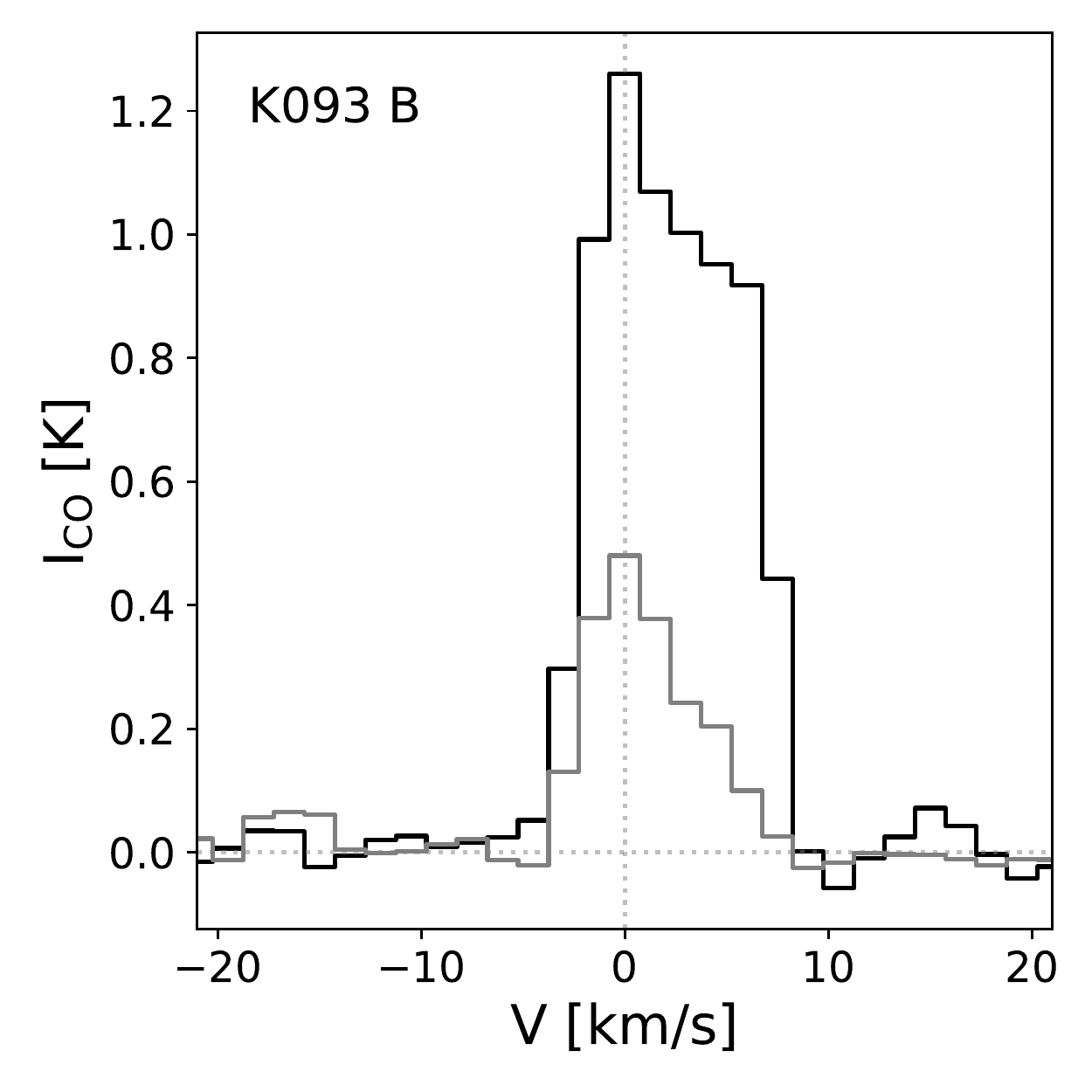}{0.25\textwidth}{}
          \fig{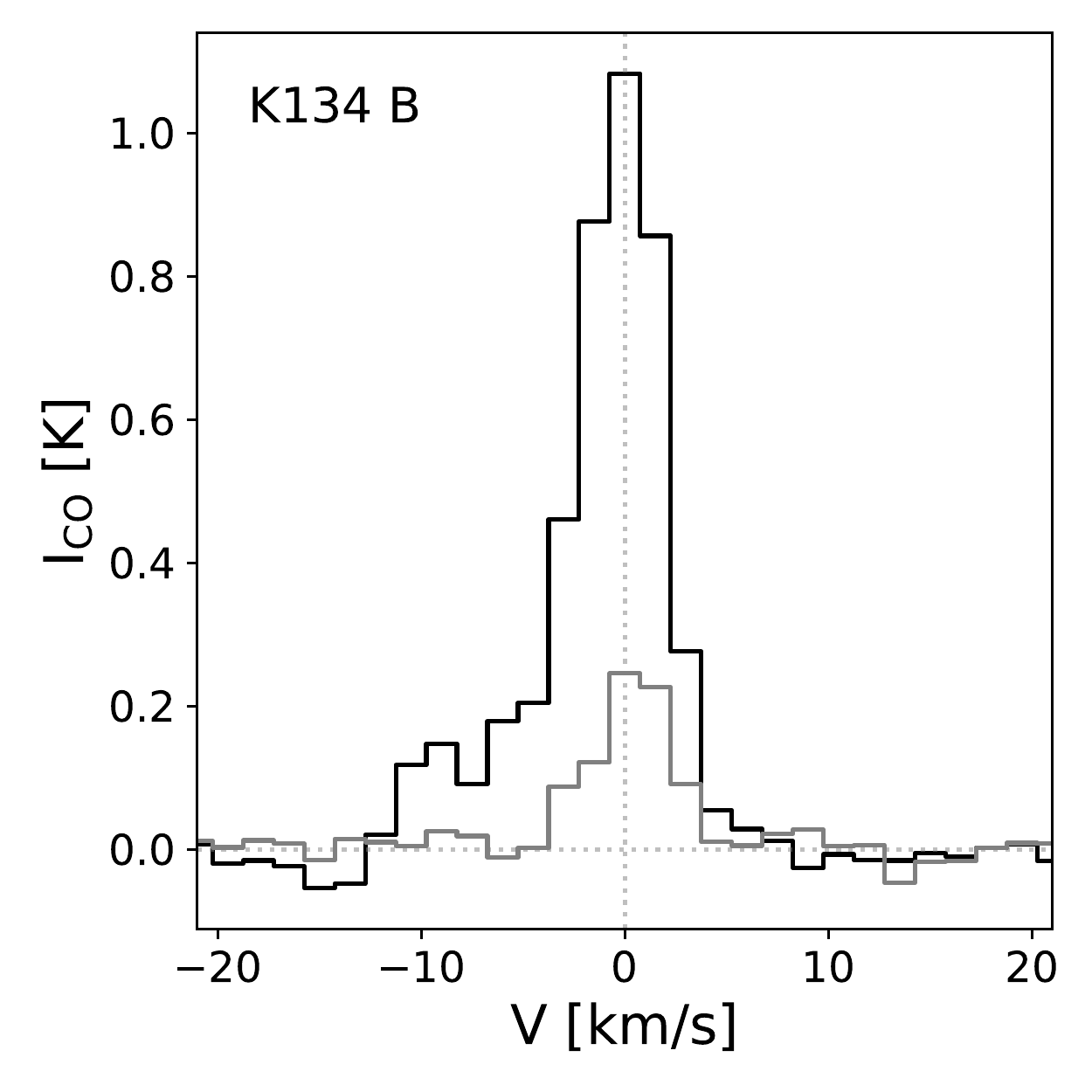}{0.25\textwidth}{}
          \fig{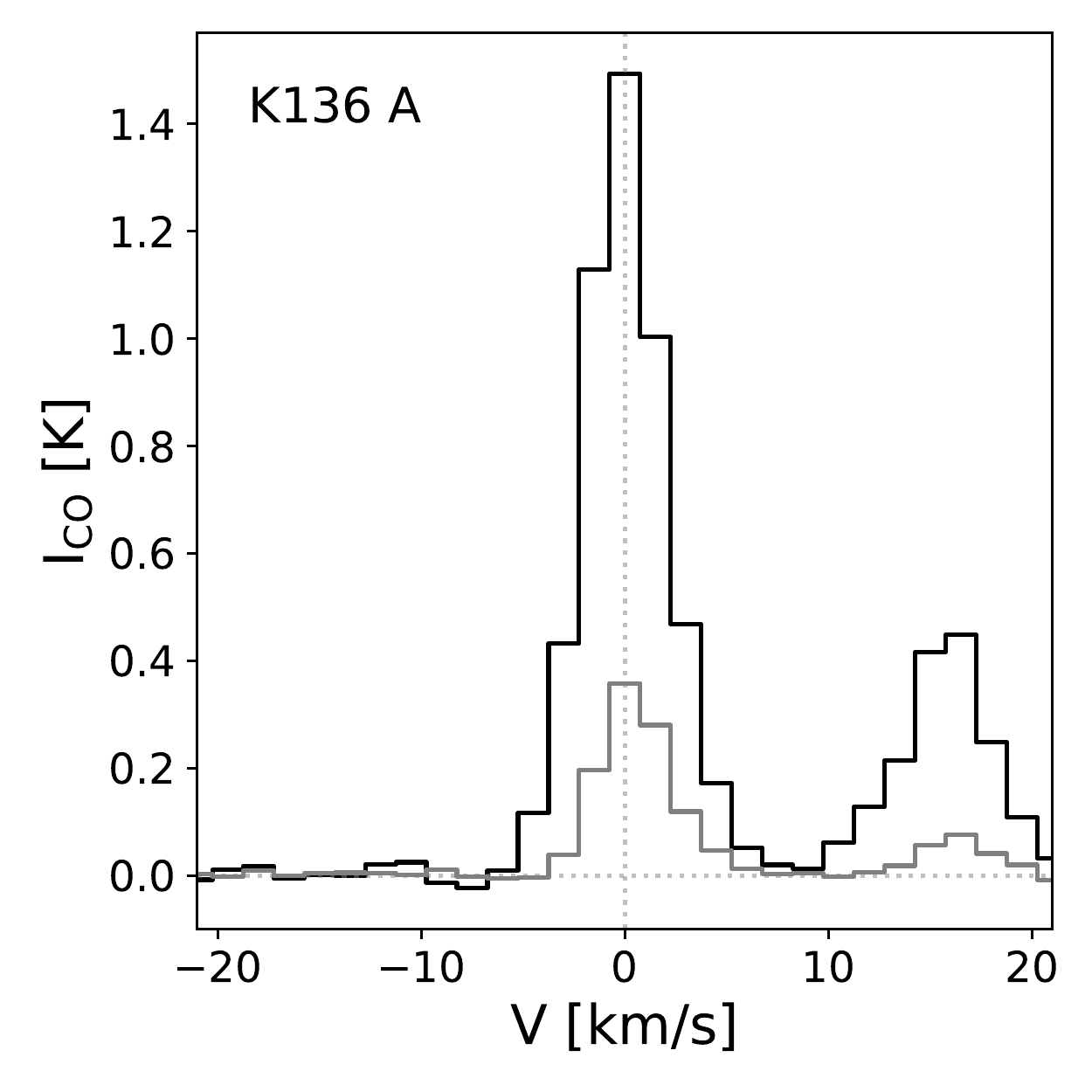}{0.25\textwidth}{}
          }
\gridline{\fig{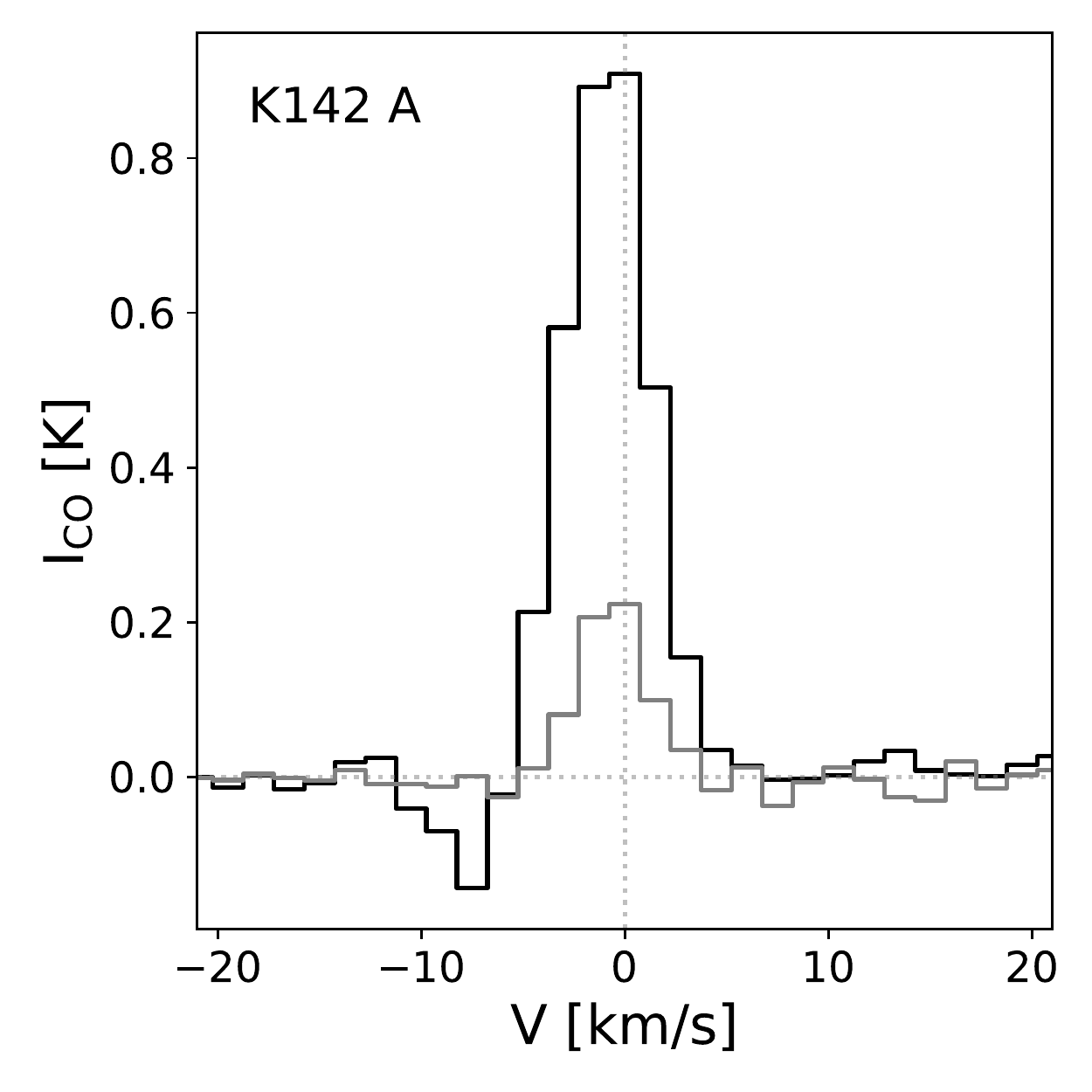}{0.25\textwidth}{}
          \fig{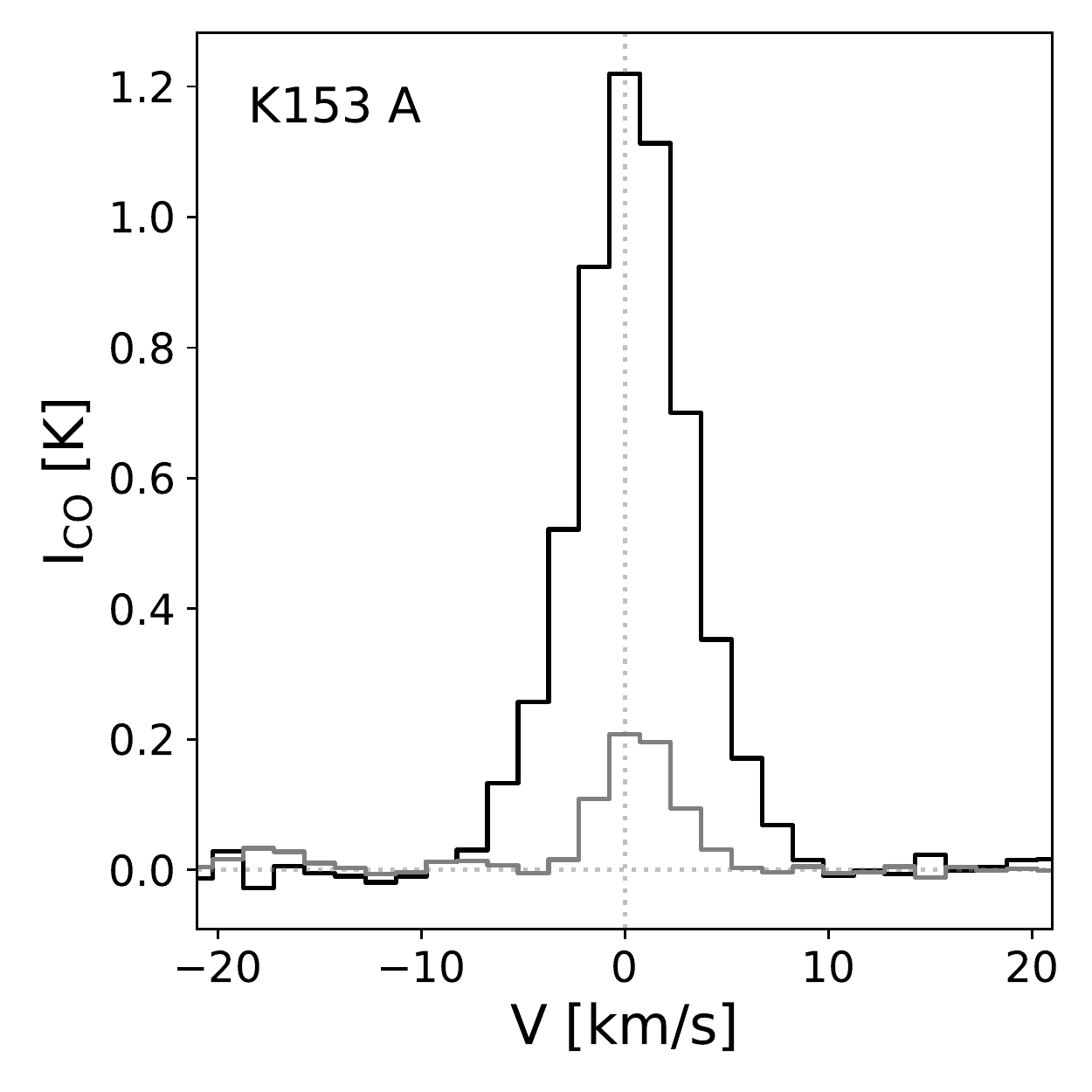}{0.25\textwidth}{}
          \fig{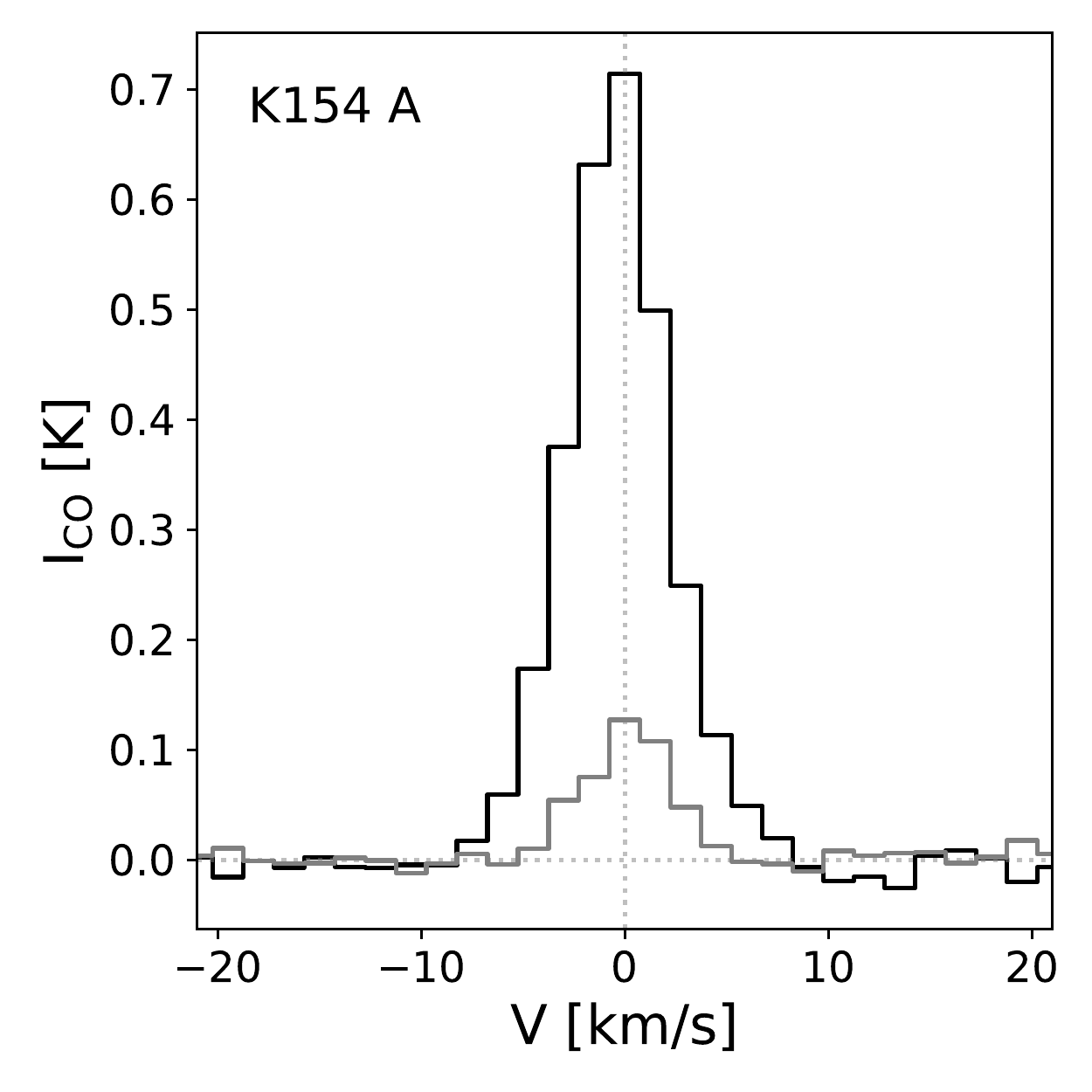}{0.25\textwidth}{}
          \fig{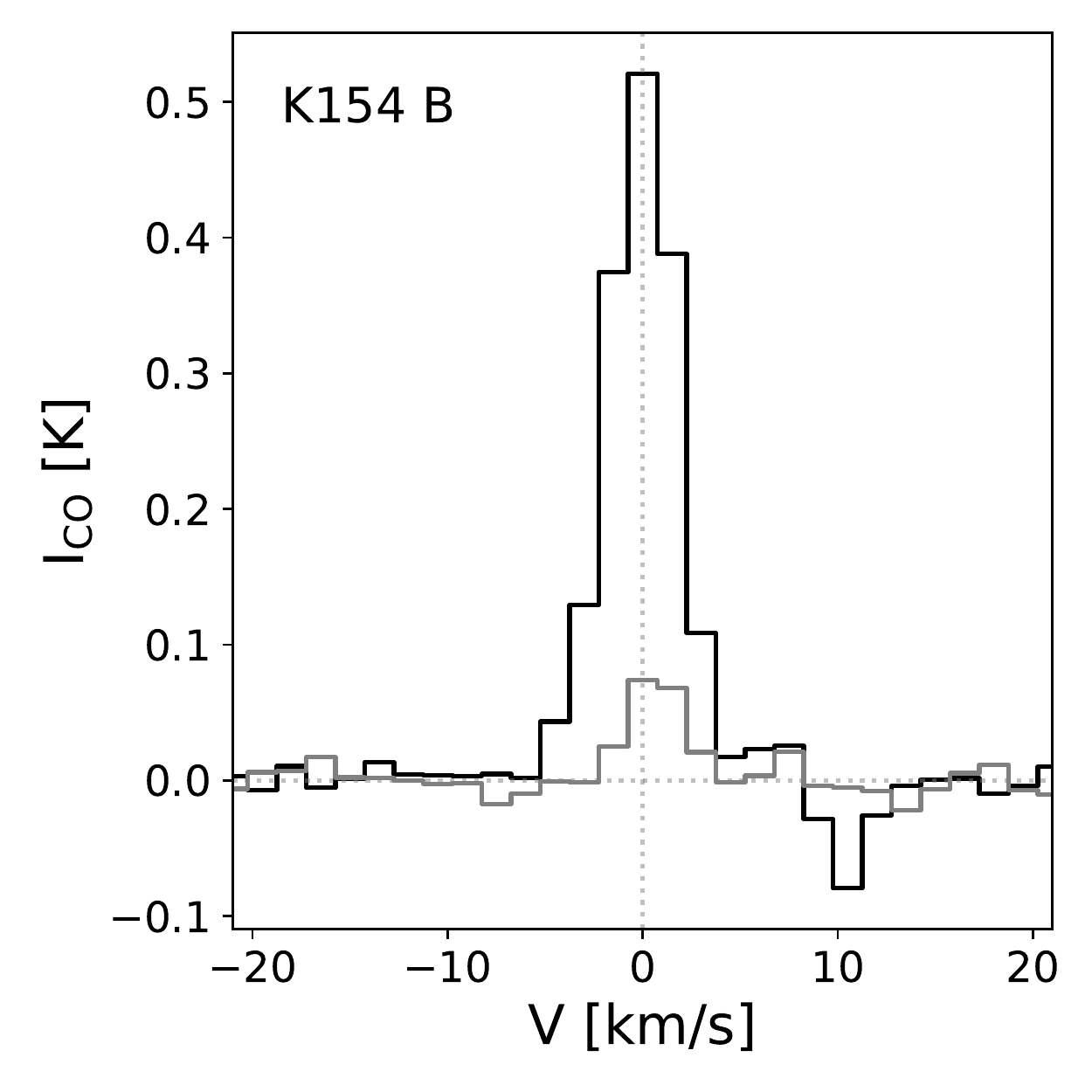}{0.25\textwidth}{}
          }
\gridline{\fig{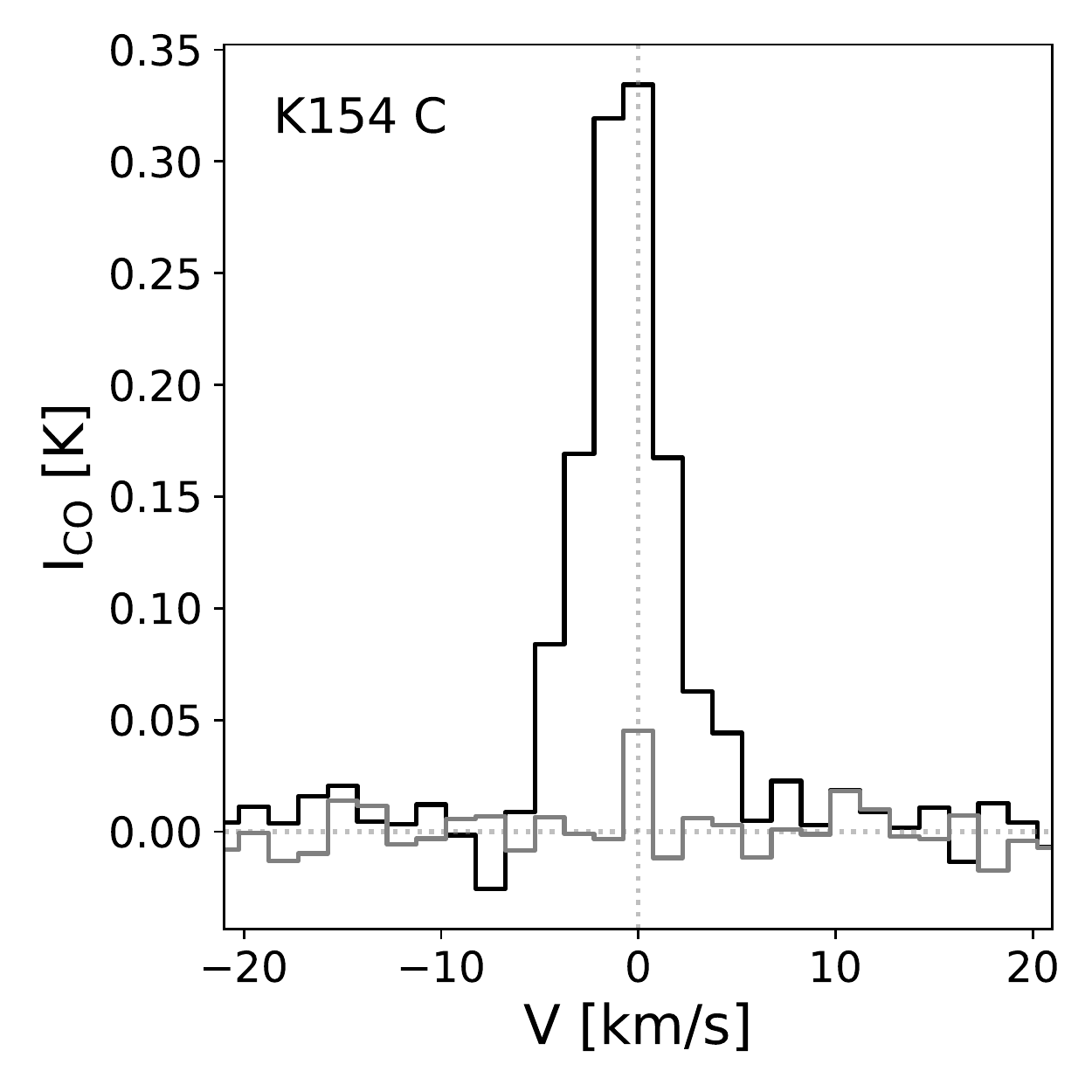}{0.25\textwidth}{}
          \fig{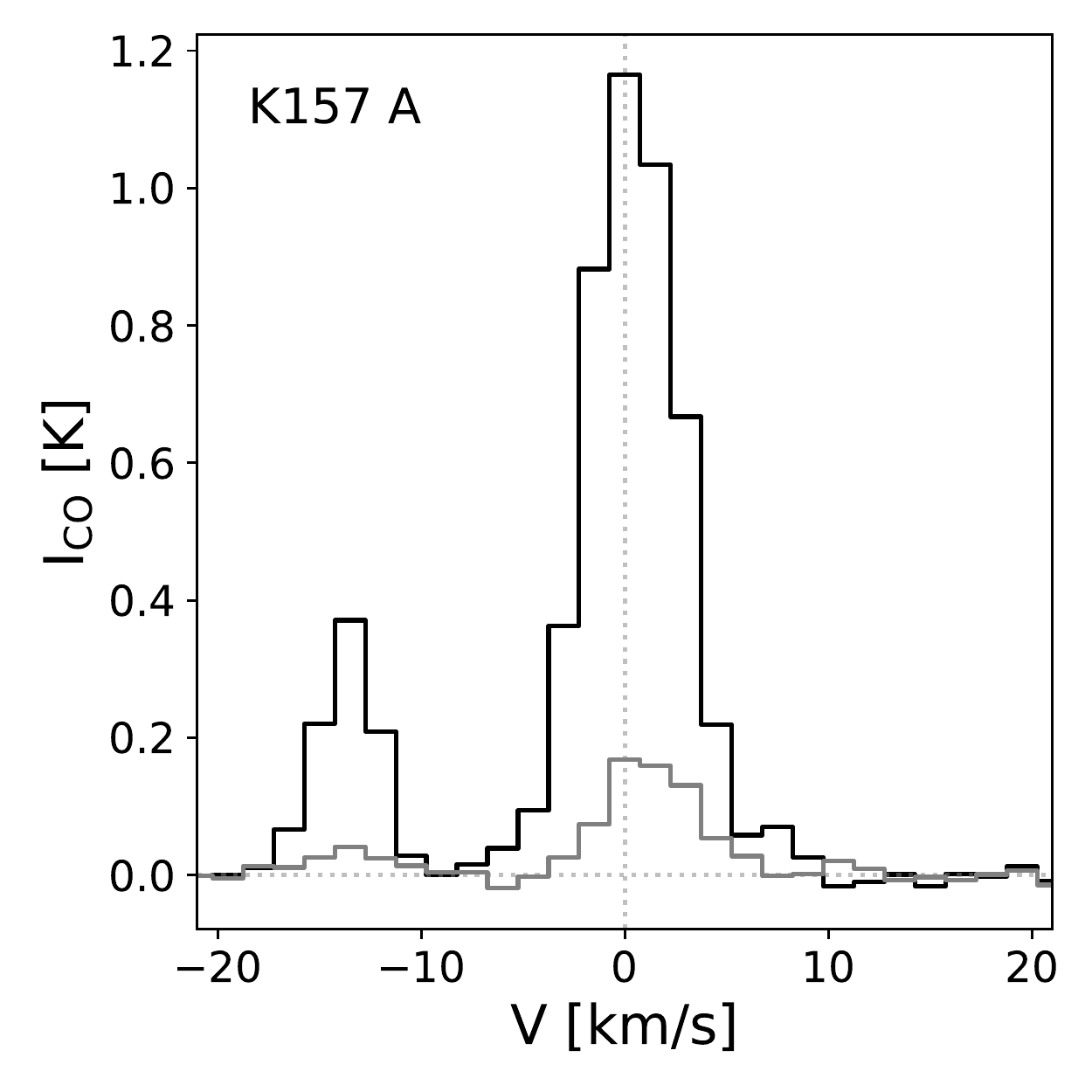}{0.25\textwidth}{}
          \fig{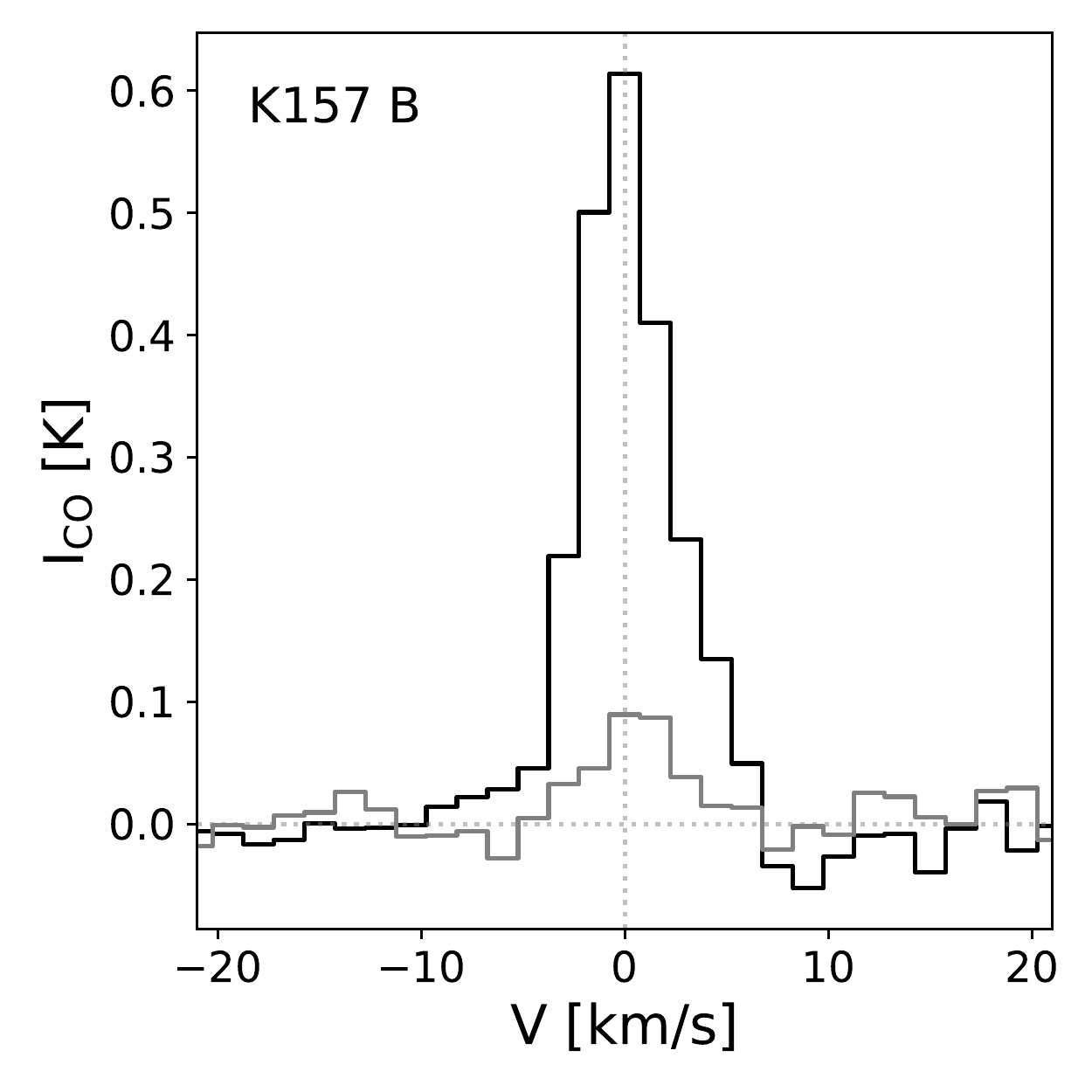}{0.25\textwidth}{}
          \fig{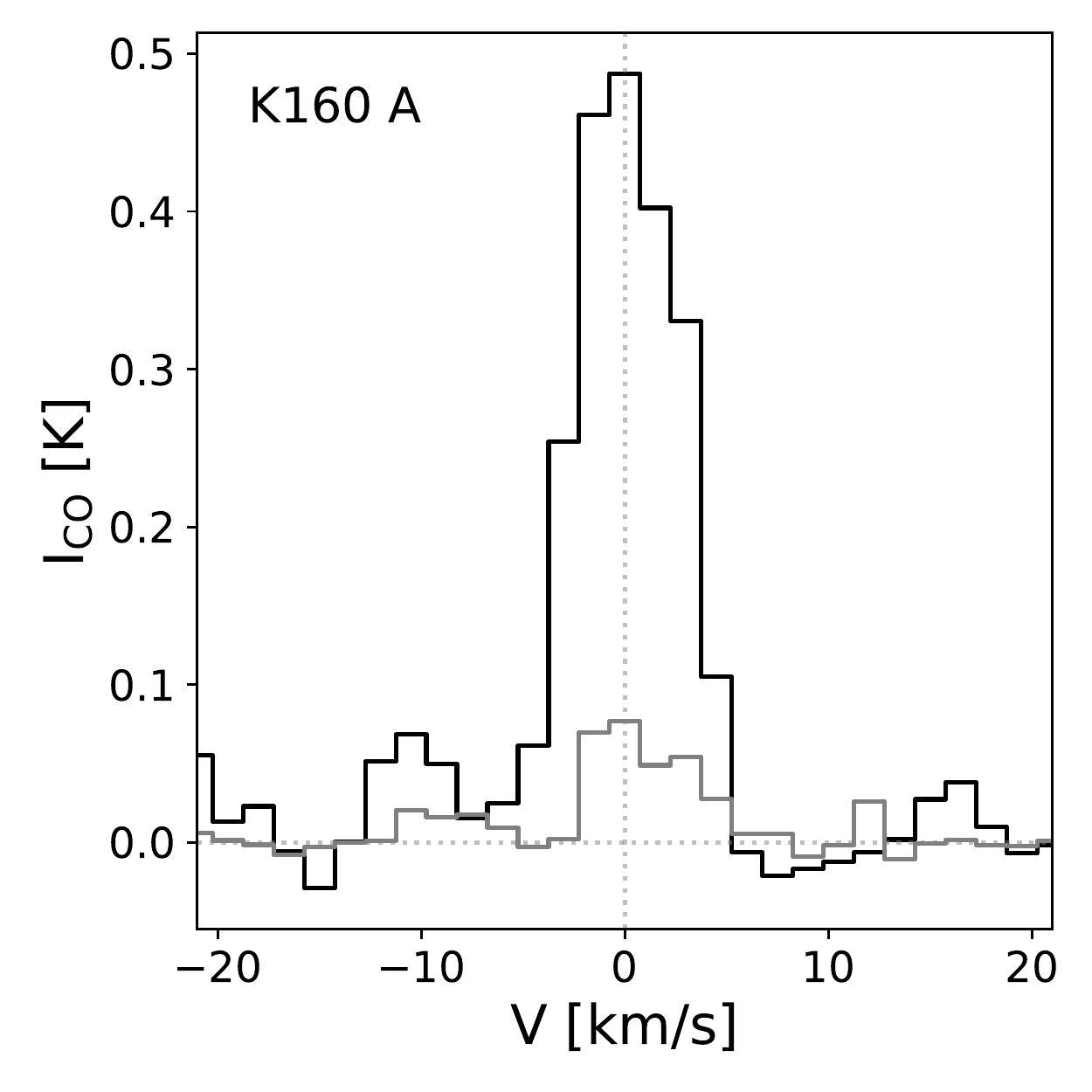}{0.25\textwidth}{}
          }
\figurenum{B.1}
\caption{Spectra of the individual clouds. The black line corresponds to the $^{12}$CO emission line.  The grey line is $^{13}$CO. \label{fig:spectra}}
\end{figure*}

\begin{figure*}
\gridline{\fig{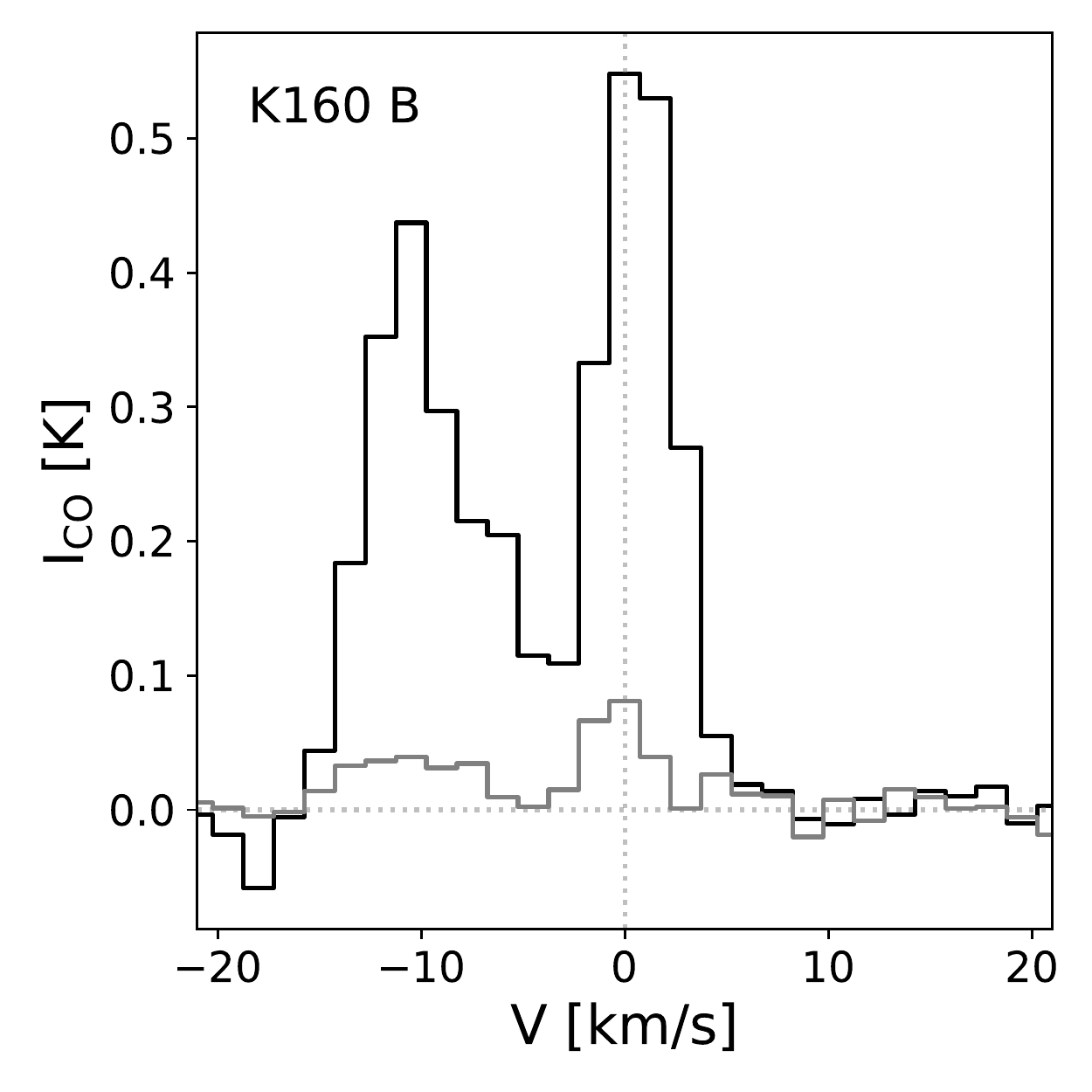}{0.25\textwidth}{}
          \fig{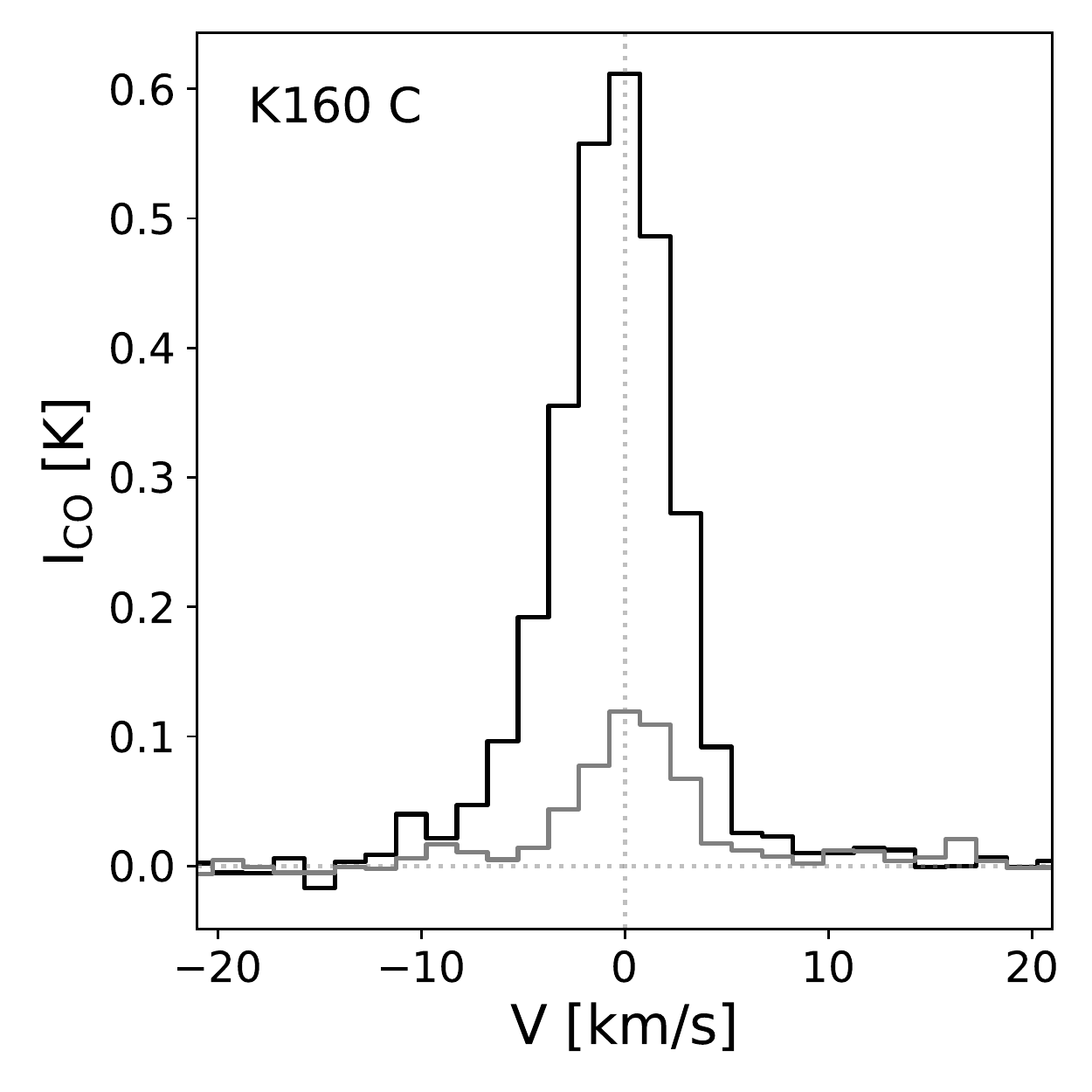}{0.25\textwidth}{}
          \fig{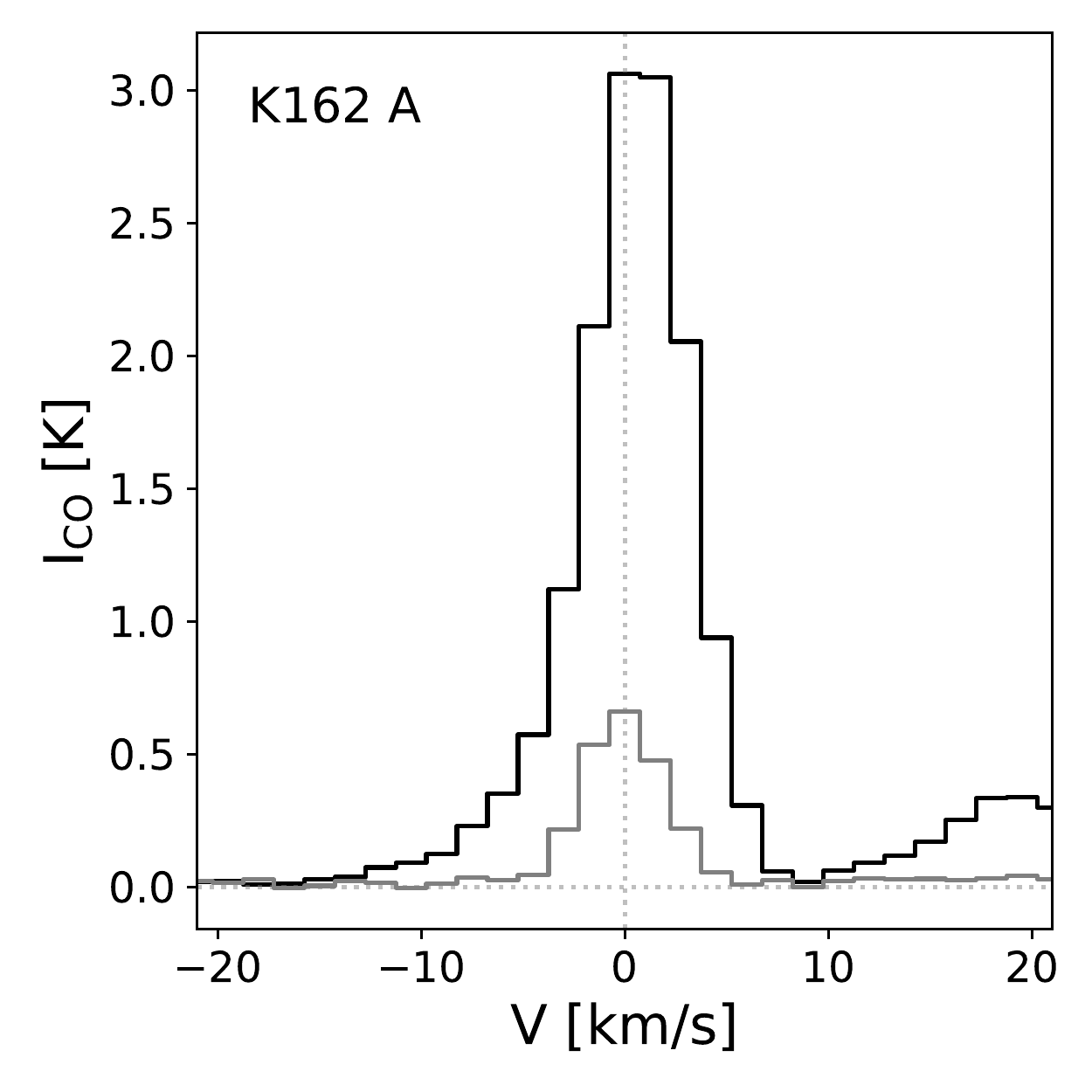}{0.25\textwidth}{}
          \fig{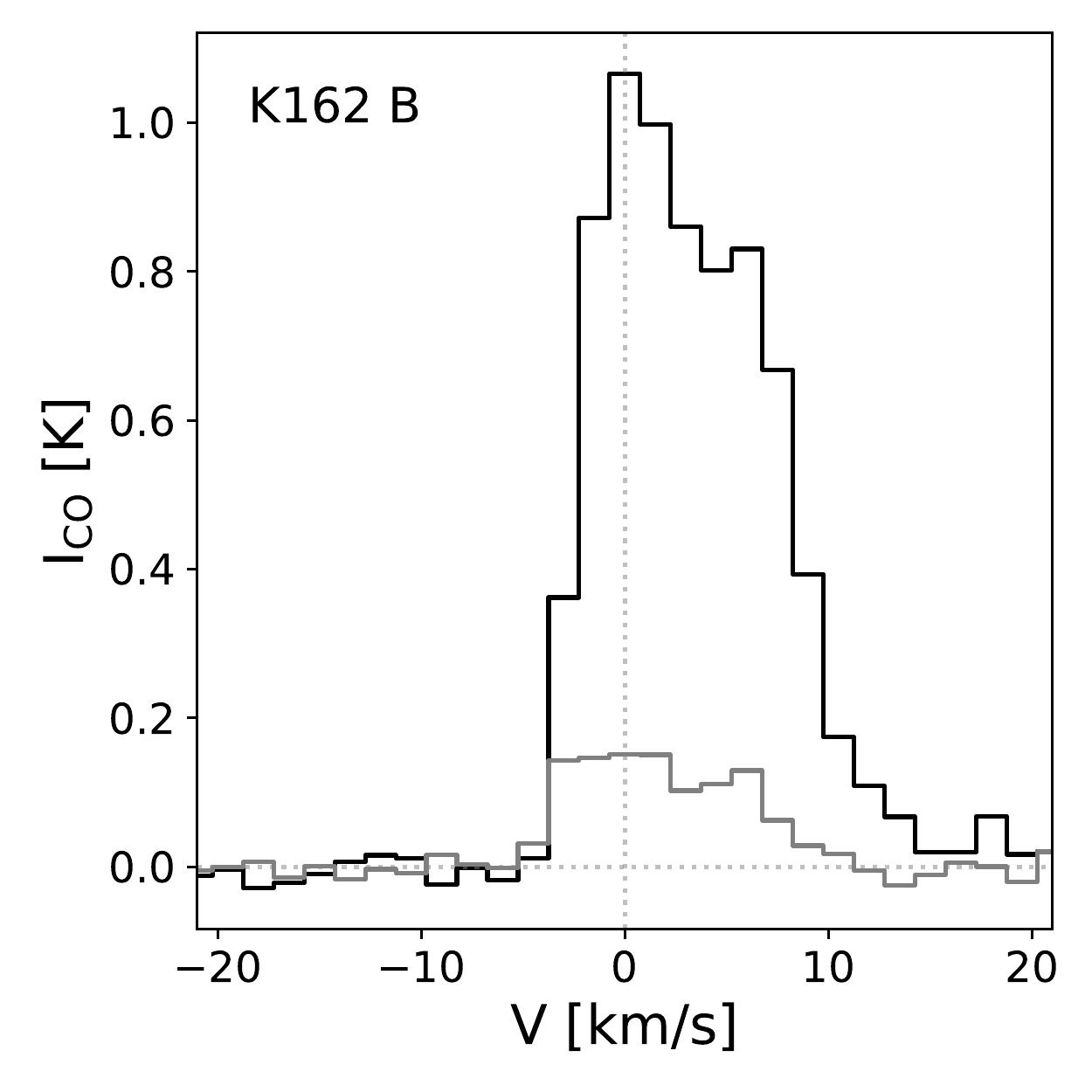}{0.25\textwidth}{}
    }
\gridline{\fig{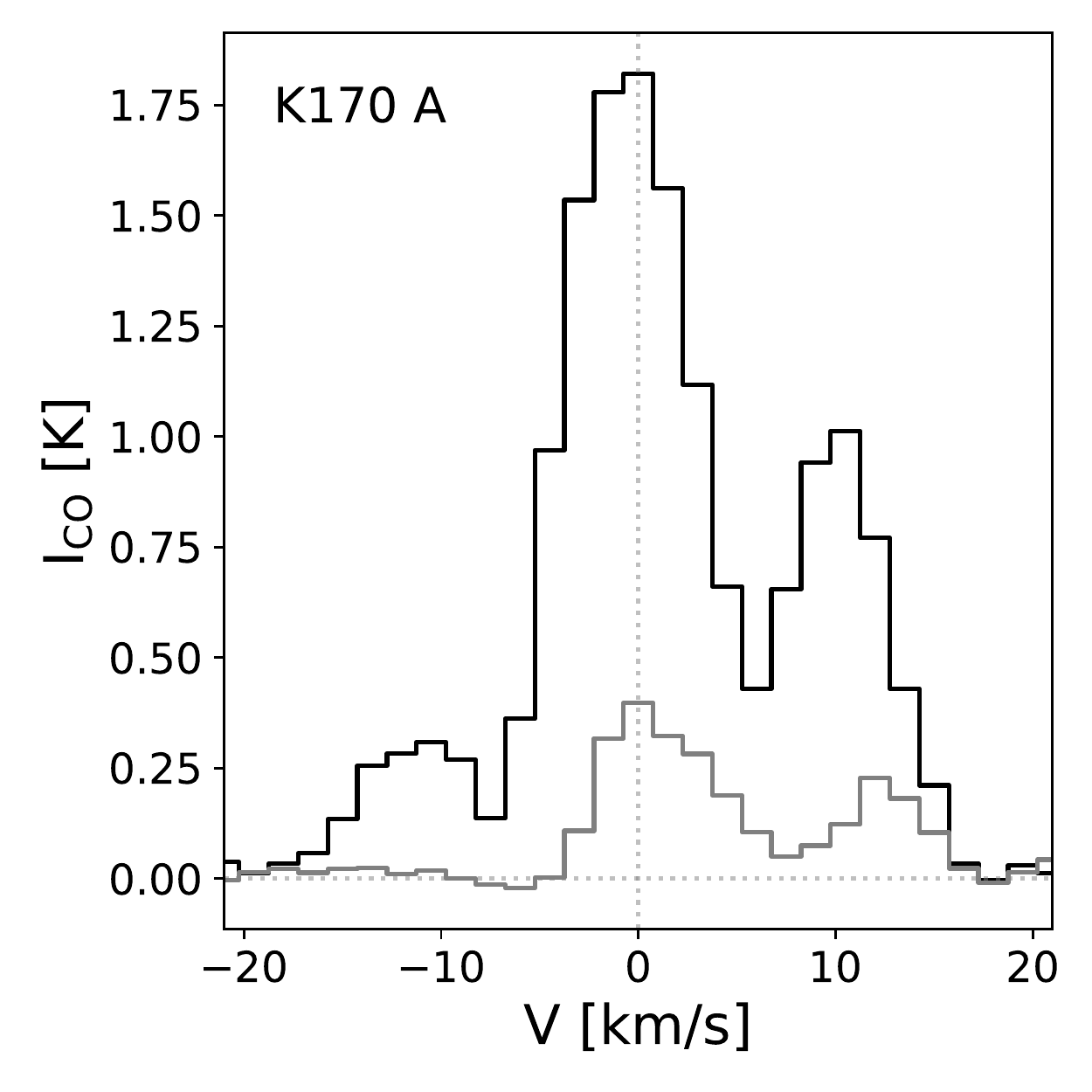}{0.25\textwidth}{}
          \fig{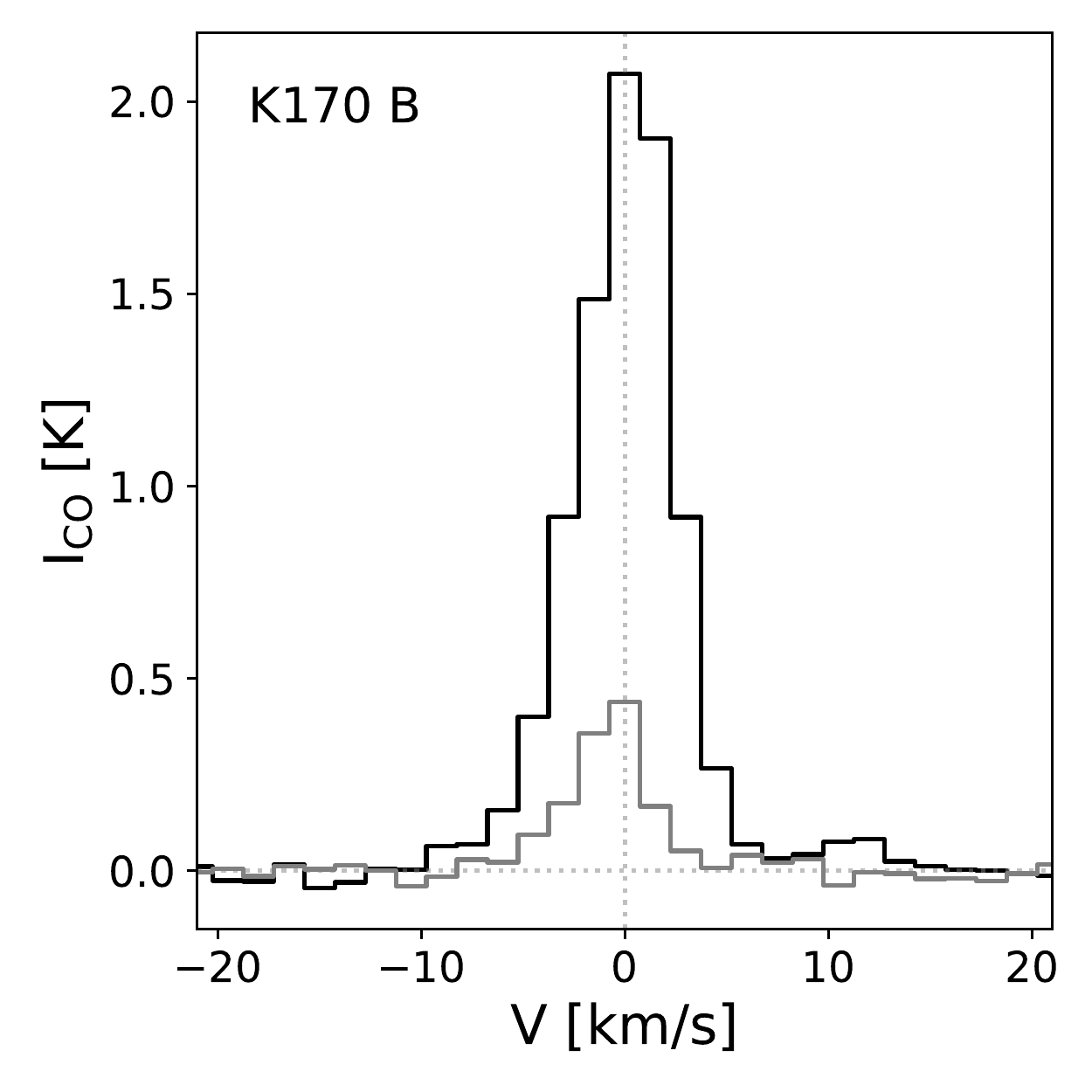}{0.25\textwidth}{}
          \fig{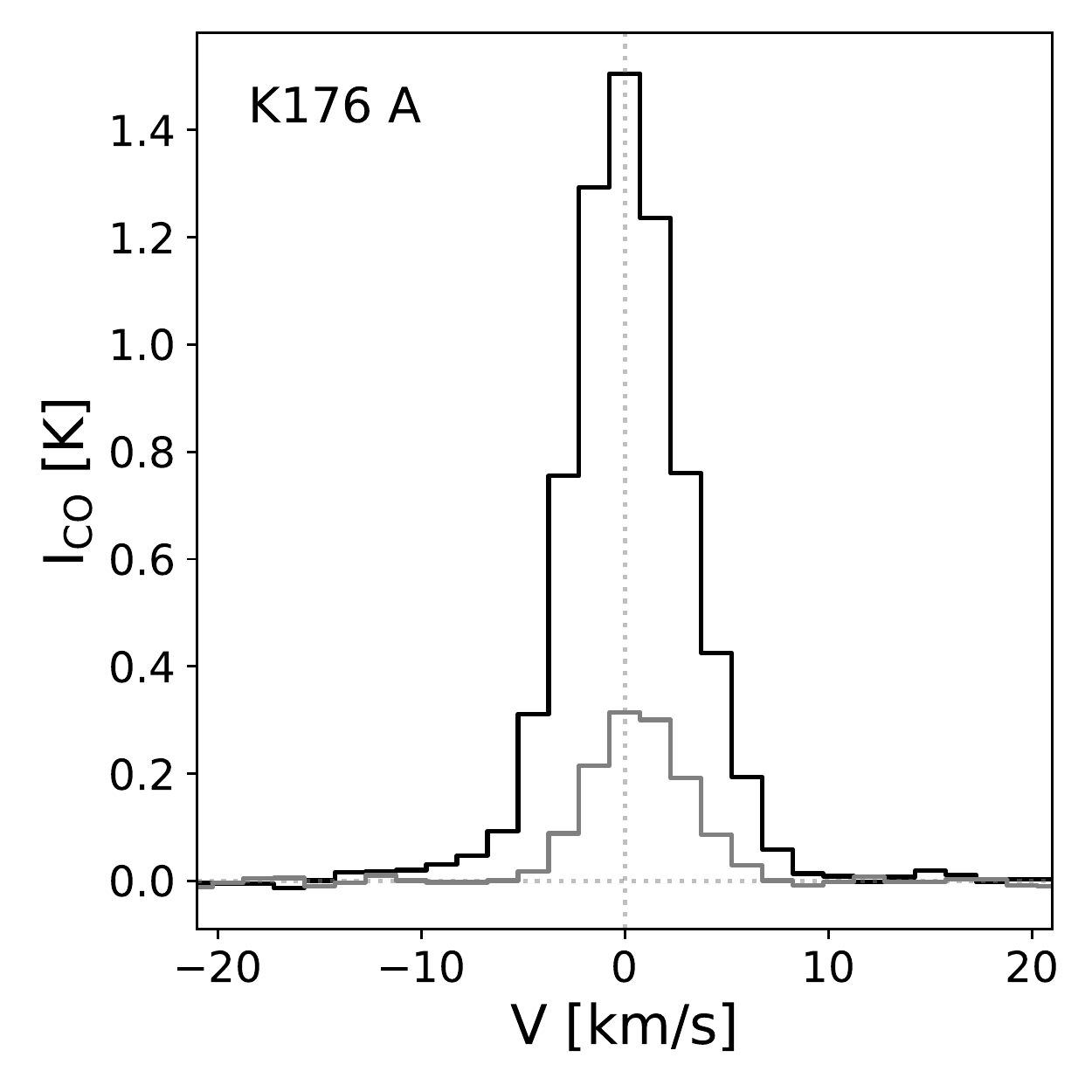}{0.25\textwidth}{}
          \fig{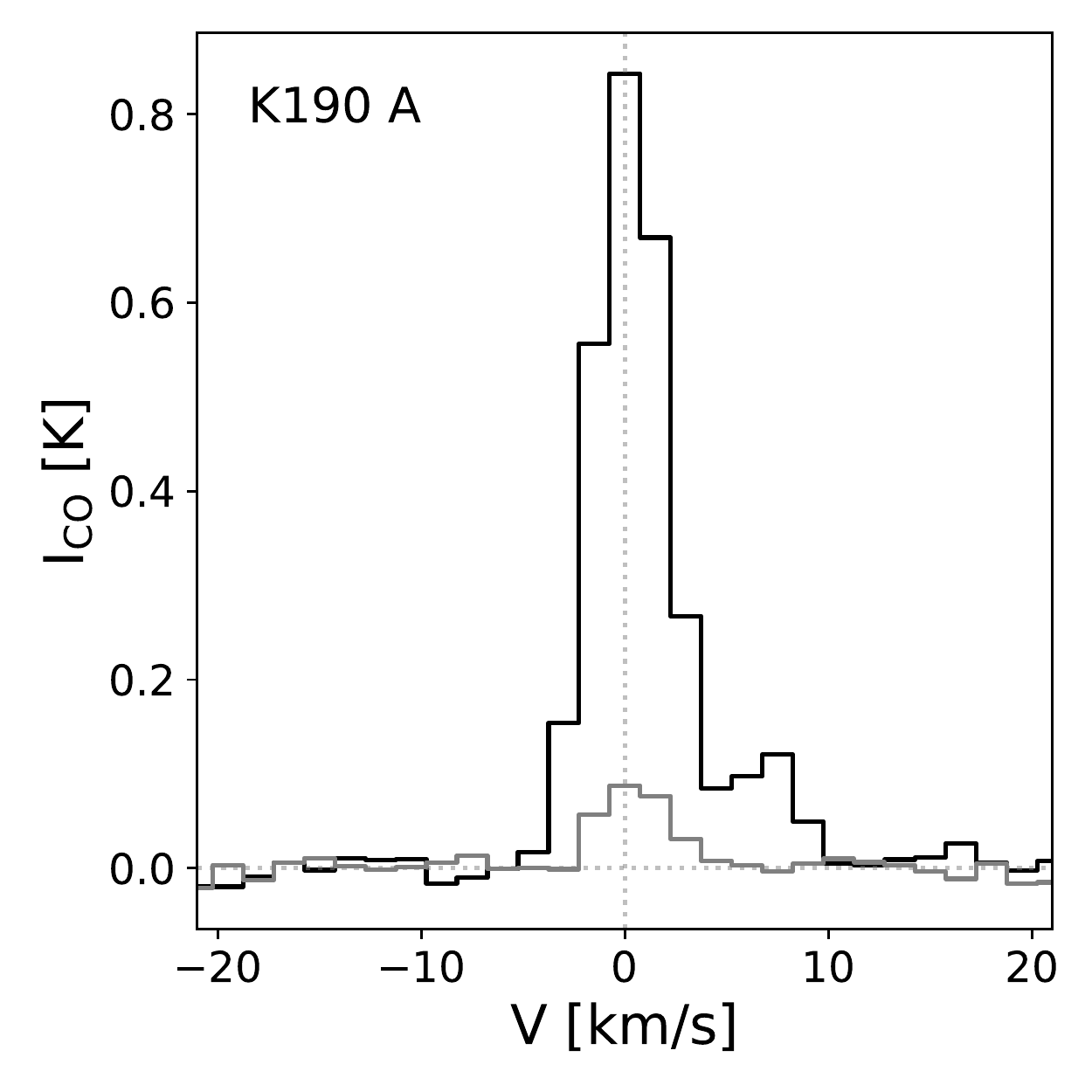}{0.25\textwidth}{}
          }
\gridline{
          \fig{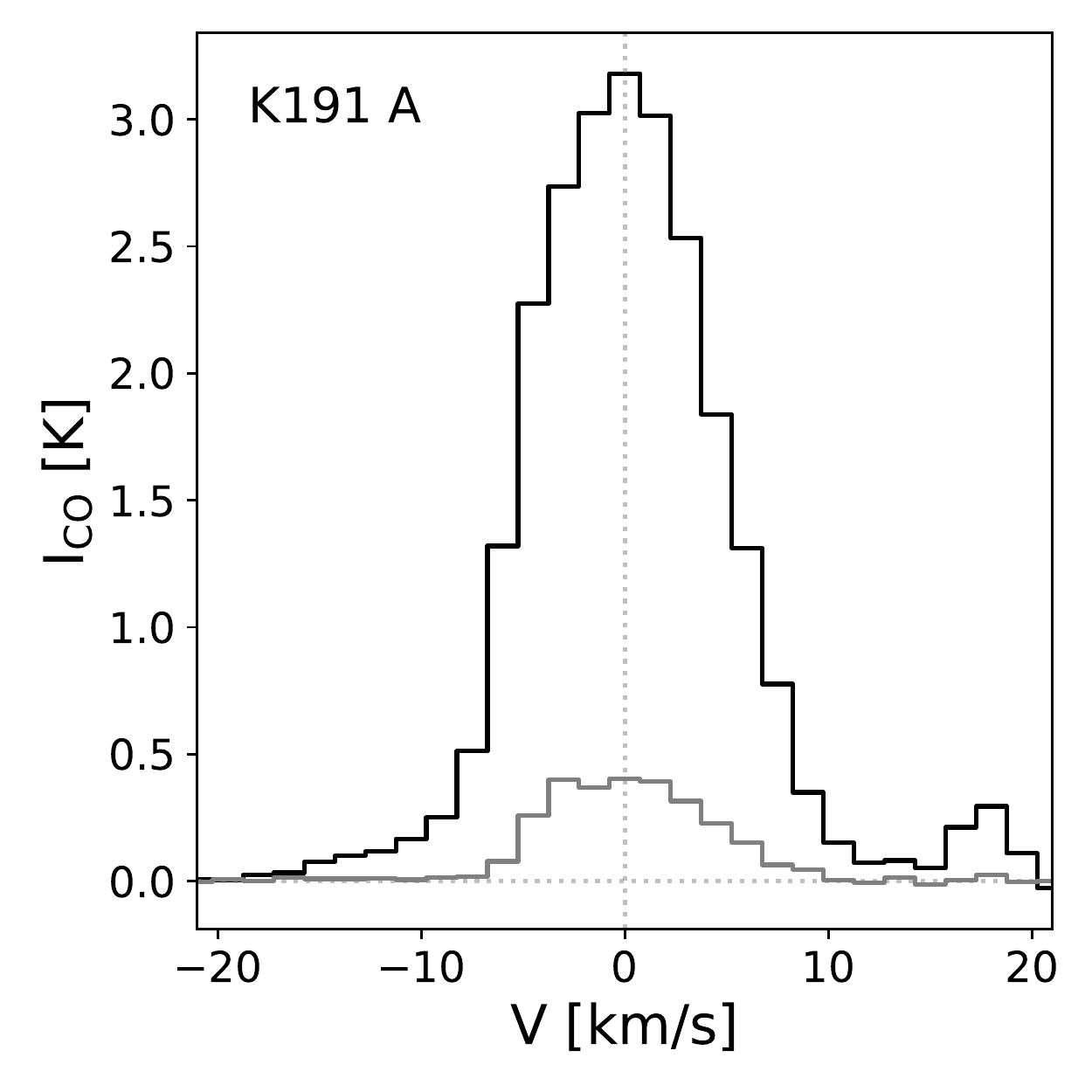}{0.25\textwidth}{}
          \fig{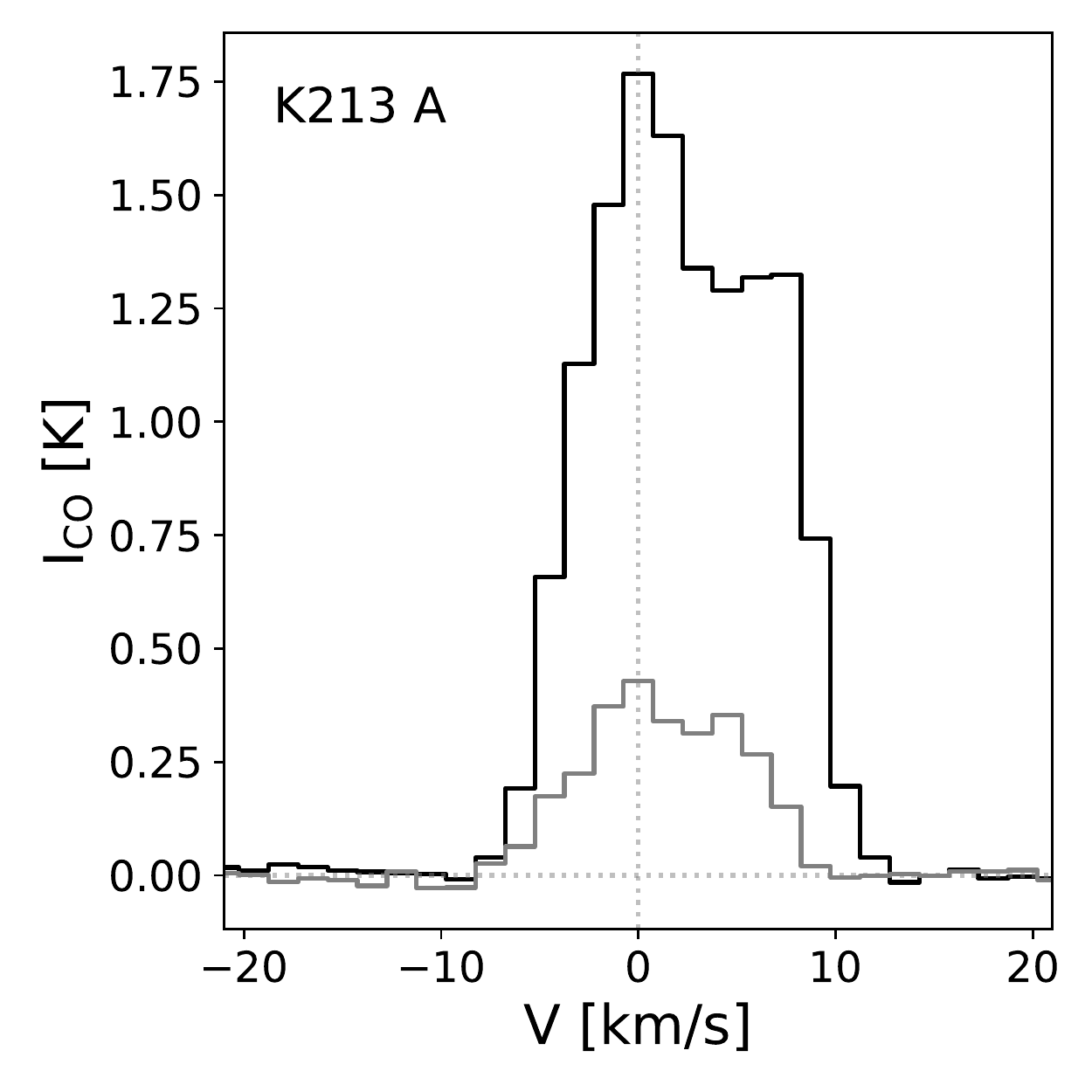}{0.25\textwidth}{}
          \fig{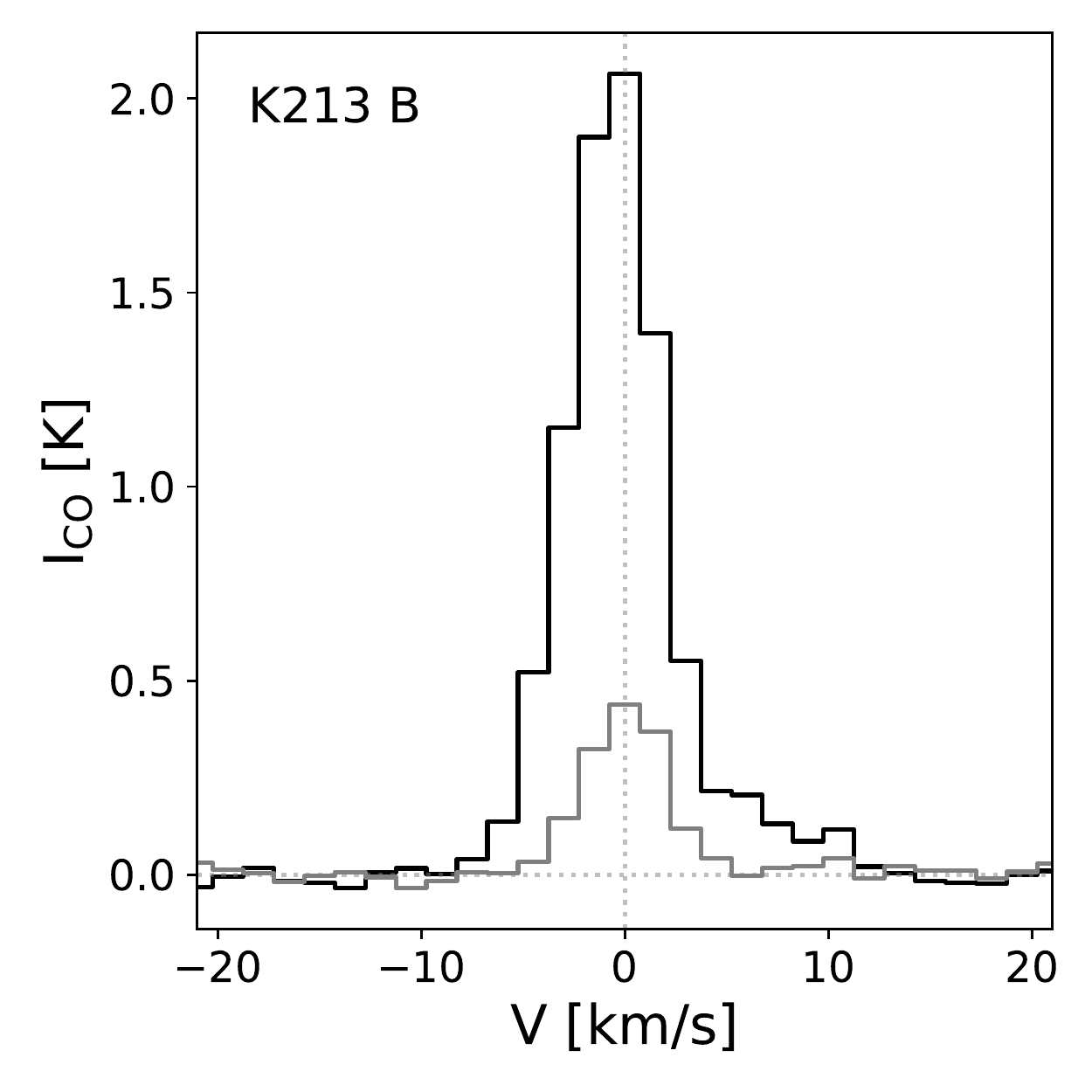}{0.25\textwidth}{}
          \fig{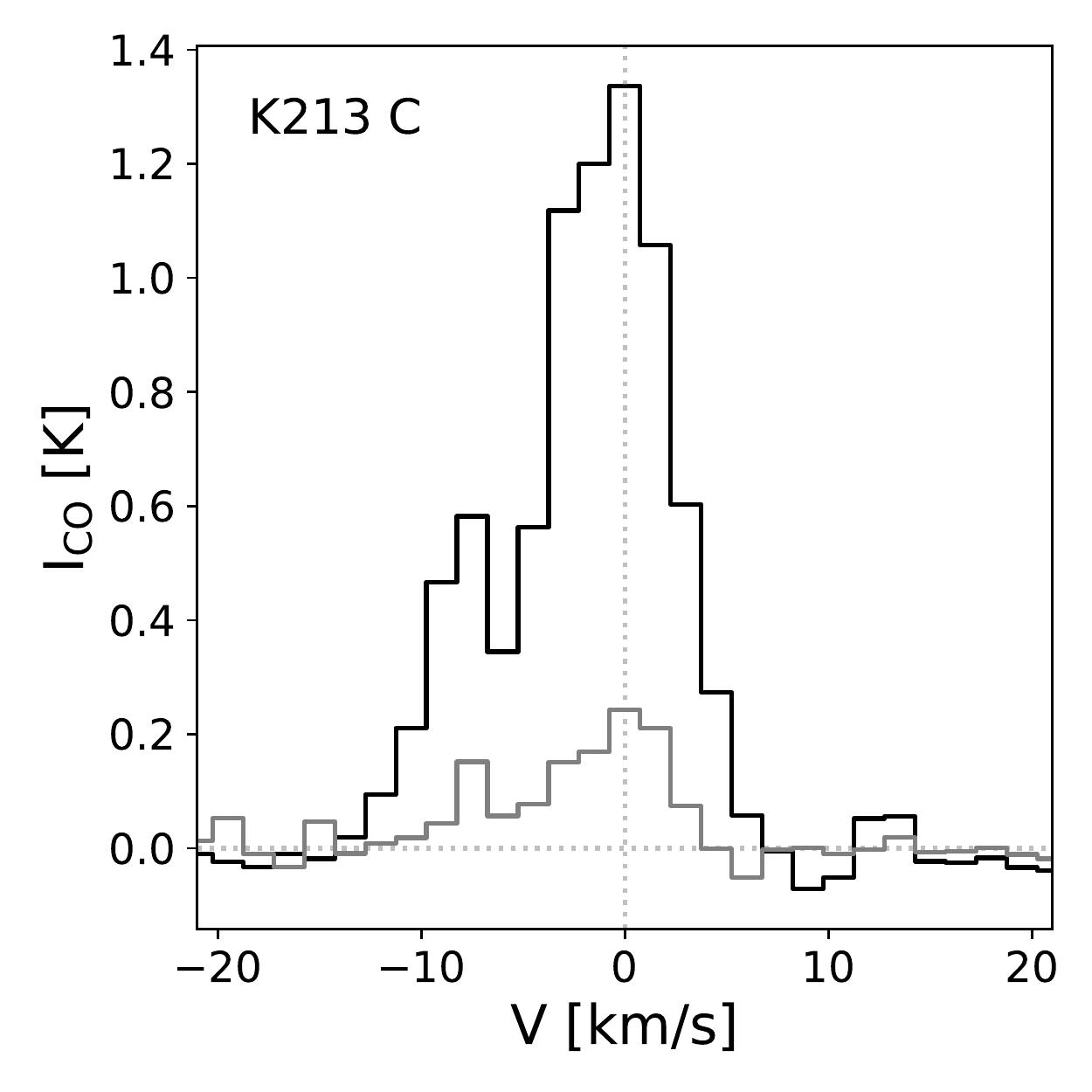}{0.25\textwidth}{}
          }
\gridline{\fig{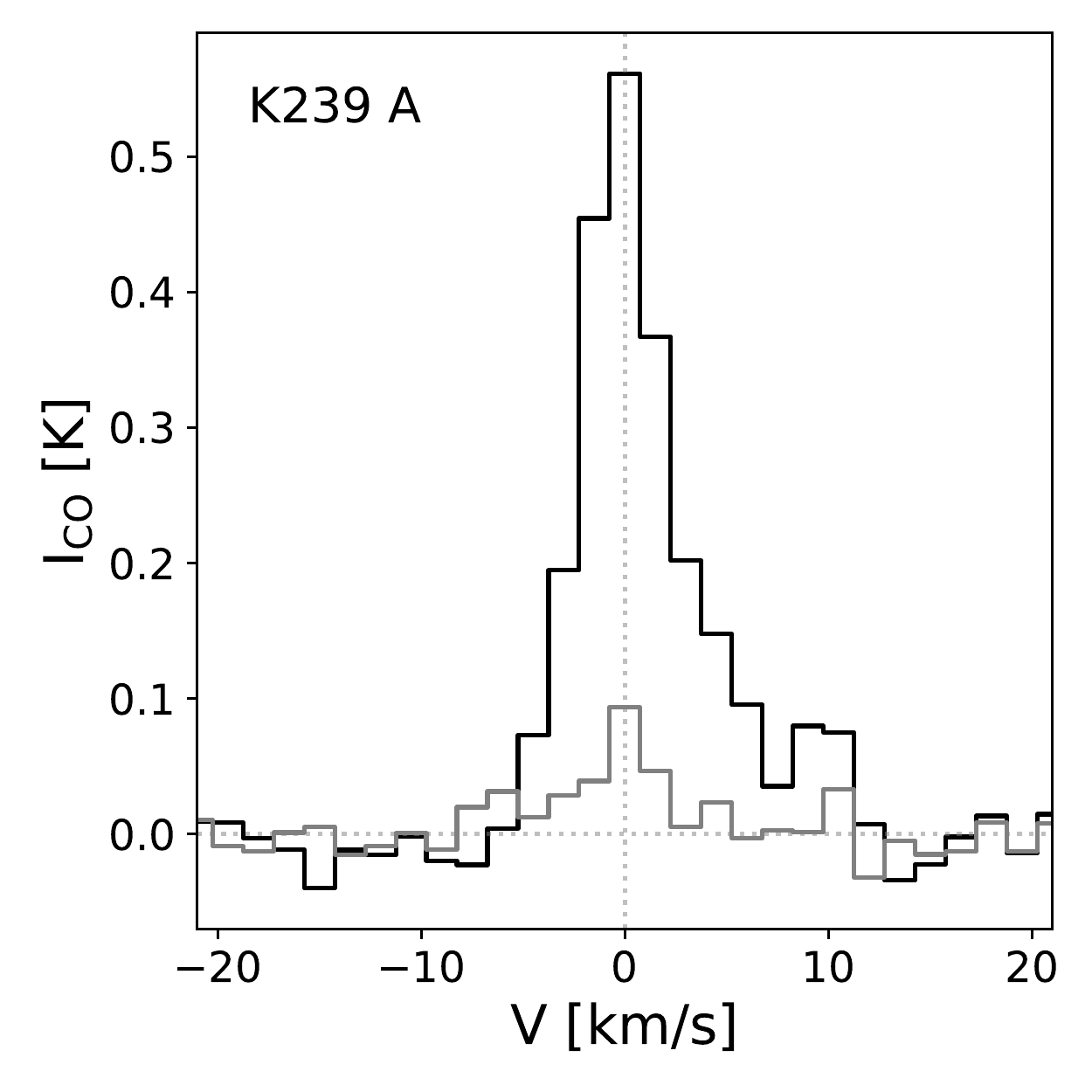}{0.25\textwidth}{}
          \fig{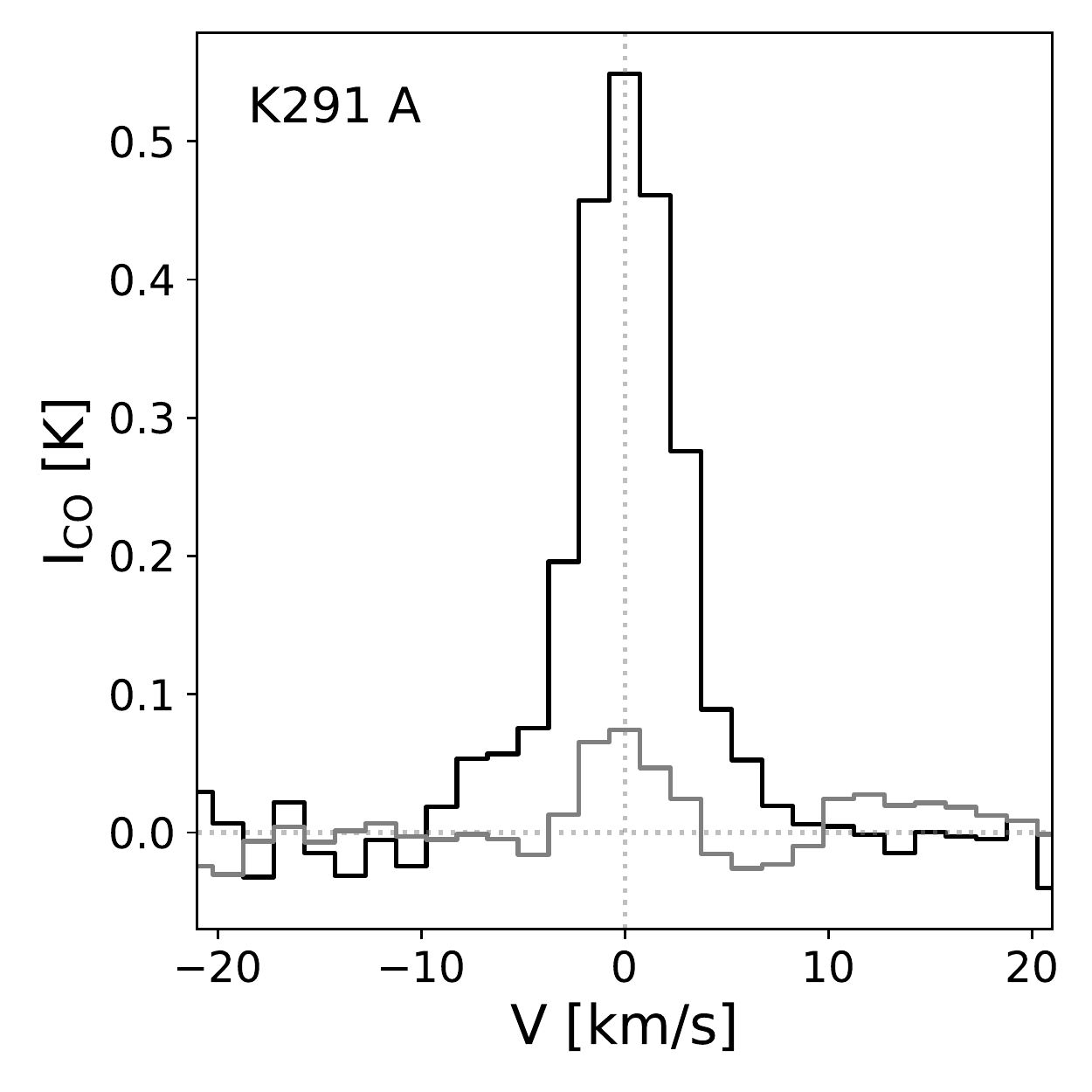}{0.25\textwidth}{}
          \fig{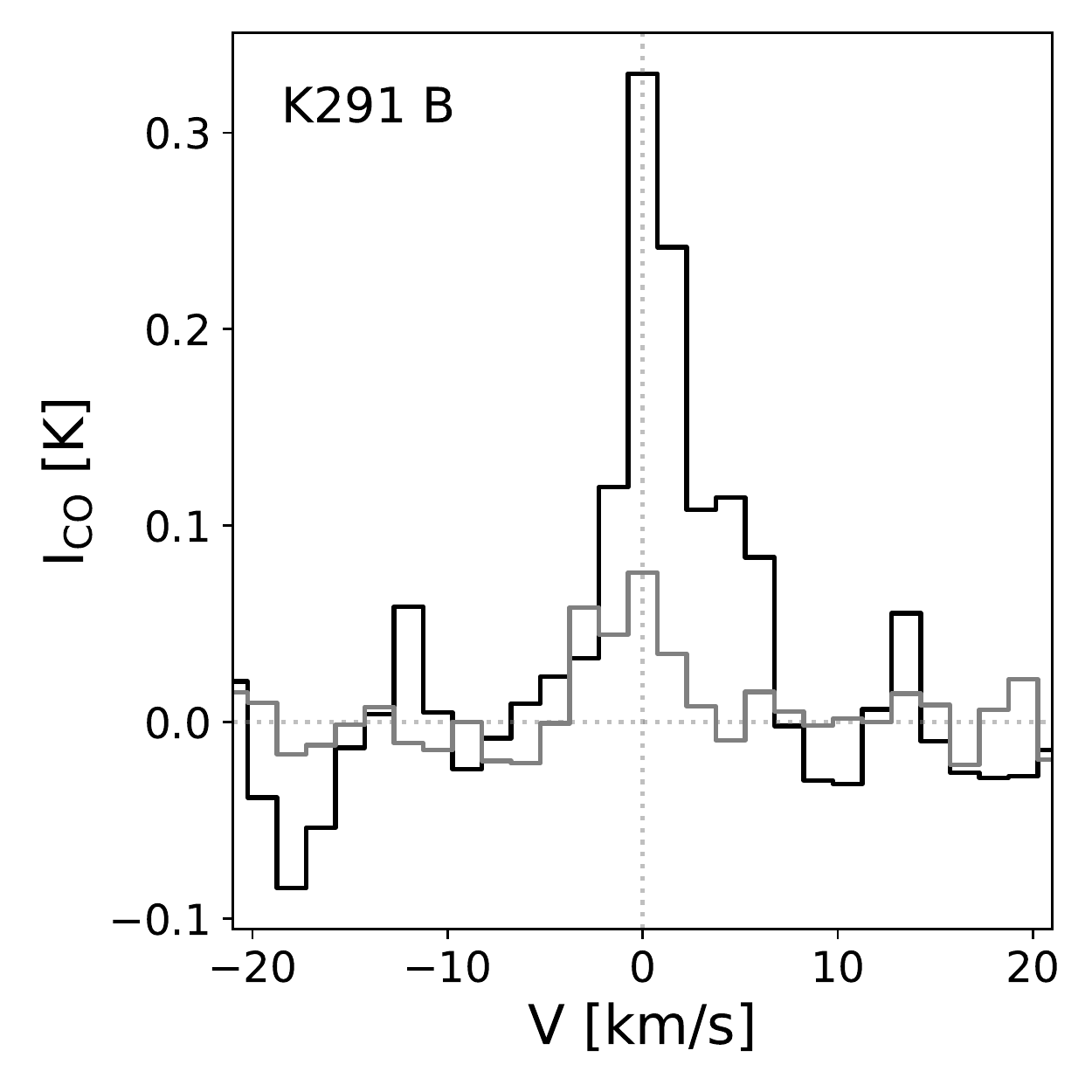}{0.25\textwidth}{}
          \fig{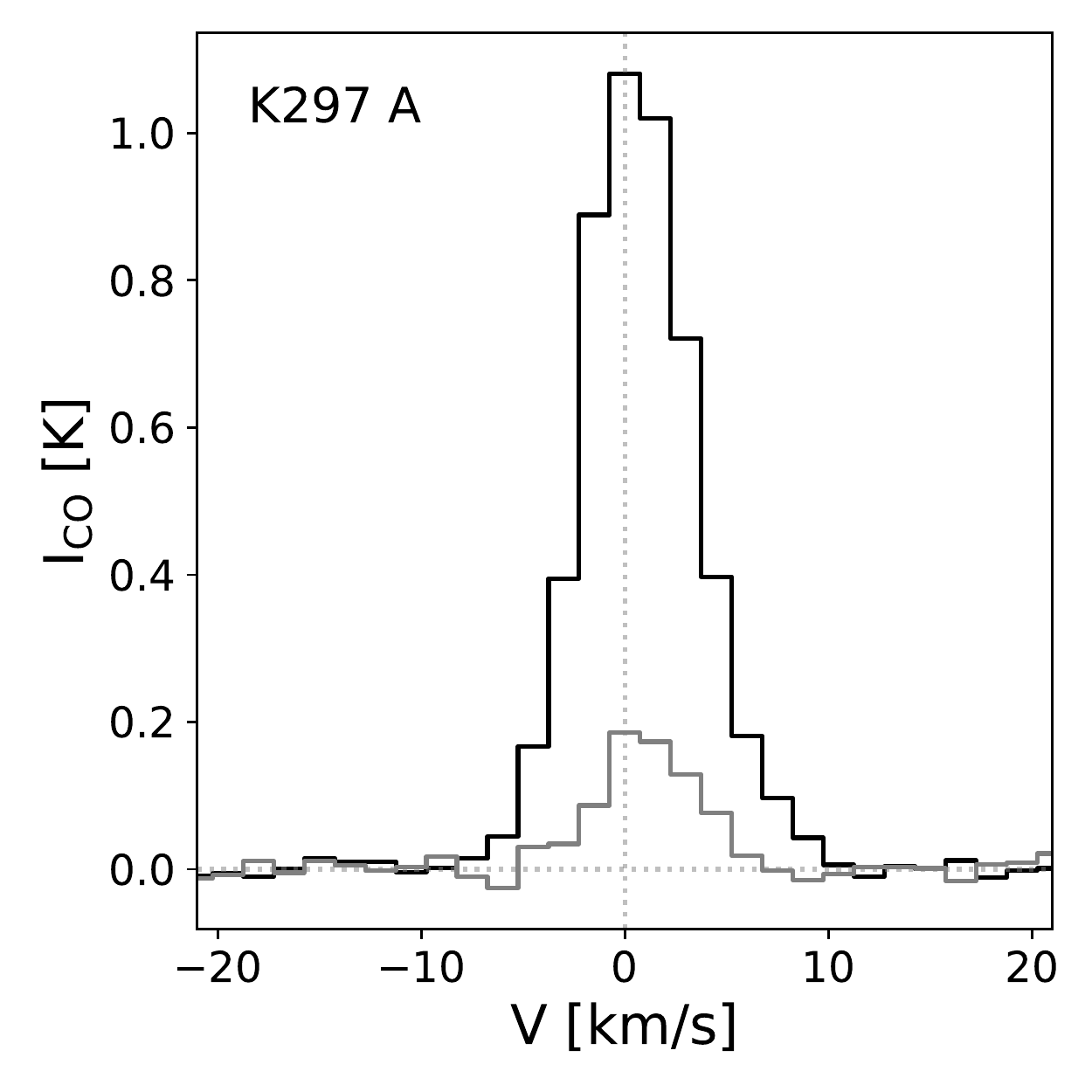}{0.25\textwidth}{}
          }
\figurenum{B.1}
\caption{continued}
\end{figure*}



\end{document}